\documentclass[%
 aip,
 amsmath,amssymb,
 reprint,%
]{revtex4-1}

\usepackage{wrapfig,lipsum,booktabs}
\usepackage{graphicx}% Include figure files
\graphicspath{{figures}}

\usepackage{dcolumn}% Align table columns on decimal point
\usepackage{bm}% bold math
\usepackage[utf8]{inputenc}
\usepackage[T1]{fontenc}
\usepackage{mathptmx}
\usepackage{etoolbox}

\makeatletter
\def\@email#1#2{%
 \endgroup
 \patchcmd{\titleblock@produce}
 {\frontmatter@RRAPformat}
 {\frontmatter@RRAPformat{\produce@RRAP{*#1\href{mailto:#2}{#2}}}\frontmatter@RRAPformat}
 {}{}
}%
\makeatother

\usepackage{epsfig}
\usepackage{tikz}
\usepackage{pgfplots}
\usepgfplotslibrary{groupplots,dateplot}
\usetikzlibrary{patterns,shapes.arrows}
\pgfplotsset{compat=newest}
\newlength\figH
\newlength\figW
\usepackage{xpatch}
\usepackage[percent]{overpic}
\usepackage{pict2e}
\usepackage{amssymb}
\usepackage{gensymb}
\usepackage{siunitx}

\usepackage[T1]{fontenc}
\usepackage[english]{babel}
\usepackage{scrhack}
\usepackage[utf8]{inputenc}

\usepackage[hidelinks]{hyperref}
\hypersetup{
  colorlinks   = true, %Colours links instead of ugly boxes
  urlcolor     = black, %Colour for external hyperlinks
  linkcolor    = blue, %Colour of internal links
  citecolor   = blue %Colour of citations
}

\begin{document}
\preprint{AIP/123-QED}

%%%%%%%%%%%%%%%%%%%%%%%%%%%%%%%%%%%%%%

\title{Characterisation of a self-similar turbulent boundary layer near separation measured using high-spatial-resolution two-component -- two-dimensional particle image velocimetry}

\author{Muhammad Shehzad}
\email{Muhammad.shehzad@monash.edu}
\author{Bihai Sun}
\author{Callum Atkinson}
\author{Julio Soria}
\affiliation{Laboratory for Turbulence Research in Aerospace \& Combustion (LTRAC), Department of Mechanical and Aerospace Engineering, Monash University, Clayton, 3800, Victoria, Australia}

\date{\today}% It is always \today, today,
  % but any date may be explicitly specified

%%%%%%%%%%%%%%%%%%%%%%%%%%%%%%%%%%

\begin{abstract}

Adverse pressure gradient (APG) turbulent boundary layer (TBL) flow is an important flow among the family of wall-bounded turbulent flows because it is encountered in numerous engineering applications including diffusers, turbine blades, aircraft wings, wind turbine blades as well as marine vessel flow, and has the potential to result in flow separation resulting in significant losses such as loss of lift and increase in pressure drag, which is typically accompanied with highly undesirable unsteady surface pressure forces. In order to minimise and/or control the flow separation due to APG environments, It is important to understand the physics of the APG-TBL flow to allow appropriate designs of efficient aerodynamic surfaces and/or the development and application of control methodologies to delay or prevent separation. This paper report on the design of an experiment to establish an experimental self-similar APG-TBL at the verge of separation in order to characterise its statistical structure and the contribution of the Reynolds stresses to the wall shear stress via quadrant analysis decomposition in conjunction with the Renard \& Deck (RD) decomposition \citep{renard2016theoretical} of the mean skin friction. In this study high-spatial-resolution (HSR) two-component -- two-dimensional (2C-2D) particle image velocimetry (PIV) is employed to measure the self-similar APG-TBL flow. The measurements of the self-similar APG-TBL at the verge of separation, which is referred to as `{\em strong APG-TBL}' is compared with similar HSR 2C-2D PIV measurements of a TBL under zero and mild APG to characterise the effect of the APG on the TBL. In the strong APG-TBL, the outer scaled profiles of the first- and second-order statistics, as well as the defect velocity exhibit self-similarity over the measured streamwise domain of $3.3\delta$. The quadrant decomposition of the velocity fluctuations shows that the formation of the outer peak in the Reynolds shear stress profile under APG is due to the energisation of sweep motions of high-momentum fluid in the outer region of the boundary layer. The RD decomposition of the mean skin friction shows that the skin friction has larger contributions from ejections than sweeps. It is also observed that the contribution of the turbulent kinetic energy in the skin friction generation is enhanced with increasing APG. This is due to the emergence of the outer peaks with a correspondingly diminishing of the inner peaks in the turbulence production profiles. The skewness and flatness profiles show that the zero skewness and minimum flatness locations of the streamwise and wall-normal velocity fluctuations collapse with the location of the maximum Reynolds streamwise stress in the TBLs irrespective of the pressure gradient imposed on the TBL.

%\textbf{keywords:} Turbulent Boundary Layer, Adverse Pressure Gradient, Zero Pressure Gradient, High-spatial-resolution, Particle Image Velocimetry

\end{abstract}

\maketitle

%#===================================================================#
\section{Introduction} \label{sec:intro}

Most wall-bounded flows found in engineering, transportation, energy production, and other industrial applications, as well as in nature are of a high Reynolds number turbulent nature, such as the the Reynolds number of the turbulent flow over the wing of a heavy transport aircraft is of the order of $10^8$, which to this day are unattainable by direct numerical simulations, and must rely on experimental investigation for understanding via empirical measurements \citep{talamelli2009ciclope}.  An important turbulent boundary layer (TBL) encountered in numerous engineering applications including diffusers, turbine blades, aircraft wings, wind turbine blades as well as marine vessel flow is one that develops through stagnation to flow reversal at the wall if it has developed for a sufficient streamwise domain  in an Adverse pressure gradient (APG) environment \citep{Oswatitsch1958,Perry75}, or if the boundary layer is subjected to an abrupt increase in the applied pressure gradient with the detached TBL separating from the surface consists of vortices within the separation bubble \citep{Chong-etal-1998}. Flow separation in TBLs is undesirable, as it may lead to catastrophic consequences and reduced performance due to a significantly altered mean pressure distribution on the surface \citep{kitsios2016direct}, as well as large pressure fluctuations \cite{FRICKE1971113} that can be an important source for fatigue failure. Therefore, for such flows  subjected to an APG environment to remain attached to the curved surface is crucial to the efficiency of many engineering systems \citep{kitsios2017direct}. Much research has been conducted on the design of aerodynamic and hydrodynamic surfaces and the controls to prevent or delay flow separation, such as the boundary layer trip devices on the lower surfaces of a glider and the vortex generators on aircraft wings. However, many of the solutions to boundary layer separation due to APG environments are based on trial-and-error with limited detailed measurements that have shed little light on the fluid physics of the APG-TBL.  In order to enable the design of more efficient active or passive techniques to prevent, delay or reduce the flow separation, significant new understanding based on well defined experiments measured with diagnostics tools that elucidated the statistical and spatial structure of the APG-TBL with sufficient spatial resolutions, especially in the near-wall region of an APG-TBL at the verge of separation. 

Few numerical and experimental investigations have studied the spatial structures, specifically the structure of coherent motions, in an APG-TBL near separation \citep{skaare1994turbulent,krogstad1995influence,Chong-etal-1998,cheng2015large,drozdz2016study,kitsios2017direct,drozdz2017experimental,eich2020large}. 
The non-equilibrium turbulent boundary layer near separation, whose turbulent structure depended on the upstream history conditions was investigated using hot-wire anemometry by \citet{drozdz2017experimental}. Their study showed that, for the same APG, the near-wall momentum increases with increasing Reynolds number at incipient detachment with an observed increase in the turbulent kinetic energy. It was found that the outer scaling laws are no longer applicable in this region. The outer inflection point in the mean streamwise velocity profile was found to corresponds to the outer peak location in the Reynolds stresses and the zero-crossing values of the skewness. The wall-normal position of this inflection point was found to vary with upstream flow history conditions. The inflection point is the result of interaction between the large and small scales, which causes enhanced convection velocities of the near-wall small scales. At the incipient detachment, the turbulence activity drops near the wall and is maximum in the outer region \citep{drozdz2016study}. 

\citet{eich2020large} showed that the dynamics of the separation line are strongly modulated in space and time by the low-frequency large-scale motions (LSMs). Using conditional sampling, they observed that high-momentum LSMs are able to shift the separation point downstream, whereas low-momentum LSMs resulted in the opposite. 

\citet{drozdz2021effect} studied a wide range of Reynolds numbers, $Re_{\delta_2}$ ranging from 4,900 to 14,600, in an APG-TBL approaching separation using hot-wire anemometry. This study found that the separation point is shifted downstream with increasing Reynolds numbers due to an increase in the near-wall convection velocity. In addition it was also observed that with increasing Reynolds number the interactions between the large and small scales, {\em i.e.} the amplitude modulation, is enhanced.

Streamwise vortices in wall-bounded flows are known transfer high momentum from the outer region to the near-wall region and visa versa with regards to low momentum \citep{10.1017/s0022112097008562}. This mechanism is used in the application of vortex generators to control APG-TBL flow separation where the vortex generators reduce separation by redirecting the reversed flow towards the mean flow direction \citep{logdberg2006vortex}. 
%\citet{logdberg2009streamwise} investigated the streamwise evolution of the longitudinal vortices in a turbulent boundary layer using smoke visualisation and three-component hot-wire measurements. The authors proposed a vortex-path model predicting the streamwise evolution of longitudinal vortices downstream of the vortex generators. 
\citet{chan2022investigation} used micro-vortex generators to produce large-scale coutour-rotating vortex pairs, which contribute to the streaks by transferring the high-momentum fluids from the outer region to the near-wall region. This produces high-speed regions with their centres at the vortex pairs and low-speed regions on the outer sides of the vortex pairs. Local skin friction decreases by up to 15\% in the low-speed region, whereas an increase in the skin friction is observed over the high-speed region. If the contributions of high- and low-momentum motions to skin friction can be accurately predicted, then the design and implementation of such flow-control methodologies could be more efficient. 

% The study of these flows is important to draw conclusions about flow physics under certain environmental conditions. In wall-bounded turbulence, the determination of the mean velocity distribution in the overlap region is a first step in the study of the interaction between the large outer scales and the anisotropic near-wall structures \citep{talamelli2009ciclope}. With the improvement in the computational resources, DNS has been able to resolve the inner and outer scales, and with every passing year, new investigations are taking place leading to new observations on the interaction of the inner and outer scales. The scale interaction becomes increasingly important with increasing Reynolds numbers. Although DNS gives valuable information, there are a number of outstanding questions in turbulence research that can only be answered by studying turbulence at high Reynolds numbers under well-controlled conditions. Furthermore, due to the increasing computational time and cost with the increasing Reynolds numbers, DNS is still unable to provide details about the nature of the physics of practical wall-bounded flows.

In order to investigate the structure of the wall-bounded turbulence at high Reynolds numbers in a region that is representative of a well-defined self-similar APG-TBL, requires a facility which has been designed specifically to produce this flow. The facility requires optical access to enable high-spatial-resolution (HSR) two-component -- two-dimensional (2C-2D) particle image velocimetry (PIV) over a large streamwsie extend of the TBL, {\em i.e.}  typically more than $2\delta$ to enable the investigation of ejections and sweeps, spanwise vortices, and uniform momentum zones\citep{thavamani2020characterisation}. The requirement for HSR often means that a single camera cannot capture the entire wall-normal range of a wall-bounded turbulent flow for the required domain size along the streamwise direction with multiple cameras required to be used for the HSR 2C-2D PIV. \citet{de2014high} and \citet{cuvier2017extensive} used an array of cameras to obtain large-field measurements of a ZPG-TBL and an APG-TBL, respectively. However, the recent development of the large CCD sensors has enabled the acquisition of PIV image pairs with sufficient spatial resolution while capturing the complete wall-normal range of the boundary layer over a streamwise domain of several boundary layer thicknesses. \citet{sun2021distortion} used a 47 MPx CCD camera to acquire the complete wall-normal range of a ZPG-TBL at $Re_\tau = 2,386$. Their measurement domain size was $2.5\delta \times 1.5\delta$ with the wall-normal resolution of 2.7 viscous units. 

This paper describes the facility used to esatblish the self-similar AGP-TBL  and provides the details of the PIV experimental set-up and PIV analysis. The characteristics of the strong APG-TBL are documented, including the determination from HSR 2C-2D PIV of the boundary layer parameters of: the relevant length and velocity scales, wall-shear stress, mean skin friction, pressure gradient parameter and shape factor, as well as the spatial variation of the similarity coefficients necessary to establish that the conditions of self-similarity are satisfied in the experimental facility. First- and second-order statistics scaled with the inner and outer units and the wall-normal distribution of the quadrant decomposition of the Reynolds shear stress are presented and compared with the ZPG-TBL and the mild APG-TBL results. Additionally, the profiles of the third-order moments, skewness, and flatness are presented and used to analyse the effect of a strong APG on the structure of a turbulent boundary layer. Finally, mean skin friction is studied in detail using the RD decomposition \citep{renard2016theoretical} to analyse the contributions of the turbulent kinetic energy, as well as the APG to mean skin friction. Furthermore, a conditional decomposition of mean skin friction is also presented to demonstrate the contributions of ejections and sweeps to the generation of mean skin friction.

%=======================================================================
\section{Experimental method}

%***--------------------------------------------***
\subsection{LTRAC APG-TBL wind tunnel}

For the experimental study of a self-similar high-Reynolds-number APG-TBL flow near separation, a wind tunnel facility is used to develop such flow. The flow passes through a contraction following a blower and several flow straightening stages, to sufficiently thin the boundary layer on the walls. Following a trip device at the entrance of the test section, the turbulent boundary layer is required to develop over a sufficient length of a flat plate to obtain the maximum possible thickness and achieve high Reynolds numbers. In order to impose an APG on the boundary layer, the cross-section of the tunnel should diverge in the streamwise direction. This can be achieved by making the roof follow a curved path. If the roof structure is rigid, that allows only a single fixed value of the imposed APG at one streamwise location. A flexible shape of the curved roof can impose different magnitudes of APG on the flow, which is essential to fine-tune the APG required to produce a self-similar turbulent boundary layer over a considerable length in the streamwise direction.
 
The second requirement is to bring the self-similar APG-TBL flow near separation. 
According to \citet{dengel1990experimental}, the mean skin friction coefficient $C_f$ may be as low as $3.5 \times 10^{-4}$ before separation. So, in the measured domain of interest (DOI), $C_f$ should be within an order of magnitude of this value. To achieve that, the approach of \citet{skaare1994turbulent} can be adopted, which imposes a strong APG at the beginning of the test section such that $dP/dx > 0$ and $d^2P/dx^2> 0$, in order to bring the flow close to separation. Then, to keep $C_f$ constant in the DOI, the APG in the downstream region is relaxed so that $dP/dx > 0$ and $d^2P/dx^2< 0$ and a stable equilibrium boundary layer is produced over a considerable length of the streamwise domain.

\begin{figure*}[!t] 
\centering
\begin{overpic}[percent,grid=false,tics=10,width=1.05\textwidth]{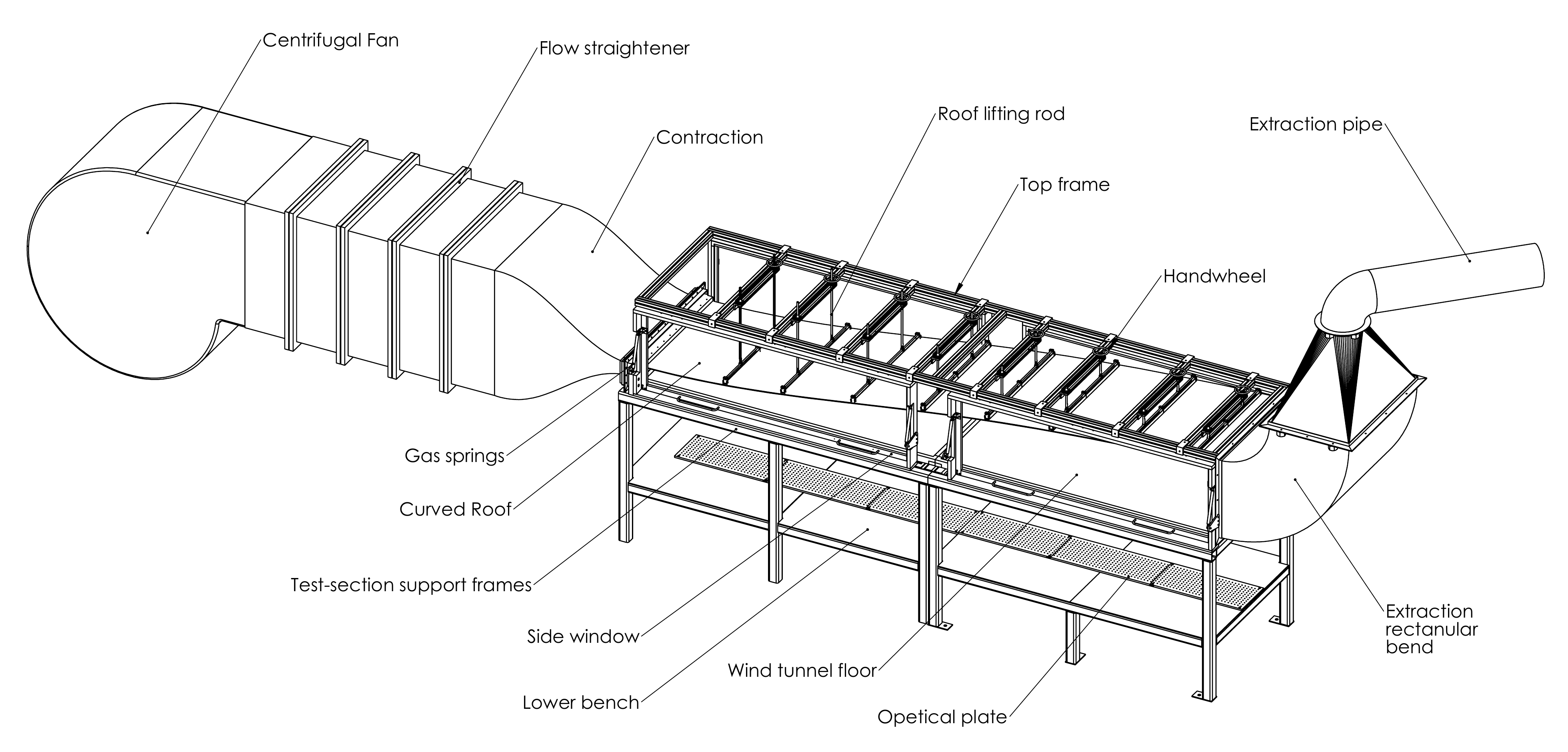}

\put(62,19){\color{blue}\rotatebox{-15}{\tiny 4.5 m}}
\put(40,24.5){\color{blue}\vector(-1,0.265){0}}
\put(40,24.5){\color{blue}\vector(1,-0.265){37}}

\put(31.8,32){\color{blue}\rotatebox{45}{\tiny 1 m}}
\put(30,31){\color{blue}\vector(-1,-1){0}}
\put(30,31){\color{blue}\vector(1,1){4.3}}

\put(30.4,26){\color{blue}\rotatebox{90}{\tiny 1 m}}
\put(30,31){\color{blue}\vector(0,1){0}}
\put(30,31){\color{blue}\vector(0,-1){8}}

\end{overpic}
\caption{A 3D model of the LTRAC APG-TBL wind tunnel.}
\label{fig:3D_model_WT}
\end{figure*}
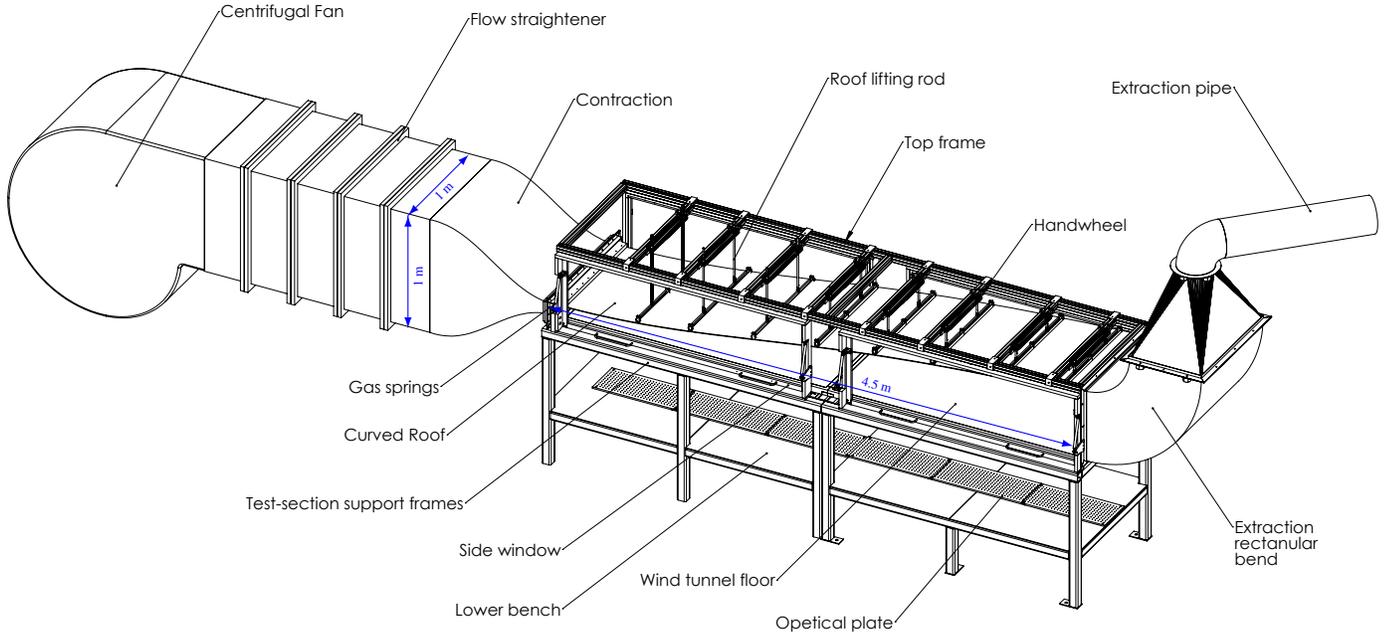

A new test section satisfying the above requirements was built on an existing wind tunnel facility in the Laboratory of Turbulence Research in Aerospace and Combustion (LTRAC), Monash University, Australia, which is reported by \citet{amili2012measuring}. A 3D model of the wind tunnel is shown in figure \ref{fig:3D_model_WT}. This is an open-type wind tunnel that consists of an inlet, a centrifugal fan, several flow straighteners, a contraction, a tripping device, a test section, and an outlet. The centrifugal fan is powered by a three-phase 5.5 kW motor. The rotational speed of the motor is adjusted to control the airflow rate. A settling chamber connected to the blower consists of honeycombs and screens which act as flow straighteners. The contraction has an aspect ratio of 10:1 with the smaller open face measuring 1000 mm $\times$ 100 mm. Based on its location and purpose, the new wind tunnel is referred to as the LTRAC APG-TBL wind tunnel.

The LTRAC APG-TBL wind tunnel is 4.5-m long, 1-m wide and 0.62-m high. The floor and sidewalls of the test section are made of clear glass fixed in aluminium frames. This allows optical access to the flow in the wall-normal as well as the spanwise directions. The side windows, when closed, are sealed with the base floor to prevent any leakage of air that may carry the seeding particles. When opened, each window is supported by two gas springs. The provision of the windows allows easy cleaning of the accumulated seeding particles from the glass to allow a clear view for the PIV cameras and to avoid irregular scattering of the laser light from the accumulated particles. The floor of the test section was levelled to the $xz$ plane.

The roof of the test section is a 4.5 mm thick polycarbonate sheet held in place using eight lifting stations, which themselves are supported on the top frame. Each lifting station has two vertically hanging screw rods that can be winded in and out of the two bearings fixed on the top frame to change the roof height at a particular streamwise location. The two winding wheels on the top are interconnected using a chain and sprocket mechanism to apply uniform lifting on both lateral ends.

The extraction of the wind tunnel consists of a 90$\degree$ elbow duct, a lofted connection from rectangular to circular shape, and a pipe of 300 mm diameter. The extraction parts are made of aluminium sheet metal. 

The aluminium parts of the window and floor frames, the top frame, and the lifting stations are black anodised to avoid reflections from the laser sheet. The 90$\degree$ elbow duct of the extraction is painted black from the inside for the same purpose.

%***--------------------------------------------***
\subsection{Development of the self-similar APG-TBL flow near separation}

%***--------------------------------------------***
\subsubsection{Importance of self-similarity} 

\noindent Continuous changes in pressure gradients on a curved surface make an APG-TBL flow difficult to study, as they are affected by the history of upstream conditions \citep{kitsios2017direct}. To study the effect of pressure gradient on turbulent statistics, it is important to make the APG-TBL independent of the influence of the upstream conditions. This is achieved by making the flow self-similar. A self-similar APG-TBL is the simplest canonical form of flow to investigate \citep{kitsios2017direct}, and is defined as the flow in which the relative contribution of each term in the governing equations does not depend on the streamwise location \citep{simpson1977features, mellor1966equilibrium}.

%***--------------------------------------------***
\subsubsection{The roof profile for self-similarity}
As the roof of the LTRAC wind tunnel can be adjusted to impose a variable pressure gradient, an optimum roof profile that enables the flow to be self-similar in a domain spanning several boundary layer thicknesses in the streamwise direction is desired. The wall tufts were fixed in a matrix of 50 $\times$ 15 on the ceiling and 50 $\times$ 5 on the floor to visualise any flow separations while adjusting the roof profile.

\begin{figure*}
\centering
 \setlength{\figH}{0.35\textwidth}
 \setlength{\figW}{0.95\textwidth}
 \input{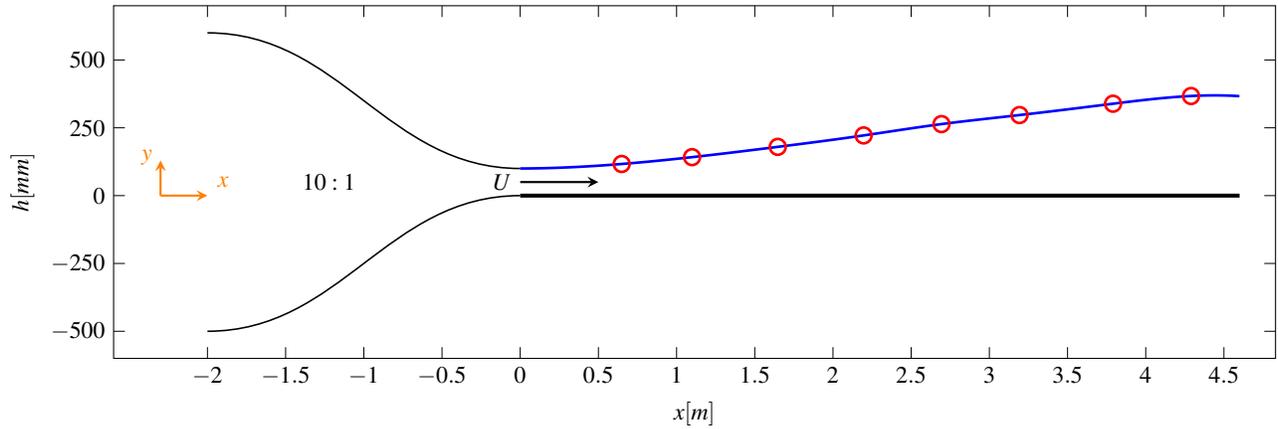}
\caption{Roof profile of the LTRAC APG-TBL wind tunnel is shown by the dashed blue line. The red circles show the locations of the lifting stations on the roof, whereas the black solid line represents the floor.}
\label{fig:roof_profile}
\end{figure*}

An estimate of the roof profile which satisfies the self-similarity conditions is shown in figure \ref{fig:roof_profile}. This is governed by the equation:
\begin{equation} 
\begin{gathered}
h= 100 - 5.1798 x + 51.9825 x^2 - 13.0411 x^3 + \\ 7.2 \times 10 ^{-12} x^4 + 0.5559 x^5- 0.0719 x^6
\end{gathered}
\end{equation}
where the height $h$ and the streamwise position $x$ are in metres and millimetres, respectively. To achieve this profile, a strong APG with $dP/dx > 0$ and $d^2P/dx^2> 0$ was imposed in the region where $x<0.95$ m, to bring the flow close to separation. Beyond $x=0.95$ m, the APG was relaxed so that $dP/dx > 0$, $d^2P/dx^2< 0$ and $C_f$ is constant and a stable equilibrium boundary layer is produced.

%***--------------------------------------------***
\subsection{Experimental setup for 2C-2D PIV measurements} 

%\subsubsection{Laser sheet creation and alignment}
A NewWave Solo 200XT Nd:YAG pulsed laser was used to illuminate the seeding particles in the 2C-2D PIV experiments. It is a dual-cavity laser that provides two pulses, each with a pulse width of 3$-$5 ns and maximum energy of 200 mJ at the maximum frequency of 10 Hz. The beam diameter at the exit of the laser head is approximately 6 mm. A laser arm was used to direct the laser beam from the laser head to the laser optics set-up. 

For 2C-2D PIV, the laser beam was expanded using a cylindrical lens of $f = -100$ mm (where $f$ is the focal length) that was installed 1,300 mm upstream of the FOV. The thickness of the laser sheet is reduced to be minimum at the middle of the FOV, using a $f=1,500$ mm focal length cylindrical lens, installed at 1,500 mm upstream of the middle of the FOV. 

The laser sheet formed by the $f=-100$ mm lens expands slowly before entering the wind tunnel from underneath the glass floor near the end of the test section and is reflected towards the FOV using a large mirror with the reflecting surface measuring 150 mm $\times$ 38 mm. All optics are mounted on an optomechanical cage system, except for the large mirror, which is placed inside the tunnel and supported on the tunnel floor. The laser sheet thickness was measured using a low-cost in situ method presented in \citet{Shehzad2019}. To measure the laser sheet thickness, an automated system has been developed which consists of a low-cost CCD camera, a Raspberry Pi to control the camera, and software that calculates the thickness of the laser sheet falling on the bare CCD array in real time. The thickness measured at the waist, which is in the middle of the FOV is approximately 500 $\mu$m. The thin laser sheet allows high-quality 2C-2D PIV measurements, provided that the seeding and camera focus are properly set. The laser sheet is aligned parallel to the calibration target using a novel alignment method presented by \citet{shehzad2021assuring}. 

Smoke particles are used as seeding in the current 2C-2D PIV experiments. A Compact ViCount smoke machine is used to create the smoke from a water-glycol mixture. The maximum aerosol output of the smoke machine is 166 mg/sec. It is operated using compressed nitrogen gas at a maximum operating pressure of 820 kPa. Pressure can be varied to change the amount of smoke output. The vapour density of the water-glycol mixture is approximately equal to the vapour density of water. The smoke is mixed with air at the inlet of the centrifugal fan.  

Two high-spatial-resolution CCD cameras (Imperx B6640) are used side-by-side to image a large field of view (FOV) of $3.3\delta \times 1.1\delta$ in the $xy$ plane, where $\delta$ is the boundary layer thickness at the middle of the field of view. Each camera has a large sensor that measures 36.17 mm $\times$ 24.11 mm and contains 6,600 $\times$ 4,400 pixels with a square pixel size of 5.5 $\mu$m. The cameras are used with Zeiss lenses of 100 mm focal length set at the f-stop number of 2.8. At a spatial resolution of 40 $\mu$m/pixel and with an overlap of 5 mm between two cameras, a combined FOV that measures 523 mm $\times$ 178 mm is achieved. 

The exposure and bit-depth for both cameras are set to 186 $\mu$s and 12 bits, respectively. The cameras are controlled by an external signal provided by a signal synchroniser and operated in the double exposure mode at a frequency of 2 Hz which enables the acquisition of two PIV image pairs every second. A BeagleBone Black (BBB) platform is used as the signal synchroniser. This control device is enclosed in a MED025 box which has 20 BNC ports for 14 outputs and 6 inputs. The device was developed at LTRAC \cite[see][]{fedrizzi2015application}. It is connected to a computer via a USB 2.0 cable and can be controlled using a Python program in a web-based Cloud9 IDE to generate synchronised signals at multiple BNC interfaces. The elapsed time between the two laser pulses $\Delta t$ is set at 100 $\mu$s, so that the maximum particle image displacement on the sensor is less than 15 pixels at the spatial resolution of 40 $\mu$m/pixel and the inflow velocity of 15.6 m/s.

A 40-mm wide strip of 40-grit sandpaper, affixed on top of a 60-mm wide and 1.2-mm thick sheet metal, is used as the boundary layer trip device. It is placed at the entrance of the wind tunnel test section and this location is taken as $x=0$.

%***--------------------------------------------***
\subsection{Digital 2C-2D PIV analysis}

PIV images of the strong APG-TBL were analysed using multigrid/multipass cross-correlation digital PIV (MCCDPIV) \citep{soria1996investigation} with an in-house parallel code. The parameters of the PIV analysis along with the experimental setup and the flow conditions are presented in table \ref{tab:piv_parameters_viscous}. The values of the field of view (FOV), grid spacing and the interrogation window (IW) size are presented in viscous units as well as in pixels. A thin laser sheet together with the sufficient seeding of microparticles leads to the use of a small IW size. The IWs have an overlap of 50\% in the streamwise direction and 75\% in the wall-normal direction. The latter leads to the wall-normal spatial resolution of less than a viscous unit. 

The MCCDPIV results were validated using the normalised median test \citep{westerweel2005universal} with a threshold value of $2.0$ and the maximum allowable velocity of 0.21$\times$IW$_x$ where IW$_x$ represents the size of the IW in the streamwise direction. The subpixel accuracy of the PIV displacement was obtained using a Gaussian function fit \citep{willert1991digital, soria1996investigation}.

\begin{table}
\begin{center}
\caption{Parameters related to the experimental setup and the PIV analysis.}
\begin{tabular}{lc}

\hline \hline\noalign{\medskip}
Measurement	(Units) 				& Value 			\\  
\noalign{\smallskip}\hline\noalign{\medskip}

Inflow Velocity (m/s) 					& 15.5 \vspace{0.0em}\\ 
Viscous length scale $l^+$ ($\mu$m) 	& 175\textsuperscript{1} \vspace{0.0em}\\	

Magnification $M$ ($\mu$m/pixel) 		& 40.65 \vspace{0.0em} \\
f-stop number 						& 2.8 \vspace{0.0em}\\
Camera lens focal length $f$ (mm) 		& 100 \vspace{0.5em}\\

FOV ($l^+ \times l^+$) 				& $3,076 \times 1,052$ \\
 (pixels $\times$ pixels) 				& $12,880 \times 4,400$ \\
(mm $\times$ mm) 					& $523 \times 178$ \\
($\delta \times \delta$) 				& $3.3 \times 1.06$ \vspace{0.5em} \\
	
Grid spacing ($l^+ \times l^+$) 			& $3.82 \times 0.96$ \\
 (pixels $\times$ pixels) 				& $16 \times 4$ \vspace{0.5em} \\

IW size ($l^+ \times l^+$)	 			& $7.65 \times 2.87$	 	\\
(pixels $\times$ pixels) 				& $32 \times 12$ \vspace{0.5em} \\

Max velocity ratio filter (Threshold) 				& 0.21 \vspace{0.0em} \\
Normalised median filter (Threshold) 			& 2.0 \vspace{0.0em} \\
Time interval between frames $\Delta t$ ($\mu$s)	& 100 \vspace{0.0em} \\
Velocity field acquisition frequency ($Hz$) 		& 2 \vspace{0.0em} \\
Number of samples 							& 25,000 	\vspace{0.0em} \\
Vector field size 							& $813 \times 1,091$ \vspace{0.0em} \\
\noalign{\smallskip}\hline\noalign{\medskip}
\label{tab:piv_parameters_viscous} 
\end{tabular}
\end{center}
\end{table}

\footnotetext{\textsuperscript{1}This is at the middle of the FOV at $x_o$.}

%#===================================================================#
\section{Boundary layer parameters}

\noindent A strong APG-TBL has a negative $\partial V_\infty/\partial x$ at the far-field in the free stream. Hence, $\partial U/\partial y$ at the boundary must be negative for the far-field to have a zero spanwise vorticity. This leads to the conclusion that $U$ profile must have a maximum point when plotted against $y$. Since the mean streamwise velocity profile in the wall-normal direction does not approach a constant value, the classical definitions of the displacement and momentum thicknesses ($\delta_1$ and $\delta_2$) are inappropriate. Therefore, the appropriate velocity scale (the edge velocity $U_e$) and the length scales (the boundary layer thickness ($\delta$), $\delta_1$, and $\delta_2$) will be used as defined in \citet{kitsios2017direct,spalart1993experimental}. \citet{lighthill1963boundary} first proposed this velocity scale, which is given as

\begin{equation}
U_e(x) = U_\Omega (x, y_\Omega),
\end{equation}

\noindent where

\begin{equation}
U_\Omega (x,y) = - \int_{0}^{y} \Omega_z (x, \widetilde{y}) d\widetilde{y},
\end{equation}

\noindent with $\Omega_z$ representing the mean spanwise vorticity, $y_\Omega$ is the wall-normal position at which $\Omega_z$ is 0.2\% of the mean vorticity on the wall and $\delta =y_\Omega$. The integral length scales are given by

\begin{equation}
\delta_1(x) = \frac{-1}{U_e} \int_{0}^{y_\Omega} y \Omega_z (x, y) dy, \,\,\,\,\, \text{ and}
\end{equation}

\begin{equation}
\delta_2(x) = \frac{-2}{{U_e}^2} \int_{0}^{y_\Omega} y U_\Omega \Omega_z (x, y) dy - \delta_1(x). 
\end{equation}

\begin{table}[!b]
\begin{center}
\caption{Turbulent boundary layer parameters at the middle of the FOV of the strong APG-TBL.}
\label{tab:bl_parameters} 
\resizebox{\columnwidth}{!}{%
\begin{tabular}{cccccccccc}
% \vspace{0.2em} \\
\hline \hline\noalign{\medskip}
$U_e$(m/s)	&$\delta $ (mm) 	&$\delta_1 $ (mm) 	&$\delta_2 $ (mm) 	&$H$ & $u_\tau$ (m/s) & $Re_{\delta_1}$		&$Re_{\delta_2}$ & $Re_{\tau}$ & $\beta$\\
\noalign{\smallskip}\hline\noalign{\medskip}
5.43	&158.37	 &62.04	 	&25.51	 	&2.42	& 0.084	& 22,920 	&9,660 	 & 940	&30.24 \\ \noalign{\smallskip}\hline\noalign{\medskip}
\end{tabular}
}
\end{center}
\vspace{-0.2em}
\end{table}

The measured length and velocity scales in the middle of the FOV for strong APG-TBL are presented in table \ref{tab:bl_parameters}. The current APG-TBL has $\beta=30.24$ compared to $\beta=2.27$ of the APG-TBL measured in the LMFL wind tunnel. Therefore, the former is referred to as `strong APG-TBL' and the latter as `mild APG-TBL'.

\begin{figure}[!b]
\begin{center}
\begin{tabular}{c}
\begin{overpic}[width=0.4\textwidth]{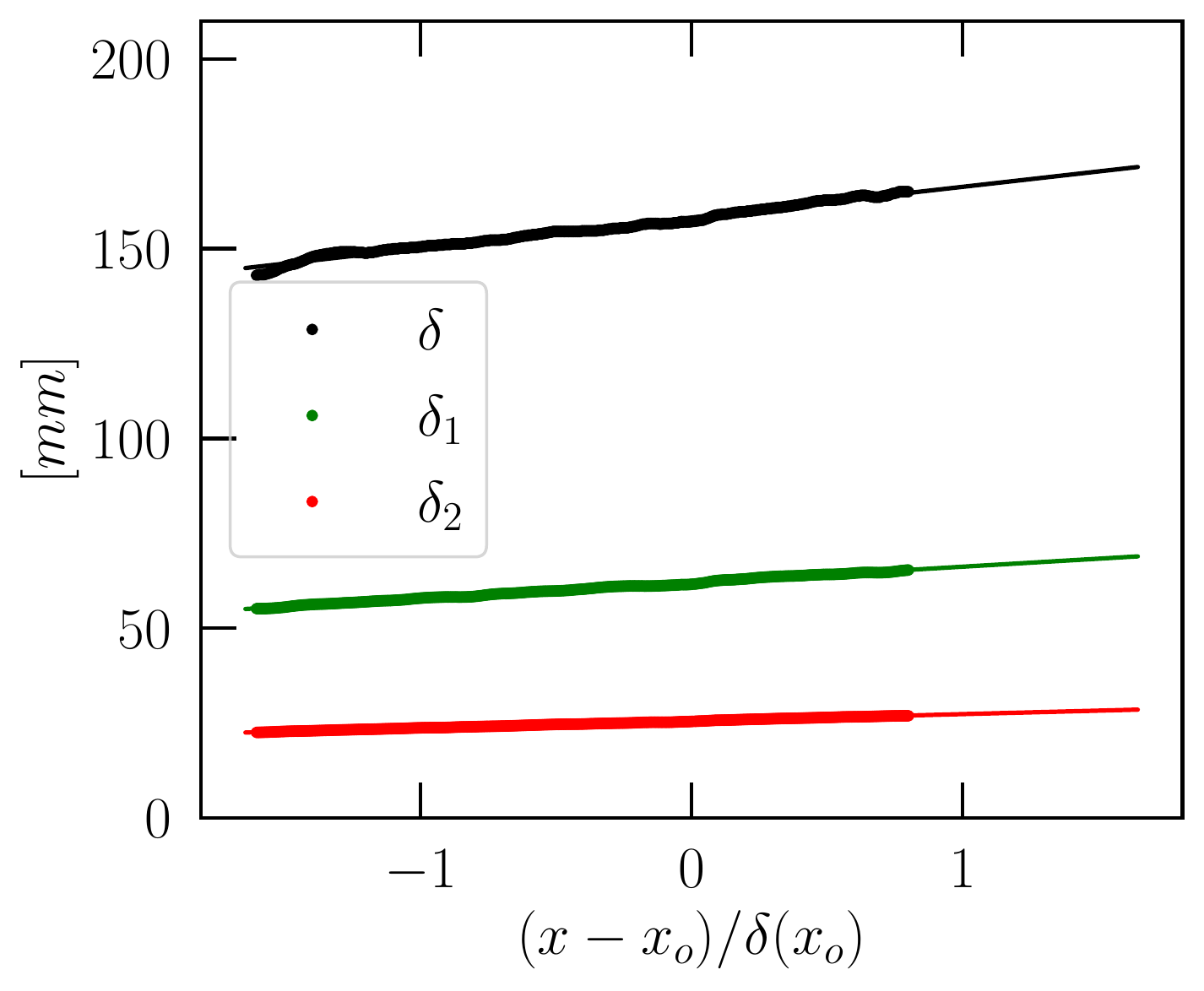} \put(25,68){(a)} \end{overpic} \\
\begin{overpic}[width=0.4\textwidth]{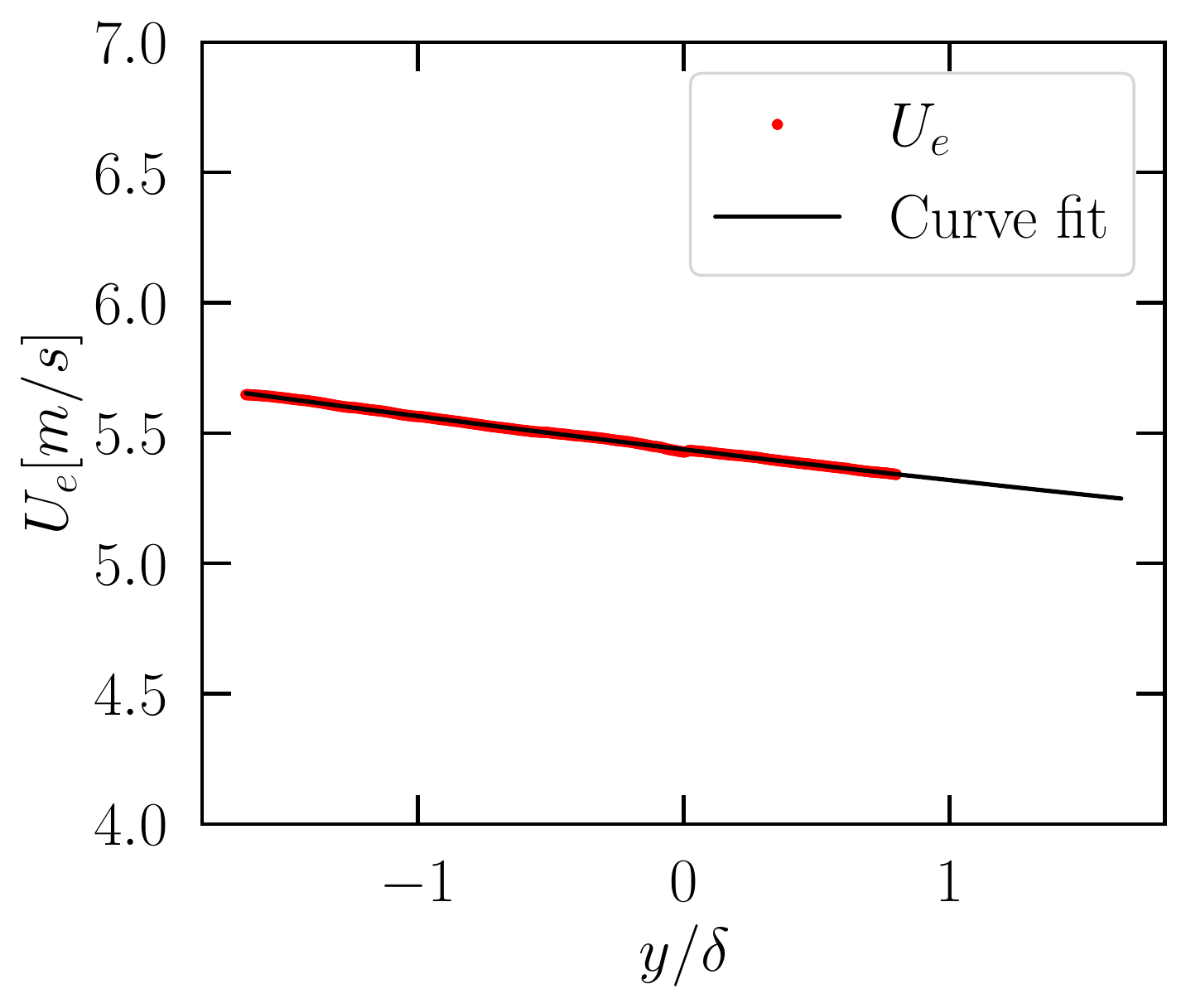} \put(25,68){(b)} \end{overpic}
\end{tabular}
\end{center}
\vspace*{-0.2in}\caption{Streamwise variation of the length and velocity scales: (a) boundary layer thickness $\delta$, displacement thickness $\delta_1$, momentum thickness $\delta_2$, and (b) the edge velocity $U_e$.
\label{fig:lengt_and_vel_scales}}
\end{figure}

The streamwise dependence of these scales is presented in figure \ref{fig:lengt_and_vel_scales}. The $\delta$, $\delta_1$, and $\delta_2$ vary linearly with $x$ for the first 70\% of the streamwise ($x$) domain. Information on vorticity is not available near the edge of the boundary layer in the latter 30\% of the $x$ domain due to the limited FOV in the wall-normal direction. It is presumed that the length scales vary linearly in this part, also. The linear fits to the length scales have been extrapolated to cover the full $x$ range. This assumption will be tested later with the collapse of the profiles of the mean streamwise velocity and the Reynolds stresses, scaled using the extrapolated outer variables $U_e$ and $\delta_1$ following the scaling in \citet{kitsios2017direct}. If the wall-normal profiles of the turbulent statistics in the latter 30\% of the $x$ domain scaled using the extrapolated outer variables are consistent with the profiles from the first 70\% of the $x$ domain, this would prove the assumption that the boundary layer is growing linearly with the streamwise distance throughout the $x$ domain.

\begin{figure}[!t]
\begin{center}
\begin{tabular}{c}
\begin{overpic}[width=0.43\textwidth]{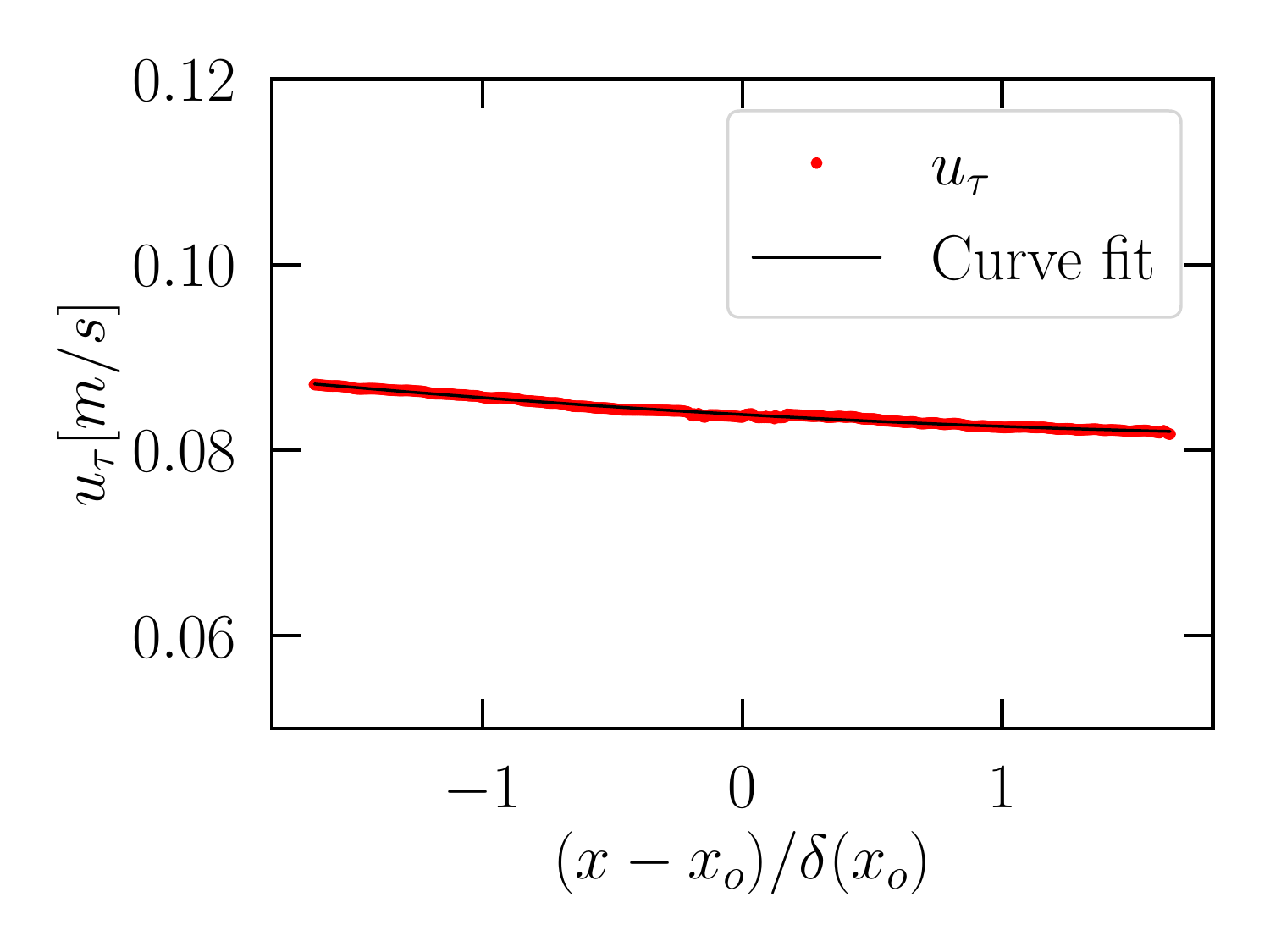} \put(25,62){(a)} \end{overpic}\\
\begin{overpic}[width=0.43\textwidth]{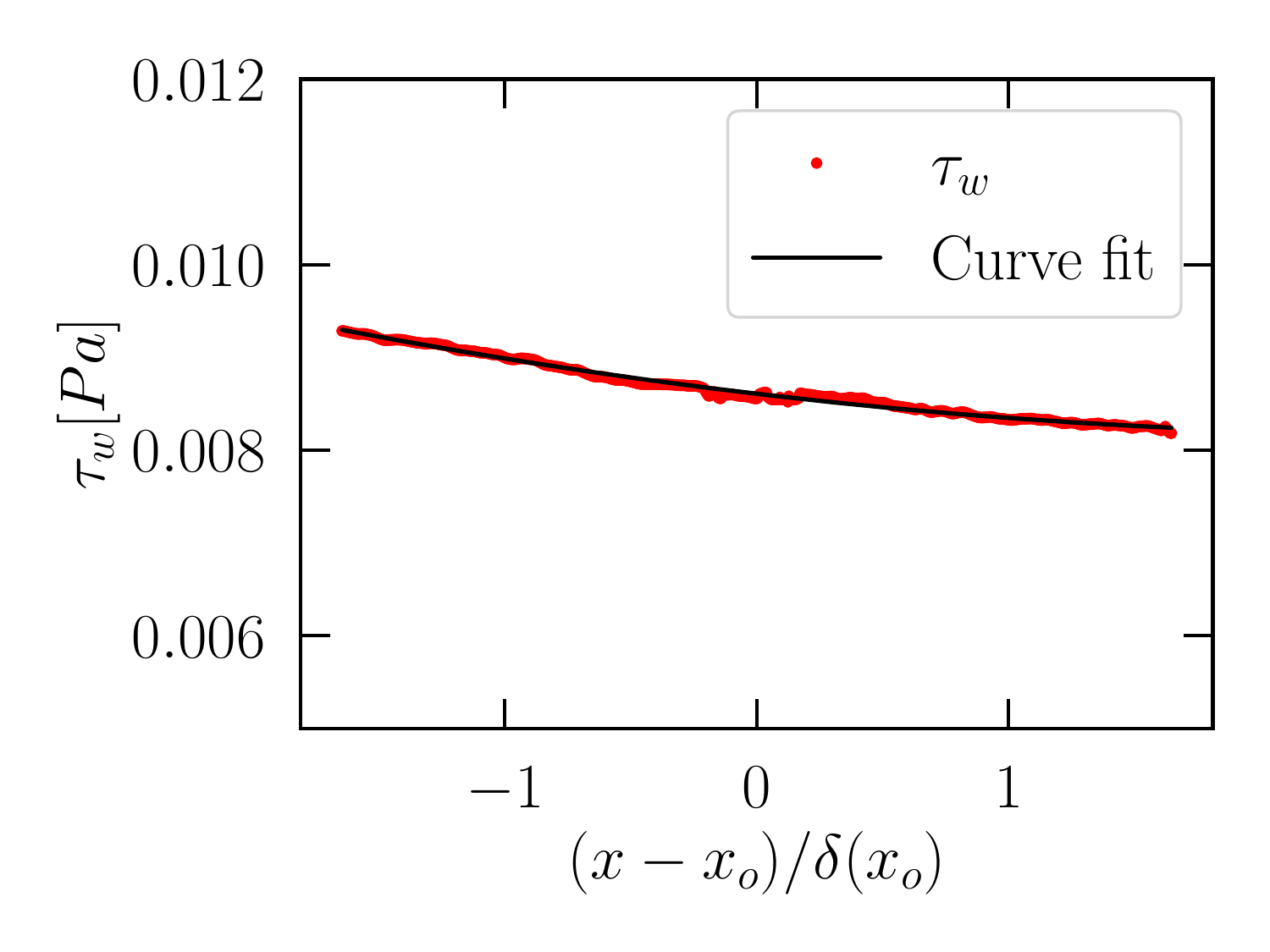} \put(25,62){(b)} \end{overpic} 
\end{tabular}
\end{center}
\vspace*{-0.2in}\caption{Streamwise variation of the (a) friction velocity, $u_\tau$ and, (b) the wall-shear stress $\tau_w$.
\label{fig:ut_and_tw}}
\end{figure}

Since the present data of the strong APG-TBL has a high spatial resolution in the wall normal direction, the wall shear stress and the friction velocity are measured at every $x$ location using the mean streamwise velocity profile within the viscous sublayer. Figure \ref{fig:ut_and_tw} shows the streamwise variation of the measured friction velocity $u_\tau$ and the wall-shear stress $\tau_w$. The decay of these velocities is weakly quadratic.

\begin{figure}
\begin{center}
\begin{tabular}{c}
\begin{overpic}[width=0.4\textwidth]{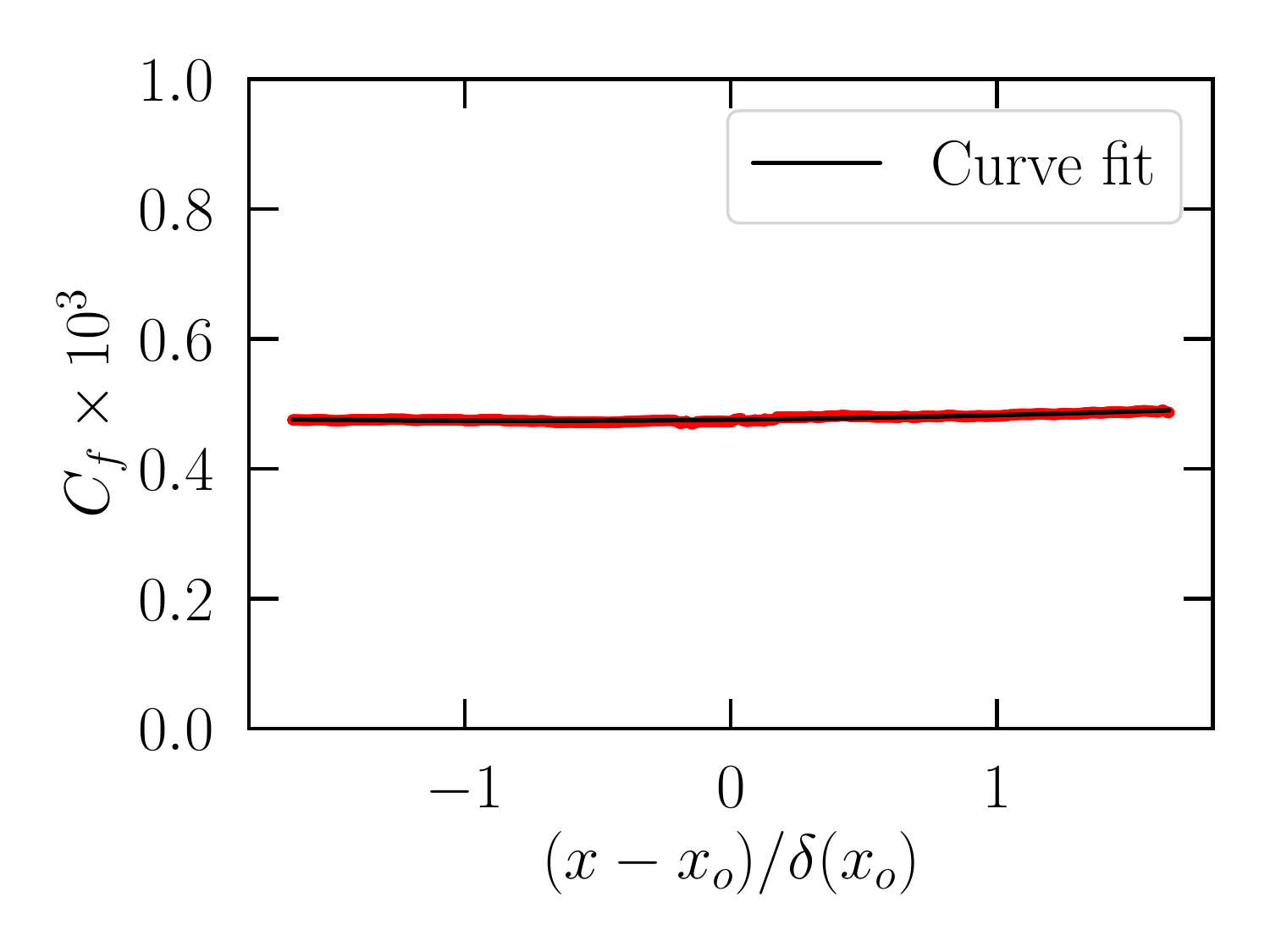} \put(25,62){(a)} \end{overpic} \\
\begin{overpic}[width=0.4\textwidth]{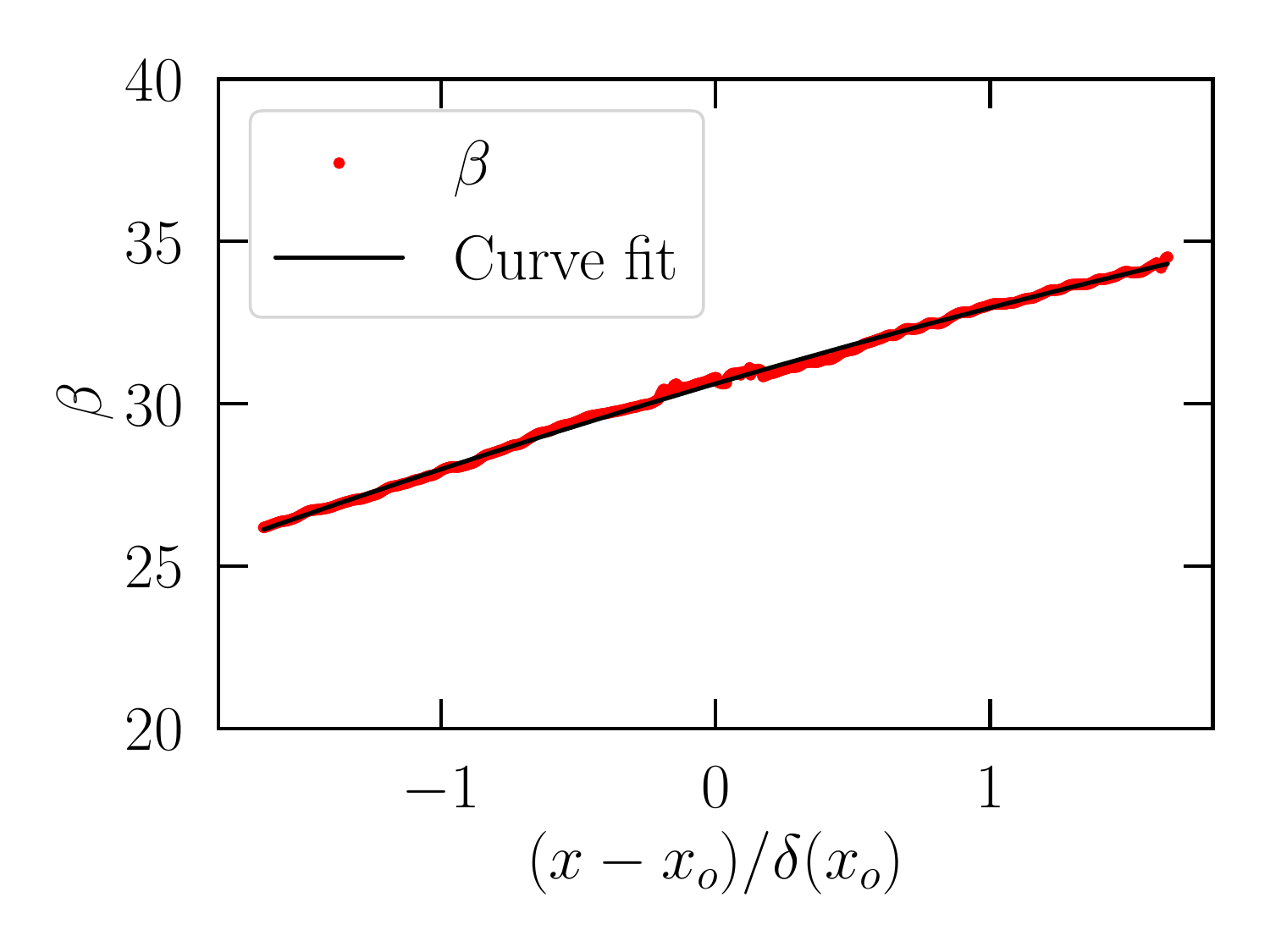} \put(21,43){(b)} \end{overpic} \\
\begin{overpic}[width=0.4\textwidth]{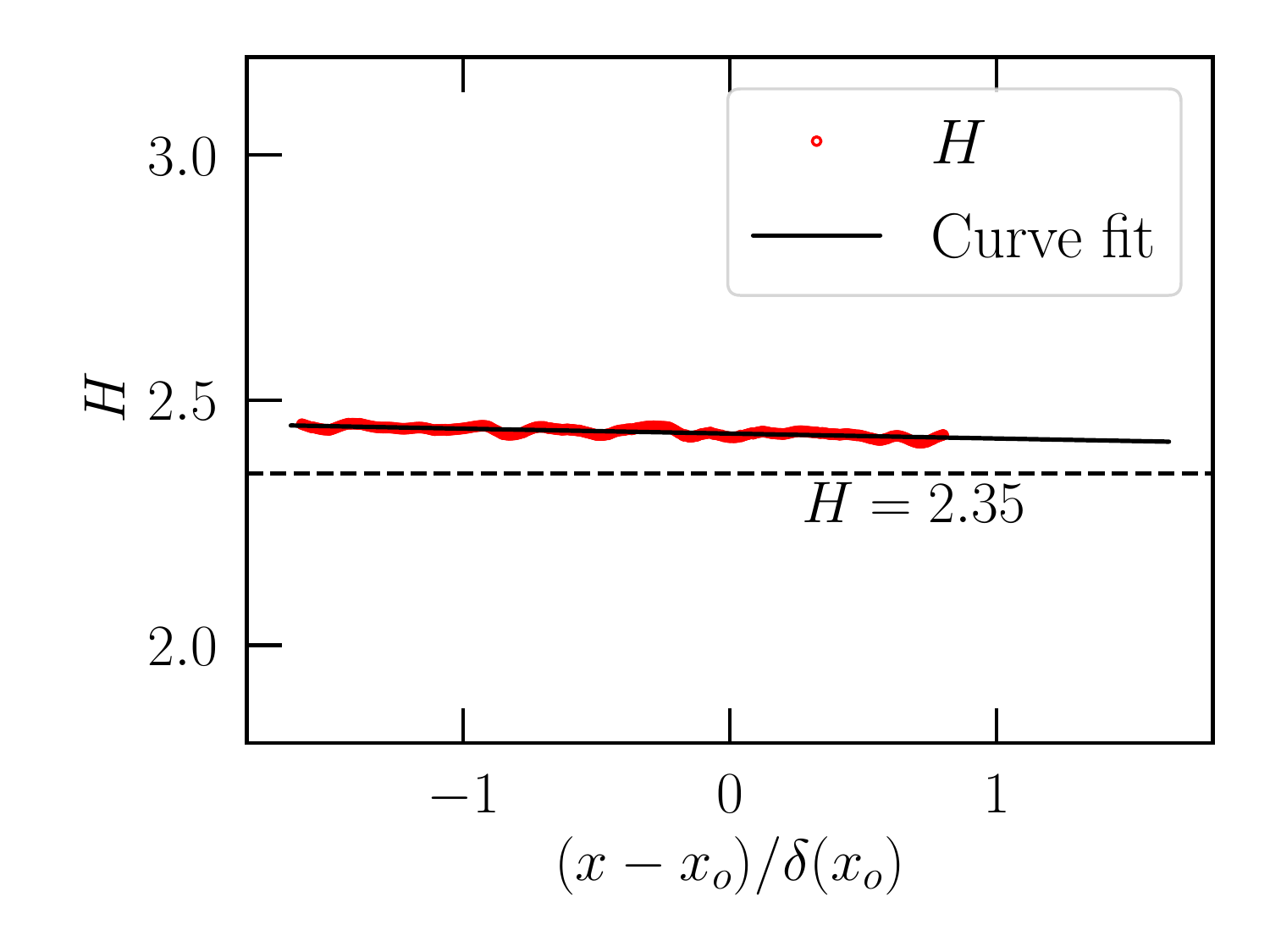} \put(23,62){(d)} \end{overpic} \\
\begin{overpic}[width=0.4\textwidth]{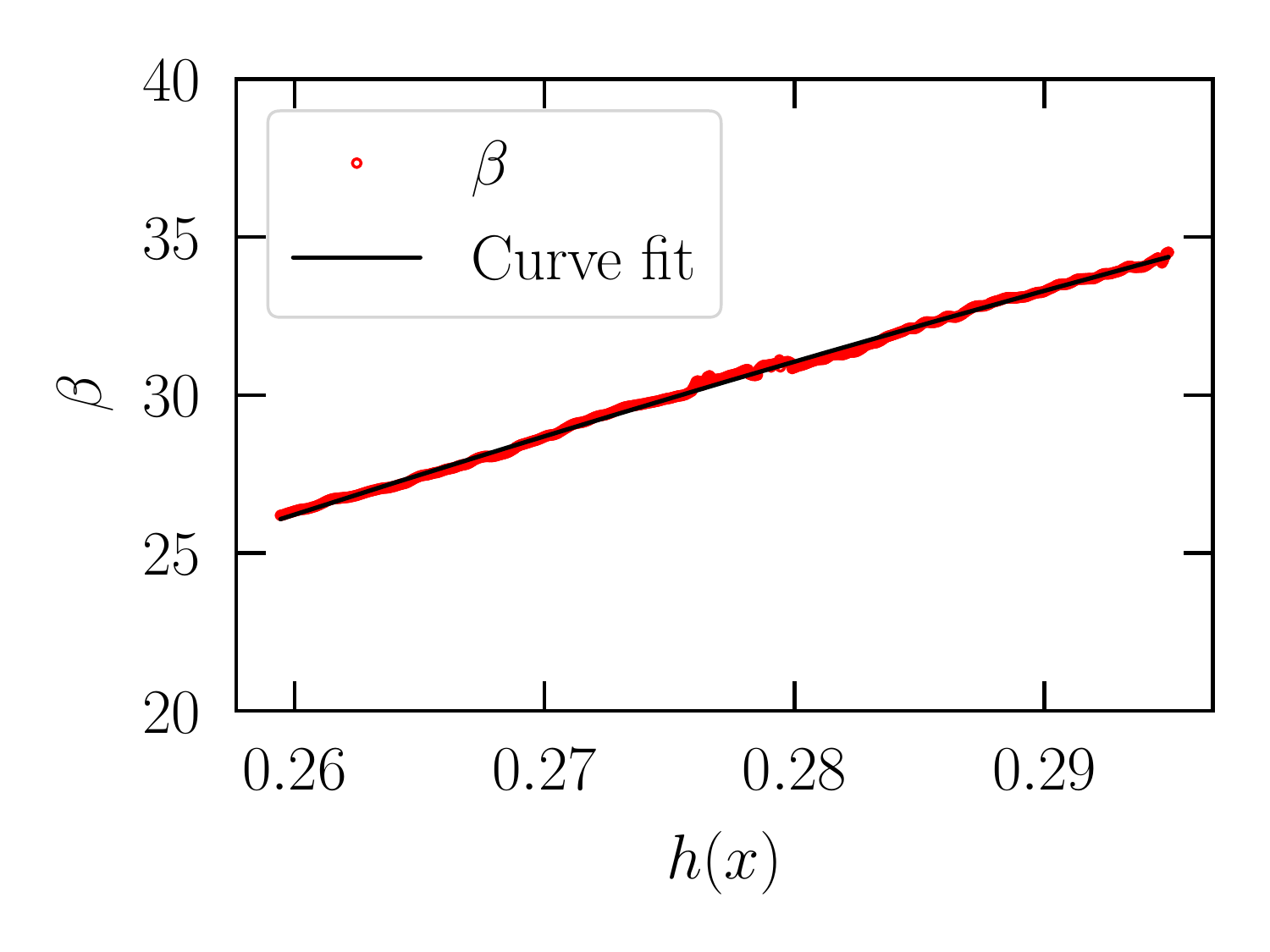} \put(27,42){(c)} \end{overpic} \\
\end{tabular}
\end{center}
\vspace*{-0.2in}\caption{Streamwise variation of (a) mean skin friction coefficient $C_f$, (b) Clauser pressure gradient parameter $\beta$, and (c) shape factor $H$. $H=2.35$ is the empirical value for an equilibrium APG-TBL reported by \citet{mellor1966equilibrium}. (d) shows $\beta$ as a function of the roof height $h$ within the measurement domain.
\label{fig:other_parameters}}
\end{figure}

Mean skin friction coefficient in turbulent boundary layer is defined as $C_f = \tau_w / (0.5 \rho U_\infty ^2) = u_\tau ^2 / (0.5 U_\infty ^2)$. Figure \ref{fig:other_parameters}(a) shows the streamwise variation of $C_f$. It has an approximately constant value of \num{4.8e-4}, which is comparable to the constant value of $\num{5.7e-4}$ for an equilibrium turbulent boundary layer near separation reported by \citet{skaare1994turbulent}. As reported by \citet{dengel1990experimental}, $C_f$ may be as low as \num{3.5e-4} before reaching the separation point.

Figure \ref{fig:other_parameters}(b) shows that $\beta$ linearly increases from 25.88 to 34.14 in the measured $x$ domain. Its streamwise average of approximately 30 is higher than the $\beta \approx 20$ in the equilibrium region of the strong APG-TBL reported in \citet{skaare1994turbulent} and less than $\beta=39$ in the DNS of a strong APG-TBL reported in \citet{kitsios2017direct}.

Figure \ref{fig:other_parameters}(c) shows that the shape factor $H$ is approximately constant at about 2.45. This is close to the empirical value of $H = 2.35$ for an equilibrium APG-TBL, as reported in \citet{mellor1966equilibrium}. It is also in agreement with the $H$ of the strong APG-TBL of \citet{kitsios2017direct}. Figure \ref{fig:other_parameters}(d) presents $\beta$ as a function of the roof height $h$. As expected, $\beta$ is linearly proportional to $h$ within the measurement domain.

%#===================================================================#
\section{Conditions of self-similarity} 

For an APG-TBL flow to be self-similar, the mean skin friction coefficient $C_f$ and the shape factor $H$ should remain constant along the streamwise direction. This is not expected in a ZPG-TBL in which both $C_f$ and $H$ decrease slowly when moving downstream \citep{townsend1980structure}. 

Much of the theoretical work to derive the conditions of self-similarity and develop velocity and length scales to collapse the profiles of turbulent statistics at different streamwise locations has been carried out by \citet{townsend1980structure, mellor1966equilibrium, durbin1992scaling, perry1995wall, castillo2004similarity}. The Reynolds-averaged Navier-Stokes (RANS) equations of continuity, streamwise momentum, and wall-normal momentum, along with some similarity ansatz, can be used to achieve conditions of self-similarity as first presented by \citet{so1993boundary} and summarised by \citet{kitsios2017direct}. Based on these conditions, one can determine that the following quantities must be independent of $x$ for a flow to be self-similar.

\begin{equation}
C_{uu} = R_{uu}/U_e^2  
\end{equation}
\begin{equation}
C_{vv} = R_{vv}/U_e^2  
\end{equation}
\begin{equation}
C_{uv} = R_{uv}/(U_e^2 \delta ^ \prime _1) 
\end{equation}
\begin{equation}
C_\nu = \nu/(U_e \delta _1 \delta ' _1)  
\end{equation}
\begin{equation}
\Lambda = - \delta_1 U '_e / (U_e \delta '_1) = \delta_1 (P') / (U_e^2 \delta'_1) = (U_p / U_e)^2 / \delta'_1
\label{eq:Lambda}
\end{equation}

\noindent where $\Lambda$ is a pressure gradient parameter defined by \citet{castillo2004similarity} and $R_{uu}$, $R_{vv}$ and $R_{uv}$ are the functions that can be determined at different $x$ locations using the integrals of $\overline{u'u'}$, $\overline{v'v'}$ and $\overline{u'v'}$ from $\zeta=0$ to $\zeta=\delta /L_0$, respectively, where $L_0(x) \equiv \delta_1(x)U_e(x)/U_0(x)$, $P'$ is the streamwise pressure gradient, $\delta_1'$ is the streamwise gradient of displacement thickness and $ U_p = \sqrt{\delta_1 P'}$ is the pressure velocity.

\begin{figure*}
\begin{center}
\begin{overpic}[width=0.8\textwidth]{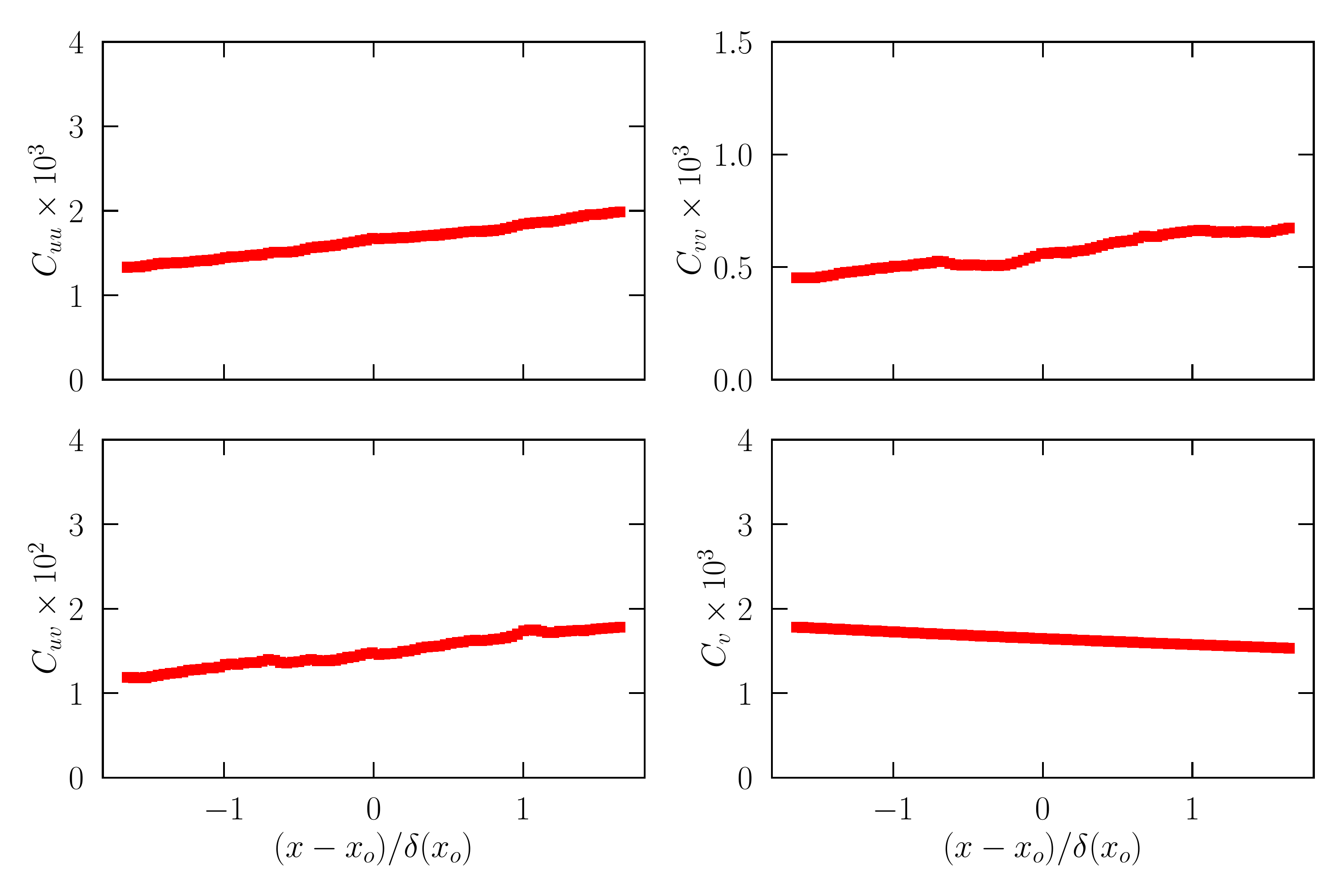} 
\put(9,60){(a)}
\put(59,60){(b)}
\put(9,30){(c)}
\put(59,30){(d)}
\end{overpic}
\end{center}
\vspace*{-0.2in}\caption{Streamwise distribution of the similarity coefficients presented in equations (1) to (4).
\label{fig:similarity_variables}}
\end{figure*}
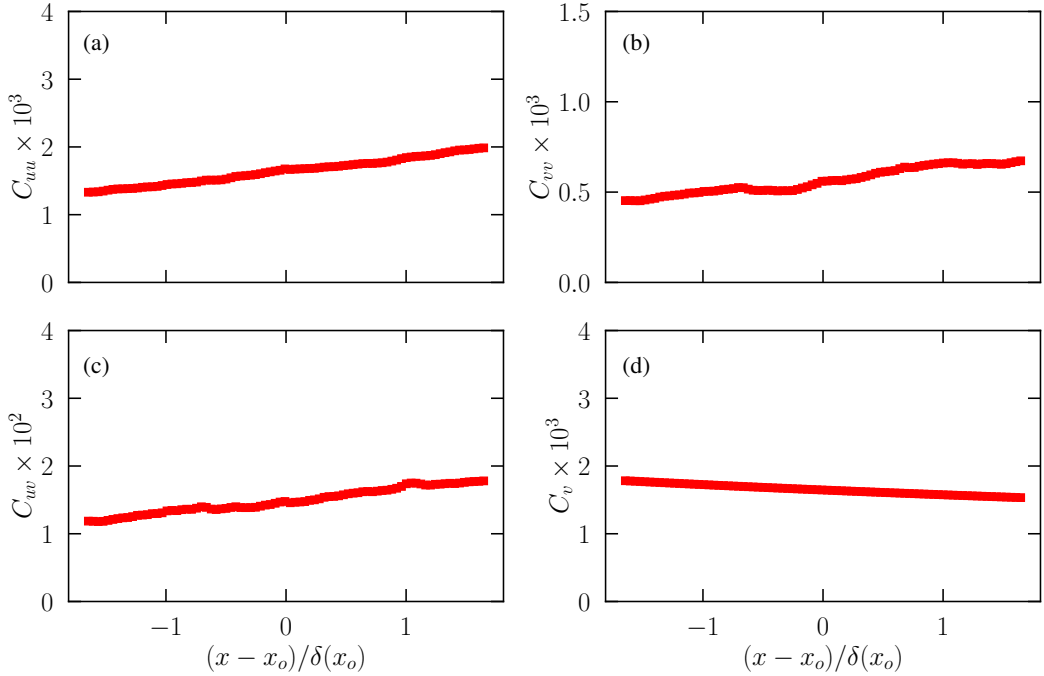

If $C_{uu}$, $C_{vv}$, $C_{uv}$, and $C_\nu$ are constant along the $x$ direction, the profiles of $\langle uu \rangle$, $\langle vv \rangle$, $\langle uv \rangle$, and $\nu \partial^2 U / \partial y^2$, respectively, show self-similarity in both the inner and outer regions. However, if only $C_\nu$ is not independent of $x$, it implies that the scaling only applies to the outer region of the turbulent boundary layer \citep{kitsios2017direct}.

The streamwise profiles of the similarity coefficients $C_{uu}$, $C_{vv}$, $C_{uv}$, and $C_\nu$ are shown in figure \ref{fig:similarity_variables} and their streamwise averages as well as standard deviations are given in table \ref{tab:similarity_variables_mean_SD}. The standard deviations of these coefficients over the measured streamwise domain size of $3.3\delta$ are comparable to corresponding standard deviations reported in \citet{kitsios2017direct}.

\begin{table}[!b]
\begin{center}
\caption{Streamwise average and standard deviation of $\beta$ and the similarity coefficients within the $x$ domain.}
\label{tab:similarity_variables_mean_SD} 
\resizebox{\columnwidth}{!}{%
\begin{tabular}{lccccc}
%\vspace{0.3em} \\
\hline \hline\noalign{\medskip}
	&	$\beta$		& $C_{uu}$		& $C_{vv}$		& $C_{uv}$		& $C_\nu$ \\
\noalign{\smallskip}\hline\noalign{\medskip}
Average			& 30.134		& \num{0.001643}		& \num{0.0005647}		& \num{0.01488}		& \num{0.00165} \\
Standard deviation	& 2.352 		& \num{0.0001911} 		& \num{7.196e-05} 		& \num{0.001827} 		&\num{7.256e-05} \\
\noalign{\smallskip}\hline\noalign{\medskip}
\end{tabular}}
\end{center}
\end{table}

\begin{figure}
\begin{center}
\begin{overpic}[width=0.45\textwidth]{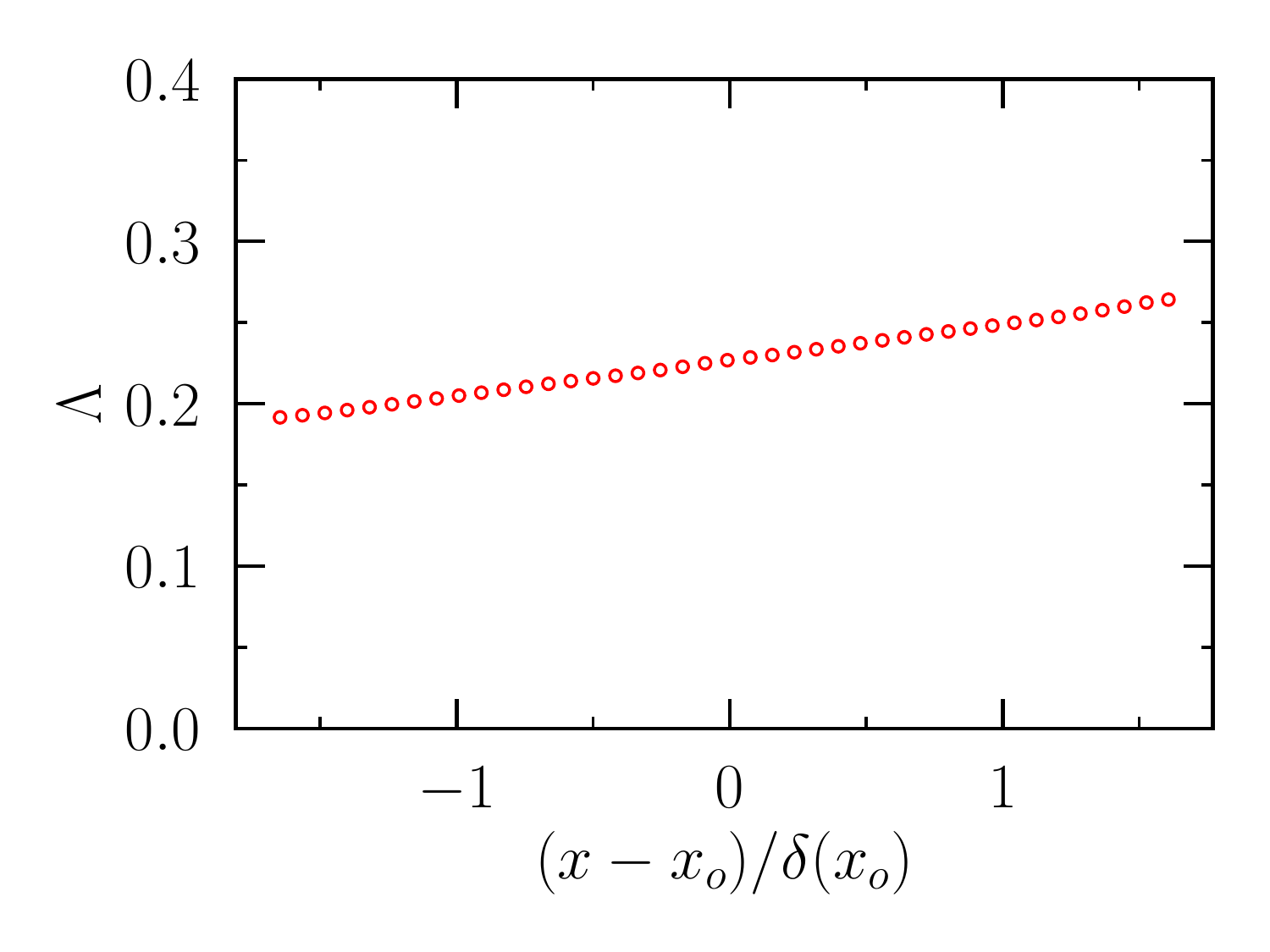} \end{overpic}
\end{center}
\vspace*{-0.2in}\caption{Streamwise variation of $\Lambda$.
\label{fig:Lambda}}
\end{figure}

The integration of equation $\Lambda \delta'_1/\delta_1 = -U'_e/U_e$, which is rearranged from the equation \ref{eq:Lambda}, can lead to the relationship $U_e \propto \delta_1^{-\Lambda}$. According to \citet{kitsios2017direct}, if a boundary layer grows linearly as in the present case, $U_e \propto (K x) ^{-\Lambda} \propto x^ {-\Lambda} \equiv x^m$ for $m = -\Lambda$. Figure \ref{fig:Lambda} shows the streamwise variation of $\Lambda$. In the middle of the FOV, the exponent $m$ is calculated using $\delta'_1 = 0.02651$, $P' = 2.8613$, $U_e = 5.4334$, and $\delta_1 = 0.06204$. The calculated value of $ m $ is $-0.23 $ which is expected for an incipient APG-TBL \citep{kitsios2017direct}.

%#***-------------------------------------------------------------------***#
\section{First- and second-order statistics}
\label{sec:first_second_order_stats}

\begin{figure}
\begin{center}
\begin{tabular}{c}
\begin{overpic}[height=2.6in,width=0.45\textwidth]{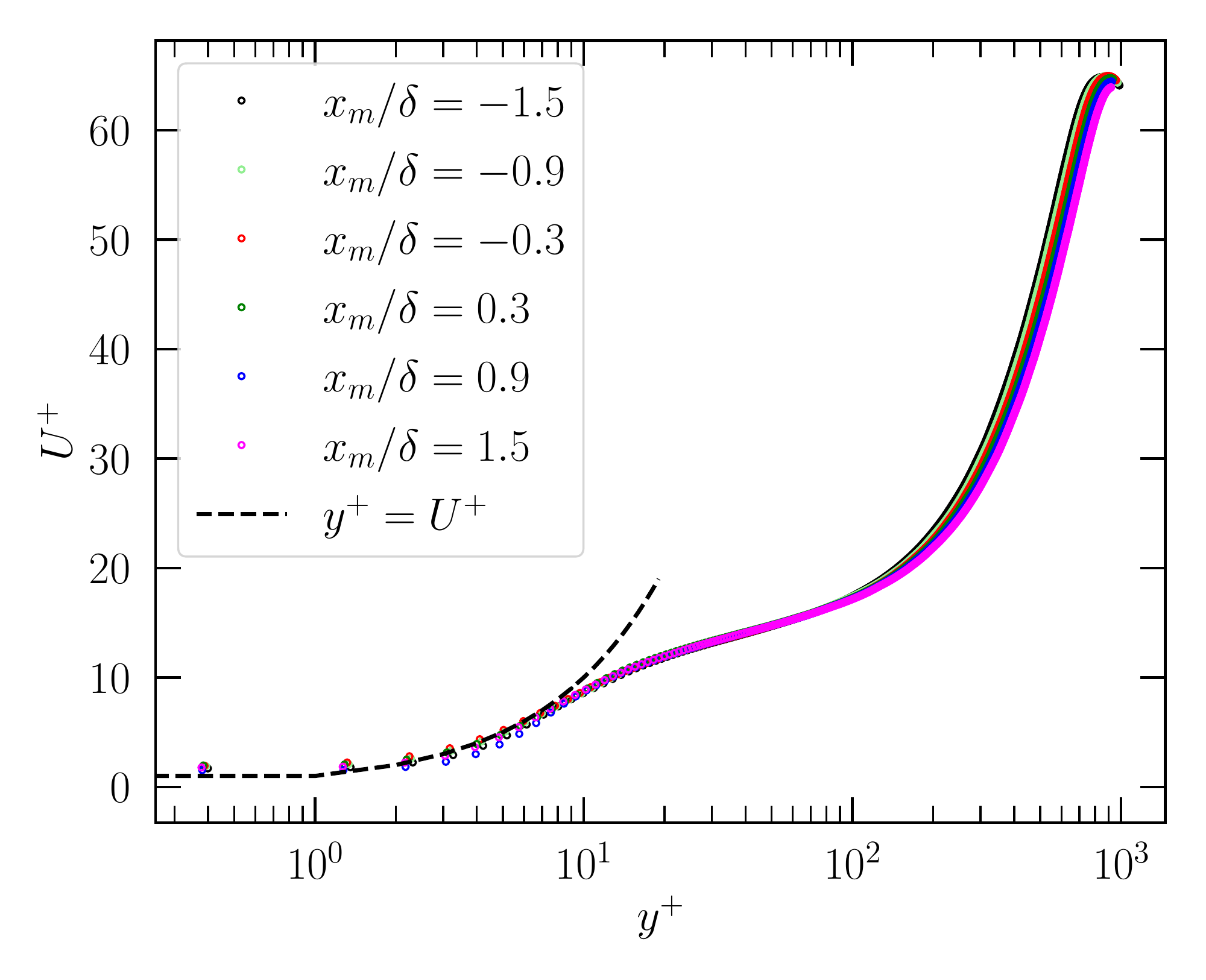} \put(60,70){(a)} \end{overpic} \\
\begin{overpic}[height=2.6in,width=0.475\textwidth]{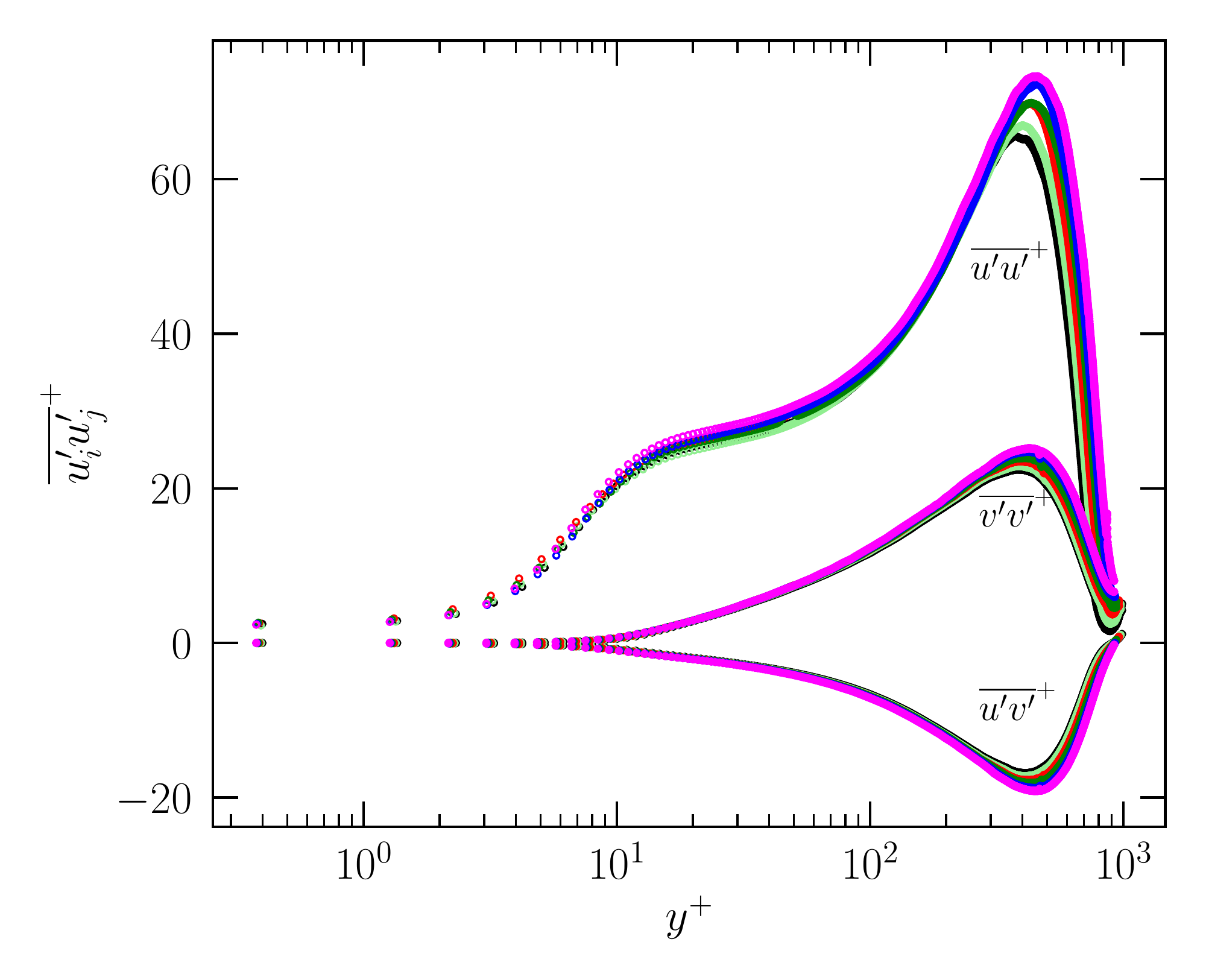} \put(30,65){(b)} \end{overpic} \\
\end{tabular}
\end{center}
\vspace*{-0.2in}\caption{(a) Inner-scaled mean streamwise velocity profiles at several equidistant streamwise locations, where $x_m = x - x_o$ and $\delta$ is the boundary layer thickness at the middle of the FOV. (b) Reynolds stress profiles at the same locations as in (a).
\label{fig:MVP_Re_stresses_inner}}
\end{figure}

Figure \ref{fig:MVP_Re_stresses_inner}(a) shows the profiles of the mean streamwise velocity scaled with the inner variable $u_\tau$ and $\nu / u_\tau$ at several equidistant streamwise positions. Here, a very good collapse is observed from the wall to the log layer. The slight spread in the wake region is the effect of the increasing Reynolds number along the streamwise direction. Figure \ref{fig:MVP_Re_stresses_inner}(b) shows the inner-scaled Reynolds stress profiles at the same streamwise locations as in \ref{fig:MVP_Re_stresses_inner}(a). The collapse in the inner region and the effect of the increasing Reynolds number in the outer region observed in the Reynolds stress profiles are similar to the mean velocity profiles. \textcolor{black}{This shows that the outer layer statistics do not scale with the viscous units in a strong APG-TBL, even if it is self-similar.}

\begin{figure}
\begin{center}
\begin{tabular}{c}
\begin{overpic}[height=2.6in,width=0.45\textwidth]{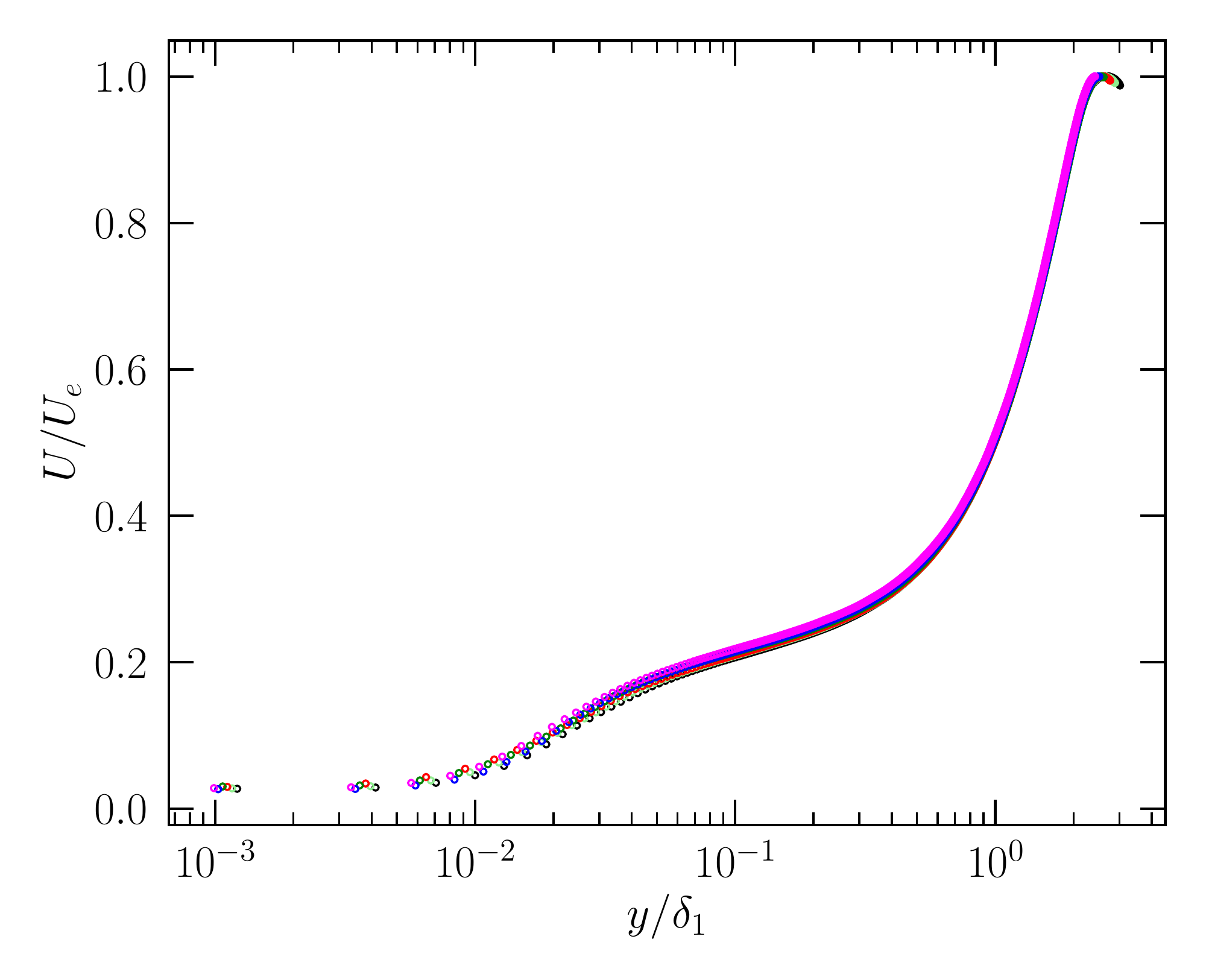} \put(20,70){(a)} \end{overpic} \\
\begin{overpic}[height=2.6in,width=0.475\textwidth]{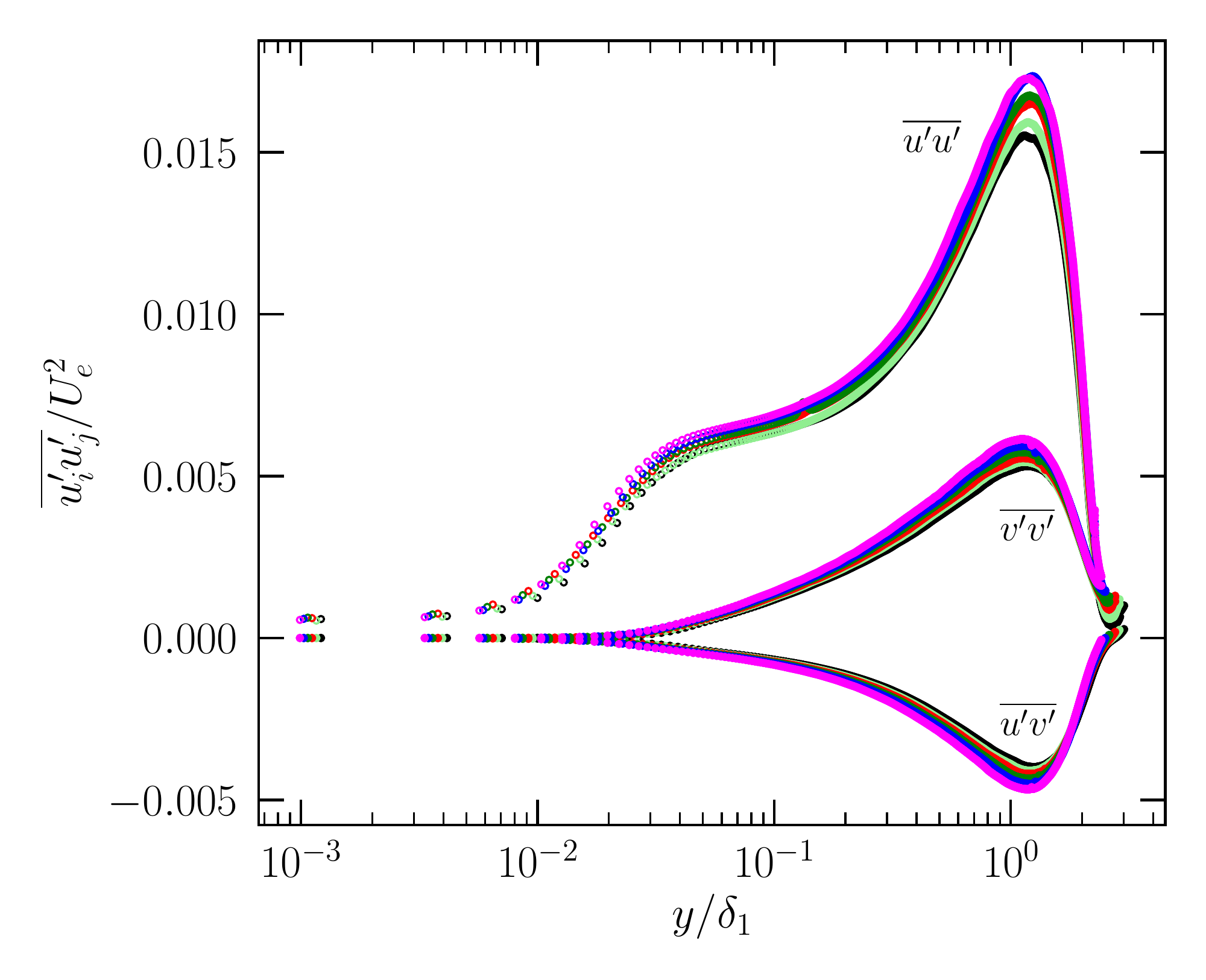} \put(30,65){(b)} \end{overpic} \\
\end{tabular}
\end{center}
\vspace*{-0.2in}\caption{(a) Mean streamwise velocity profiles, and (b) Reynolds stress profiles at the same streamwise locations as in figure \ref{fig:MVP_Re_stresses_inner}(a), scaled with the outer variables.
\label{fig:MVP_Re_stresses_outer}}
\end{figure}

 \textcolor{black}{To assess the self-similar nature of strong APG-TBL, the profiles of the mean streamwise velocity at several equidistant streamwise locations, scaled with the outer variables, $U_e$ and $\delta_1$, have been illustrated in figure \ref{fig:MVP_Re_stresses_outer}(a). Here, a very good collapse is observed in the whole boundary layer. This demonstrates the self-similar nature of the APG-TBL flow within the measurement domain.}

Reynolds stresses at the same streamwise locations as in \ref{fig:MVP_Re_stresses_outer}(a), scaled with the outer variables, are shown in figure \ref{fig:MVP_Re_stresses_outer}(b). The outer peaks in all the profiles for all Reynolds stresses are located slightly past $y=\delta_1$. These profiles also show a good collapse for each Reynolds stress, except in the region of the maximum shear stress, where they are slightly distant from each other. This is in agreement with the findings of \citet{kitsios2017direct} who also observed a spread among the peaks of the Reynolds stress profiles at several equally spaced $x$ locations. 

Since the profiles of the mean velocity and the Reynolds stress in the last 30\% of the DOI, scaled using the extrapolated length scale $\delta_1$ and the velocity scale $U_e$, are consistent with the profiles from the first 70\% of the DOI, the assumption that the boundary layer is expanding in the whole DOI is right, and the use of the extrapolated length and velocity scales is appropriate.

\begin{figure}
\begin{center}
\begin{tabular}{c}
\begin{overpic}[height=2.05in,width=0.4\textwidth]{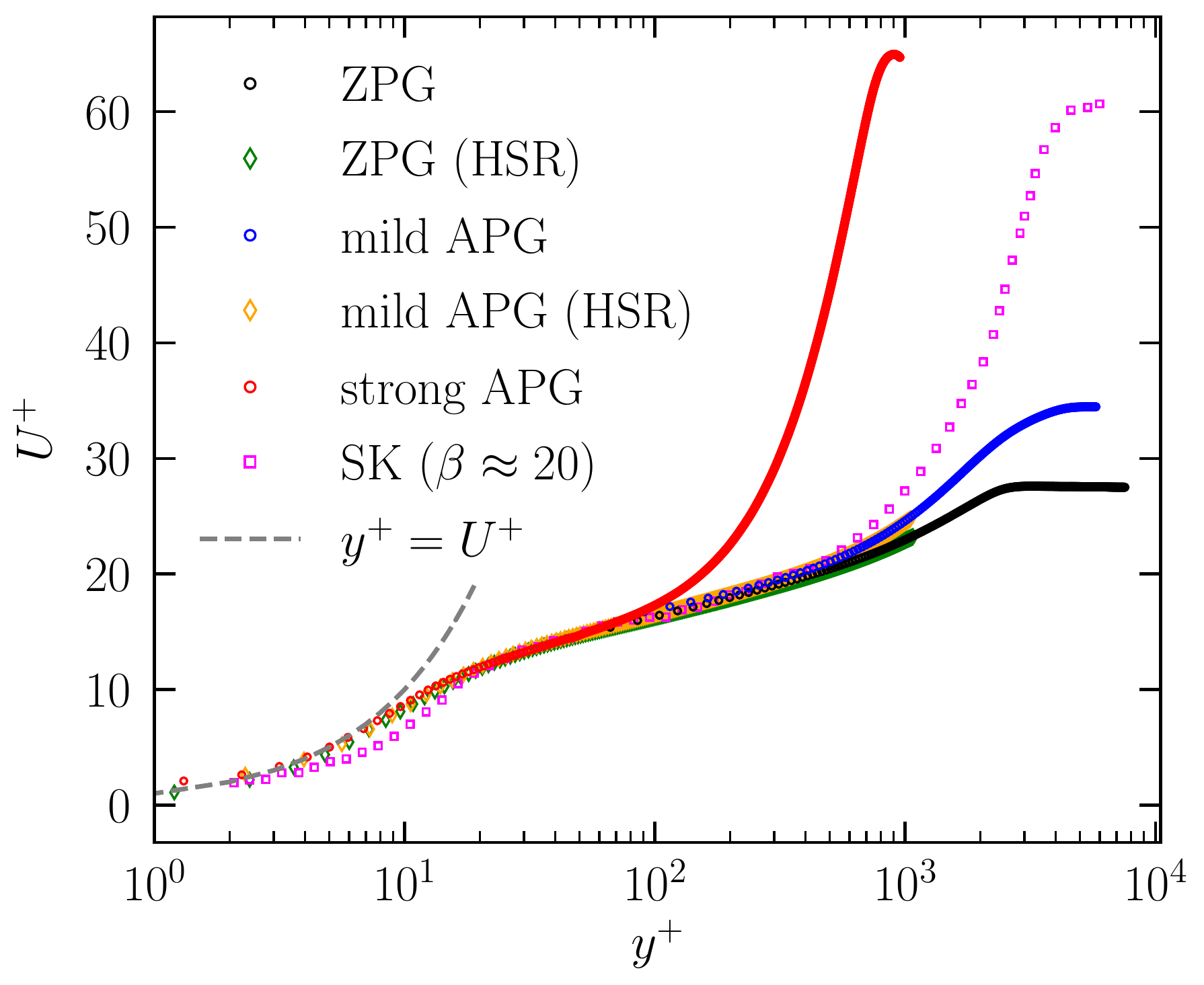} \put(80,25){(a)} \end{overpic} \\
\begin{overpic}[height=2.05in,width=0.4\textwidth]{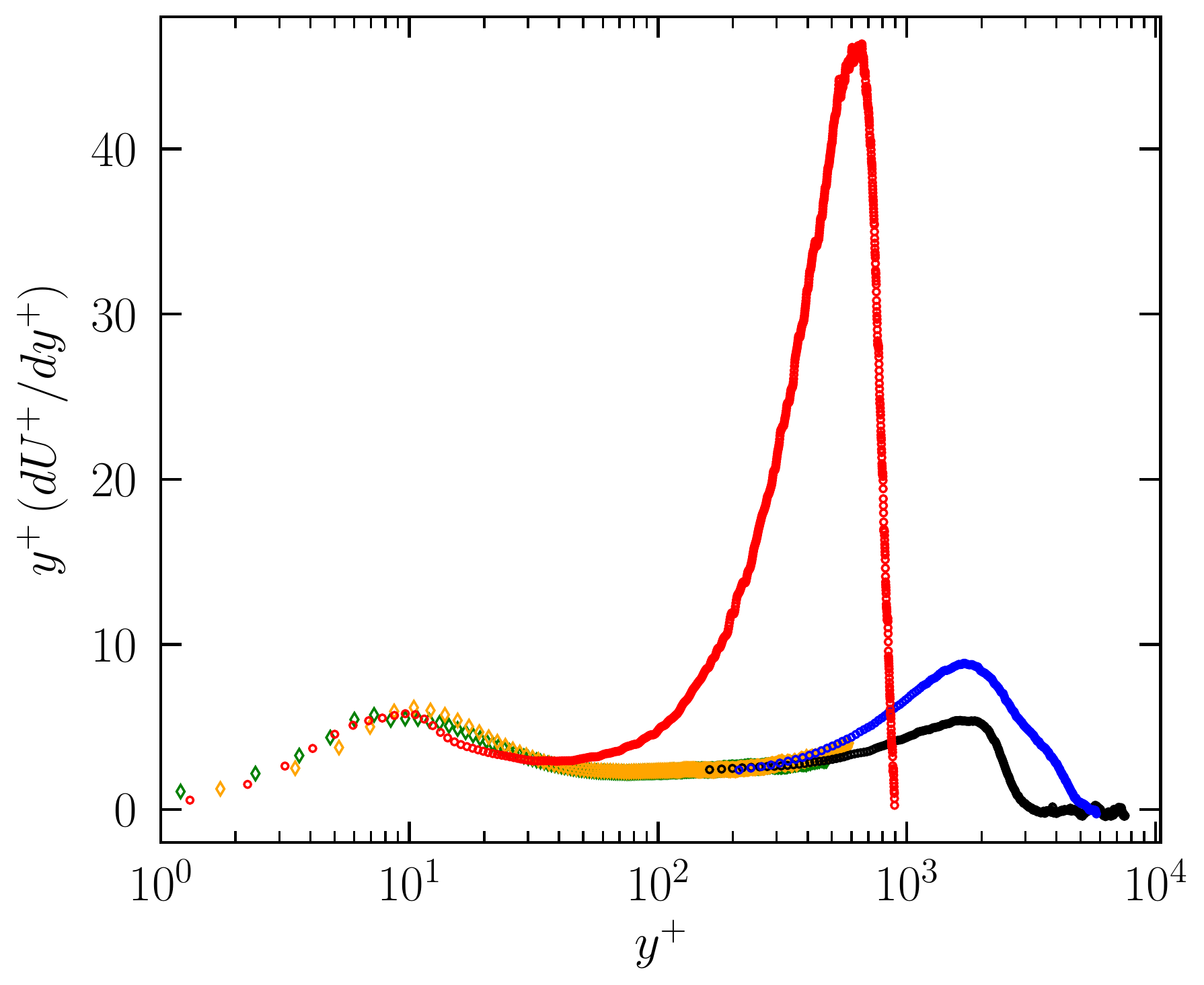} \put(80,63){(b)} \end{overpic} \\
\begin{overpic}[height=2.05in,width=0.4\textwidth]{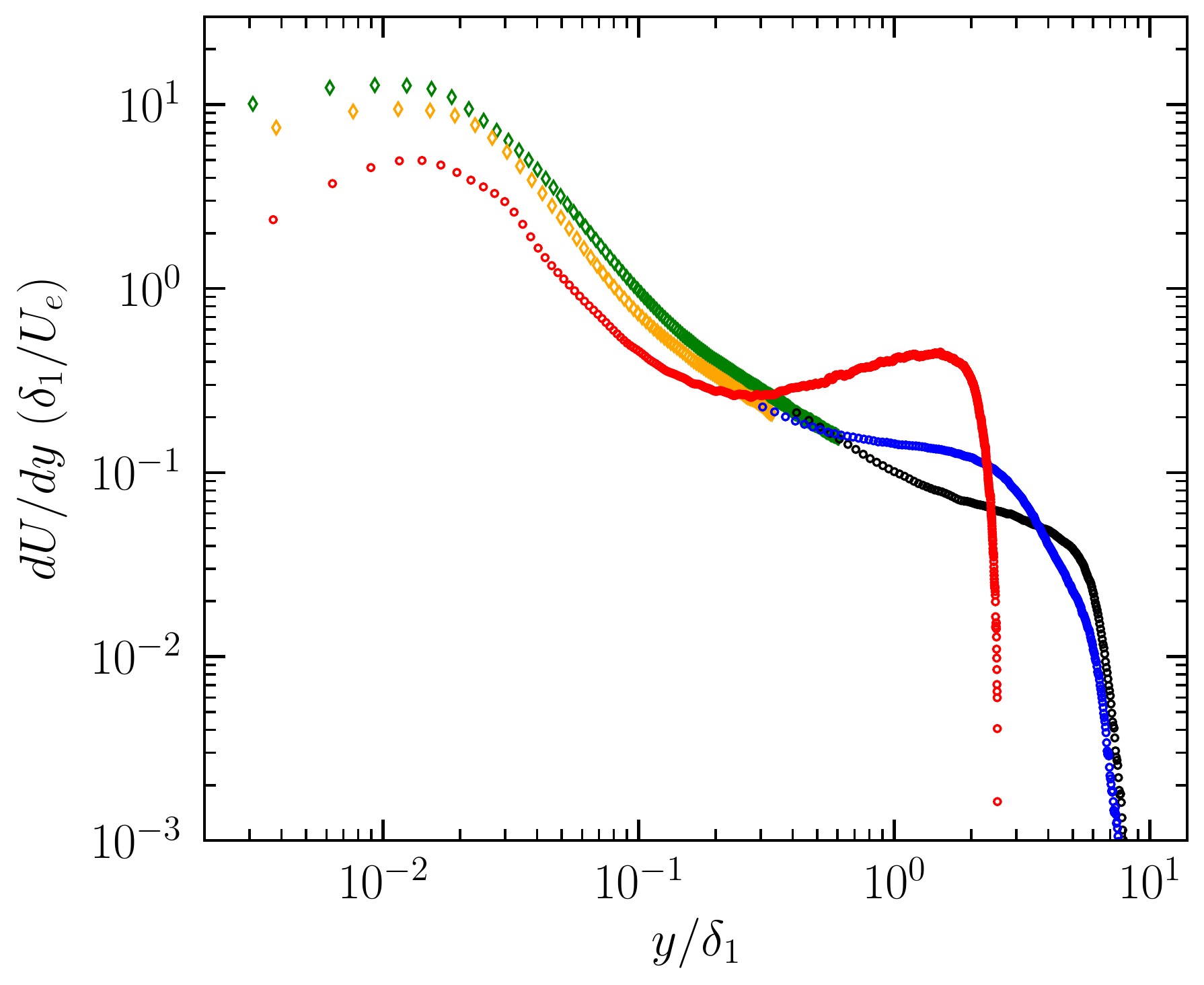} \put(80,63){(c)}\end{overpic} \\
\begin{overpic}[height=2.05in,width=0.4\textwidth]{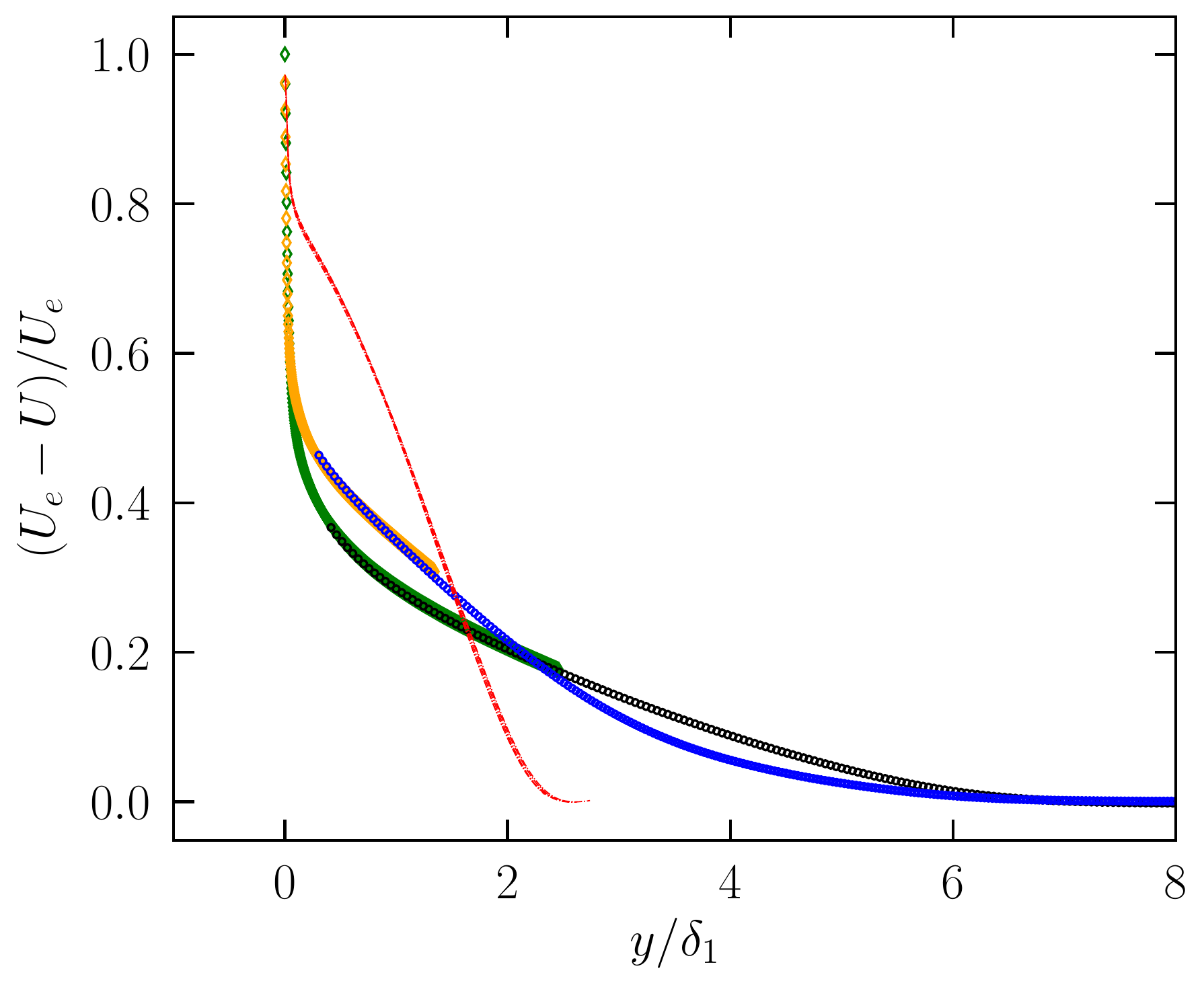} \put(80,63){(d)} \end{overpic} \\
\end{tabular}
\end{center}
\vspace*{-0.2in}\caption{(a) Mean streamwise velocity of all TBLs scaled with the viscous units, (b) diagnostic function $\Xi$ scaled with the viscous units, (c) mean shear rate scaled with the outer variables, (d) velocity defect profiles. All profiles are in the middle of the corresponding FOVs except the defect profiles of the strong APG-TBL, which are shown by the dash-dotted lines at six equidistant streamwise locations. 
\label{fig:U_MeanshearRate_defectU_inner_allTBL}}
\end{figure}

To assess the effect of the APG on the turbulent statistics, the profiles of the strong APG-TBL at the streamwise middle of the FOV are compared with the corresponding profiles of ZPG-TBL and the mild APG-TBL. Figure \ref{fig:U_MeanshearRate_defectU_inner_allTBL} illustrates the mean streamwise velocity for all TBLs scaled with the inner variables. In the ZPG-TBL and the mild APG-TBL, the outer-layer measurements are consistent with the inner-layer HSR measurements. The profile of the strong APG-TBL from \citet{skaare1994turbulent} ($\beta \approx 20$) is also included (referred to as the SK profile). As the pressure gradient increases, the extent of the wake region increases and the width of the log-layer decreases. These measurements follow the same trend as for the $U^+$ vs. $y^+$ profiles from the DNS of the four TBLs with $\beta$ values ranging from 0 to 9, reported by \citet{lee2017large}. However, their profiles in the log layer are shifted downward, with the degree of the shift increasing with an increase in APG. This is not observed in the present TBLs, including the SK profile. This may be due to the experimental uncertainty in the measurements of the boundary layer parameters of the current TBLs.

\begin{figure}
\begin{center}
\begin{tabular}{c}
\begin{overpic}[width=0.4\textwidth]{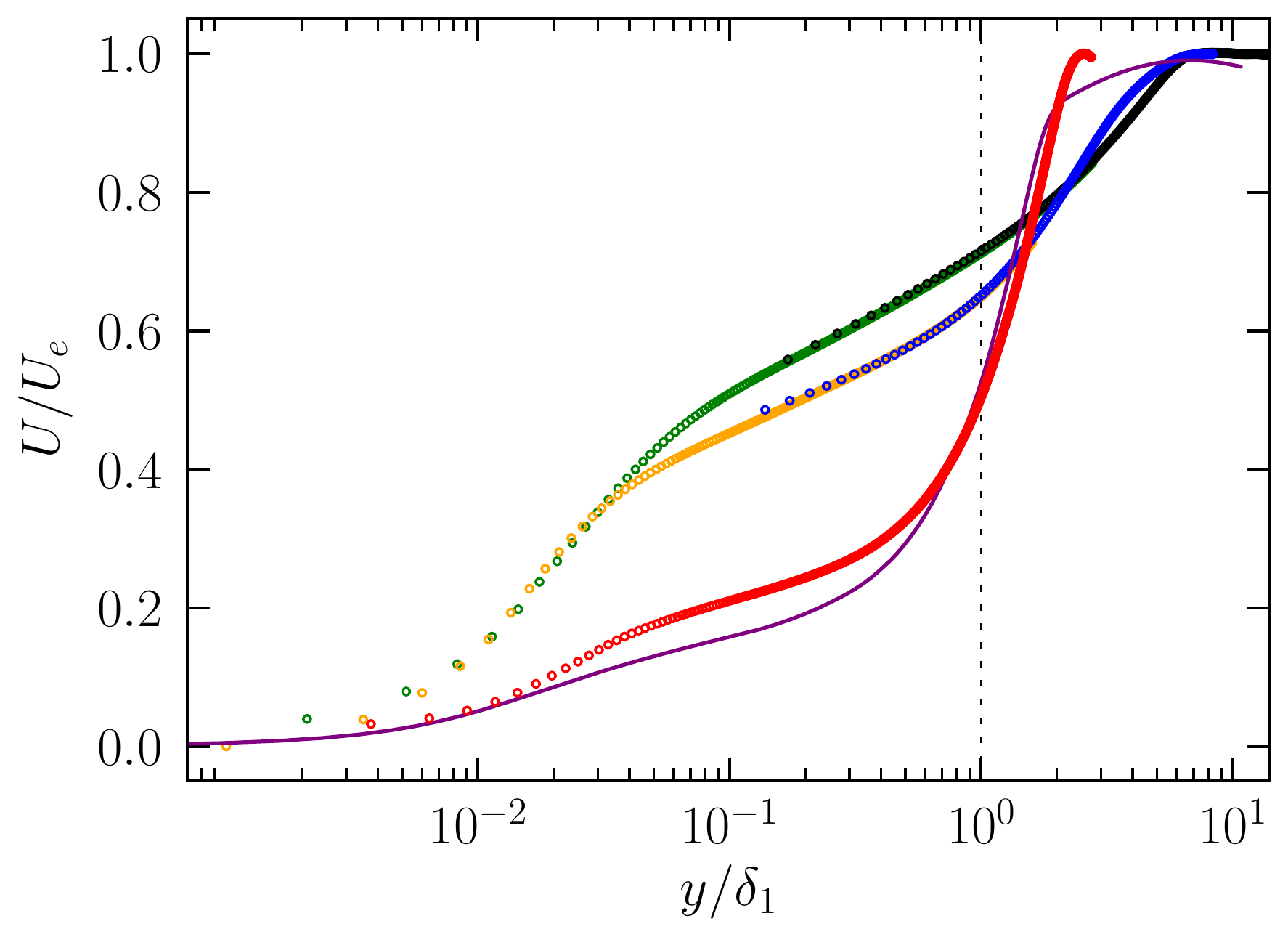} \put(22,63){(a)} \end{overpic} \\
\begin{overpic}[width=0.4\textwidth]{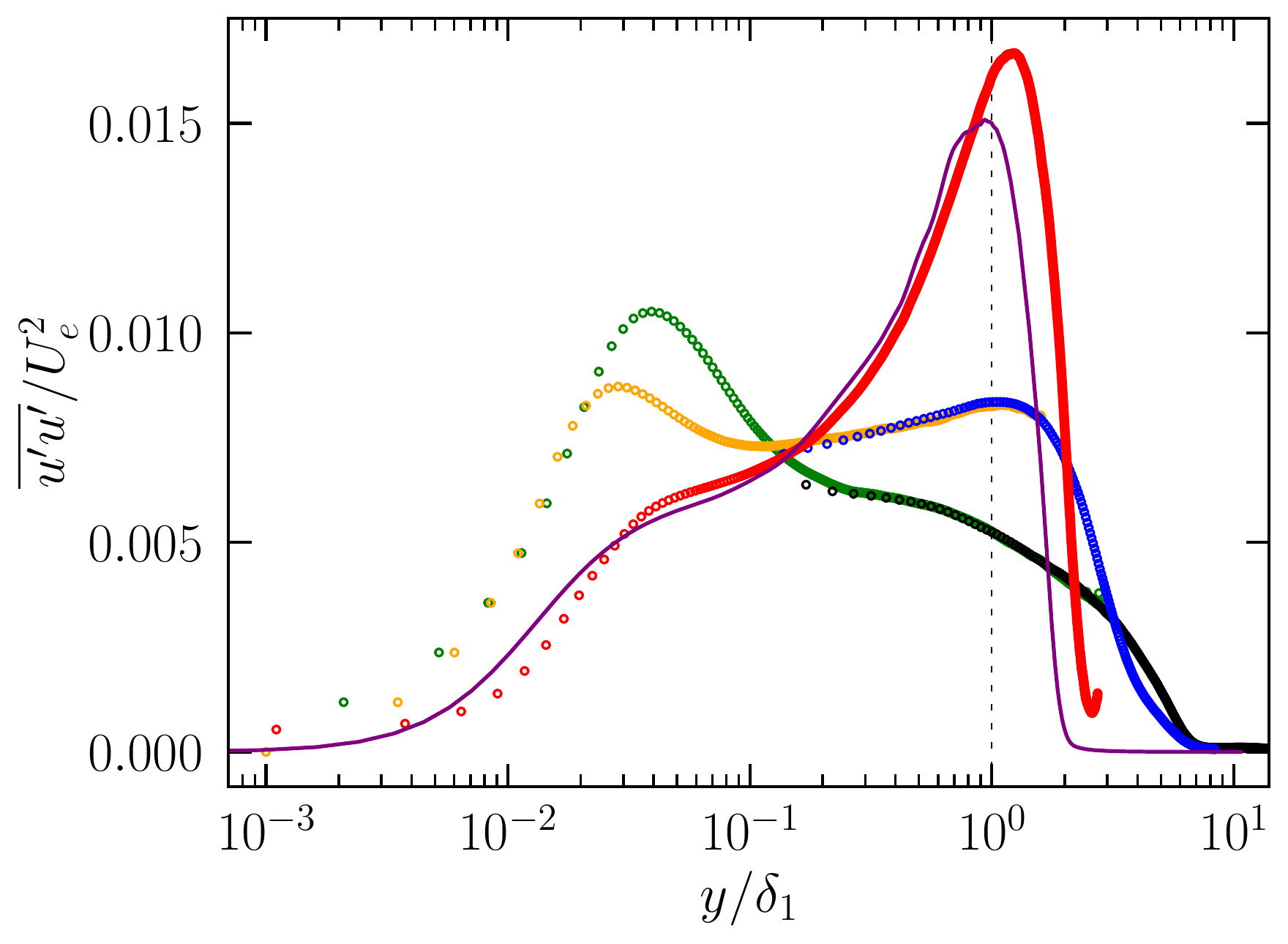} \put(24,63){(b)} \end{overpic} \\
\begin{overpic}[width=0.4\textwidth]{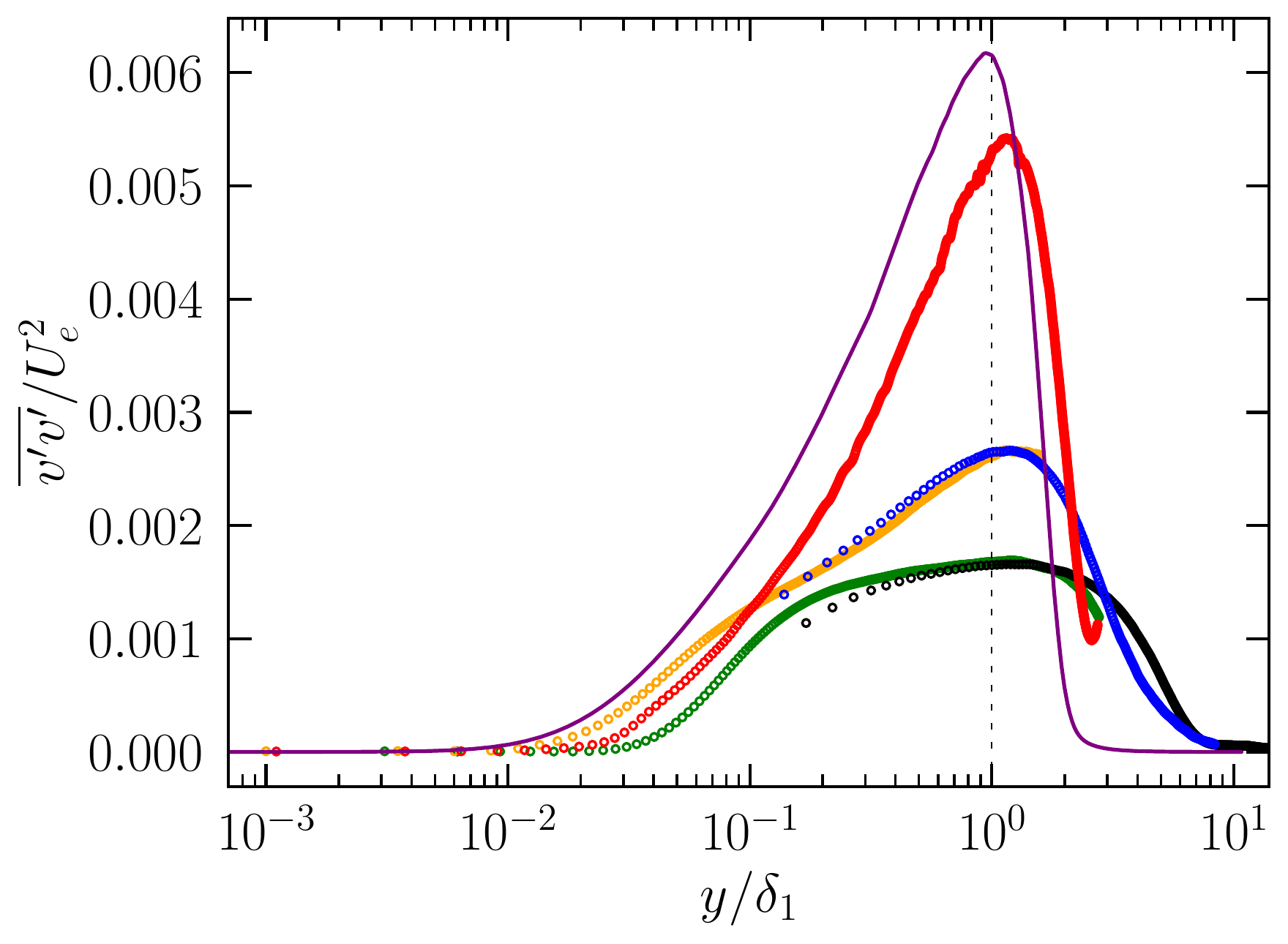} \put(22,60){(c)} \end{overpic} \\
\begin{overpic}[width=0.4\textwidth]{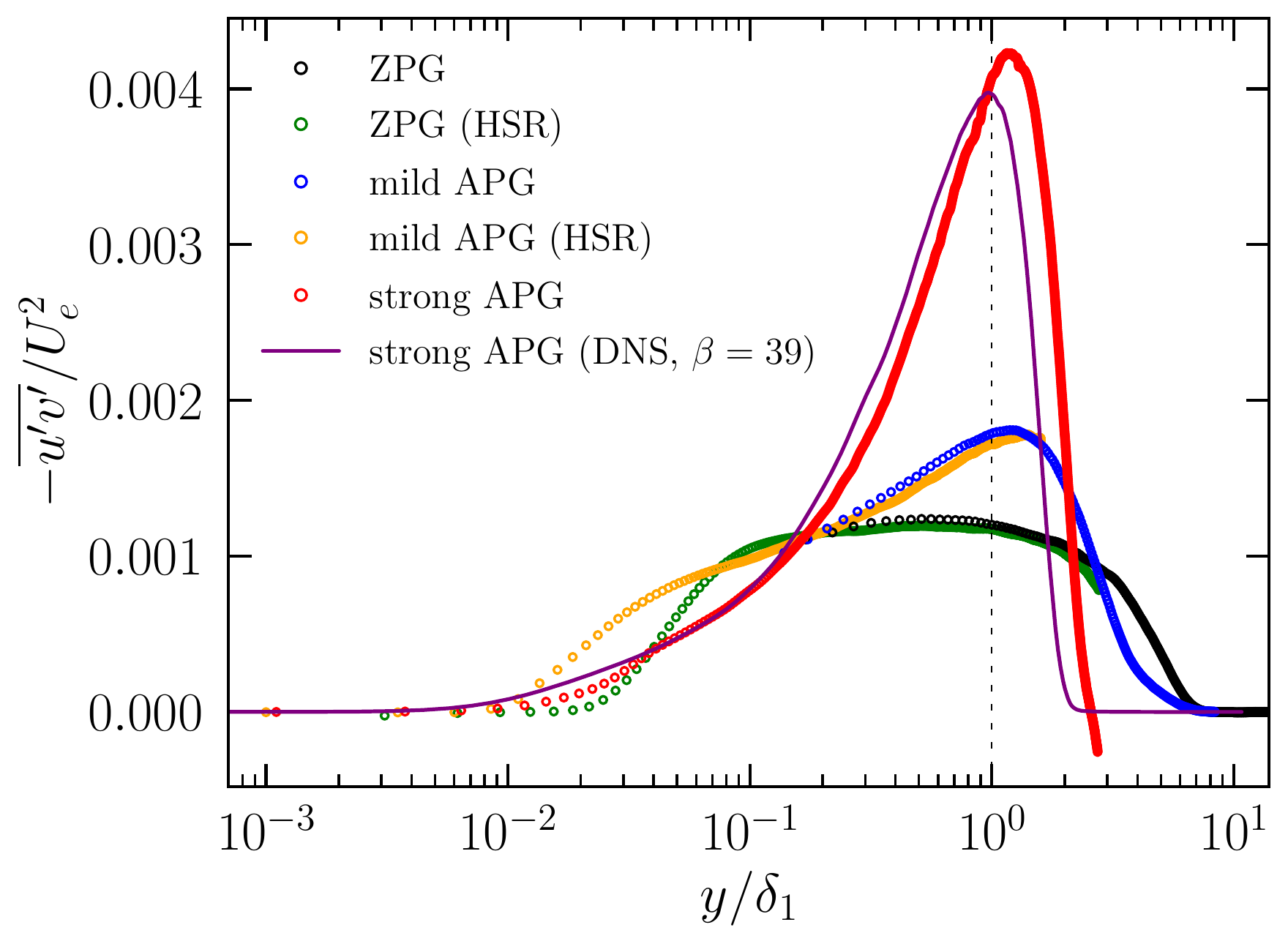} \put(22,30){(d)} \end{overpic} \\
\end{tabular}
\end{center}
\vspace*{-0.2in}\caption{Mean streamwise velocity and the Reynolds stress profiles, scaled with the outer variables, for all TBLs at the middle of the FOVs.
\label{fig:stats_comparison_allTBL}}
\end{figure}

Profiles of the diagnostic function $\Xi = y^+ (dU^+ / dy^+)$ for all TBLs are shown in figure \ref{fig:U_MeanshearRate_defectU_inner_allTBL}(b). Near the wall, $\Xi$ is observed to be almost invariant with the pressure gradient because of the law of the wall. The peaks in the outer region are caused by the inflections in the wake of the velocity profiles. The $\Xi$ profiles also indicate that with an increase in $\beta$, the log-region slope ($1/\kappa$) increases and the log-region moves closer to the wall. These findings are also consistent with those reported in \citet{lee2017large}. Figure \ref{fig:U_MeanshearRate_defectU_inner_allTBL}(c) shows the mean shear rate normalised by the outer variables. Below $y = 0.3 \delta_1$, the strength of the mean shear rate is reduced with increasing APG. Further out, the opposite is the case. Figure \ref{fig:U_MeanshearRate_defectU_inner_allTBL}(d) represents the mean velocity defect profiles of all TBL, scaled with the outer variables. Here, the profiles of the strong APG-TBL at six equally-spaced streamwise locations (as in figure \ref{fig:MVP_Re_stresses_inner}(a)) are shown as dash-dotted lines. \textcolor{black}{The outer scaling shows a remarkable collapse of the defect profiles, demonstrating the self-similar nature of the flow. The similarity among these profiles is noticeably higher than that presented by \citet{lee2017large} for various APG-TBLs while using the local friction velocity and the defect thickness $\Delta = \delta_1 (2/C_f)^{1/2}$ \citep{skote1998direct} for scaling. Near the wall, the defect velocity normalised with the local $U_e$ has larger values for larger $\beta$. The ZPG-TBL has the longest wall-normal defect region, which decreases in its extent with an increase in APG.}  

In figure \ref{fig:stats_comparison_allTBL}(a), the mean velocity profiles of all TBLs scaled with outer variables $\delta_1$ and $U_e$ are shown. Similarly, the Reynolds stress profiles of these data sets are also scaled with the outer variables and are shown in figures \ref{fig:stats_comparison_allTBL}(b-d). The DNS profiles of the strong APG-TBL ($\beta = 39$) of \citet{kitsios2017direct} are also included for comparison. The present profiles of the mean streamwise velocity for different $\beta$ are qualitatively similar to the corresponding ZPG, mild APG, or strong APG profiles reported in \citet{kitsios2017direct}. But as $\beta$ values of their mild and strong APG cases don't exactly match the $\beta$ value in the current case, the quantitative similarity between the compared profiles is not expected. 

The $\overline{u'u'}$ profile of the ZPG-TBL has only an inner peak and no outer peak. The profile of the mild APG-TBL has an outer peak that is as strong as the inner peak and is located around ${y=1.3\delta}_1$. As the APG increases, the outer peak becomes stronger and moves closer to ${y=\delta}_1$, while the inner peak becomes more and more diminished. Similarly, the outer peaks in $\overline{v' v'}$ and $-\overline{u' v'}$ profiles also become stronger with increasing APG. This indicates the movement of the turbulent activity away from the wall and a reduced area of influence of the near-wall length and velocity scales with an increase in APG. The $\overline{v' v'}$ and $-\overline{u' v'}$ profiles exhibit plateaus in the outer region in ZPG-TBL. 

Interestingly, at the wall-normal position of $y \approx 0.15 \delta_1$, the $\overline{u'u'}$ and $-\overline{u'v'}$ profiles exhibit similar values among all TBLs when scaled with the outer variables.

%#***-------------------------------------------------------------------***#
\subsection{Quadrant decomposition of the Reynolds shear stress}
\label{sec:overall_quadrant_analysis}

Ejections and sweeps are studied using the quadrant splitting method proposed by \citet{wallace1972wall} who used it in a turbulent channel flow at $Re_\tau = 187$. The four quadrants are: $Q1(+u', +v')$, $Q2 (-u', +v')$ (ejections), $Q3 (-u', -v')$, and $Q4 (+u', -v')$ (sweeps). This segmentation allows the study of the contributions of ejections and sweeps to Reynolds shear stress across the boundary layer.

\begin{figure*}
\begin{center}
\includegraphics[width=0.8\textwidth]{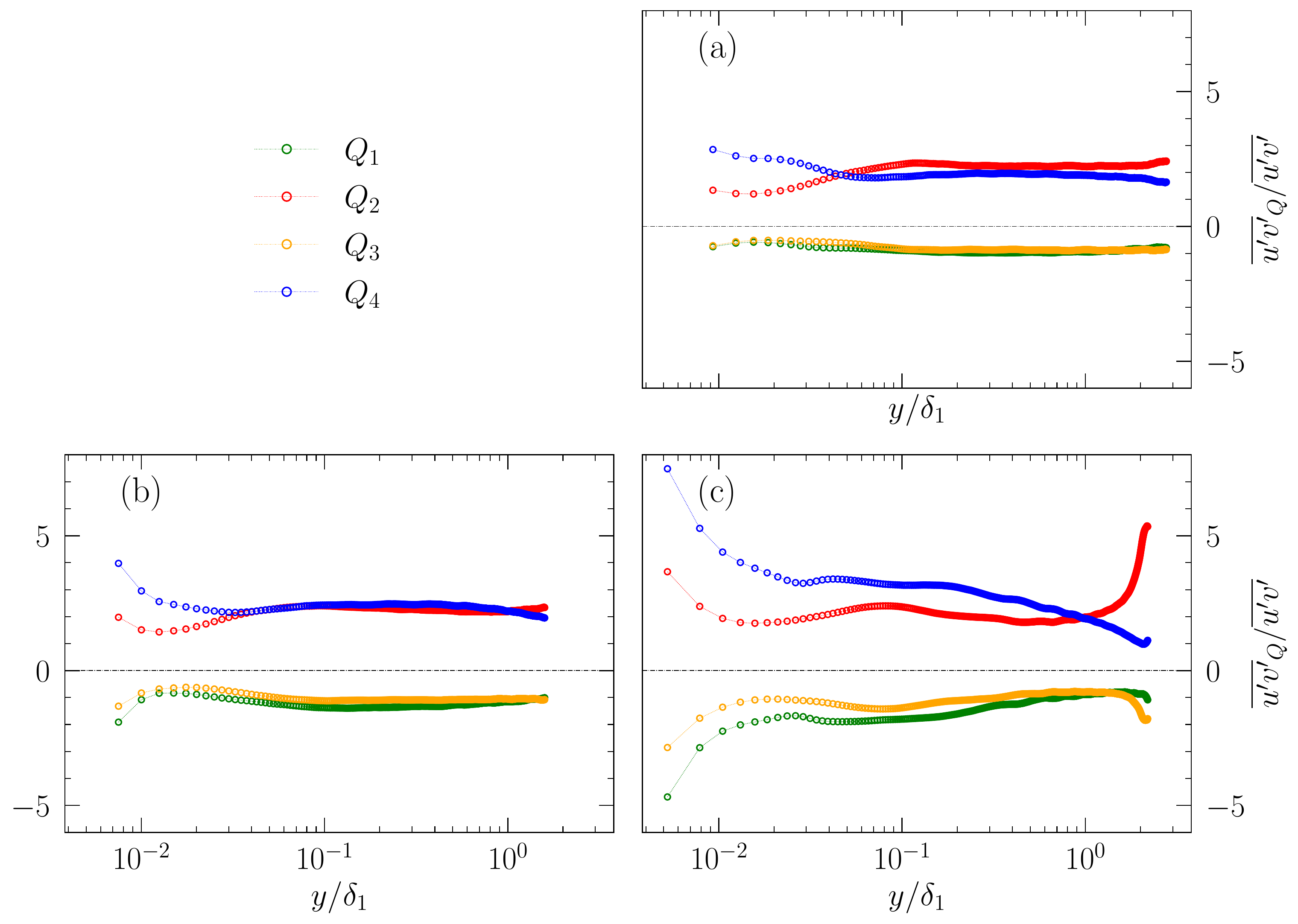}
\end{center}
\vspace*{-0.2in}\caption{Quadrant contributions to the Reynolds shear stress in (a) ZPG-TBL, (b) mild APG-TBL, and (c) strong APG-TBL.
\label{fig:quadrant_analysis}}
\end{figure*}

Figure \ref{fig:quadrant_analysis} presents the contribution of the quadrants to the Reynolds shear stress $\overline{{u'v'}}$, where $\overline{u'v'}_Q$ is the contribution of the Reynolds shear stress of an individual quadrant $Q$ and normalised by the overall $\overline{{u'v'}}$. As expected, Q2 and Q4 events have positive contributions, while Q1 and Q3 events have negative contributions from the wall to the boundary layer edge in all TBLs. In the ZPG-TBL, very close to the wall, Q4 contributes considerably more than Q2. Beyond the cross-over point at $y\approx 0.04\delta_1$, the contribution of Q2 is larger than Q4. Above $y = 0.2\delta_1$, the contributions of both ejections and sweeps are nearly constant. In the same region, Q1 and Q3 have similar contributions which are nearly 50\% of the Q4 contribution but with the opposite sign. The distribution of the quadrant contributions is qualitatively similar to the quadrant contributions in a turbulent channel flow presented in \citet{wallace1972wall}. 

In the mild APG-TBL, contributions of Q2 and Q3 are similar to their counterparts in the ZPG-TBL, but are up to 80\% more intense near the wall. Contrary to the ZPG-TBL, Q4 contributes more than Q2 from the wall to $y \approx \delta_1$ in the mild APG-TBL. Similarly, Q1 contributes more than Q3 in the same region. 

There is no significant increase in the Q2 and Q3 contributions due to the strong APG, but the Q4 and Q1 contributions are stronger than in the mild APG-TBL from the wall to $y \approx \delta_1$. In this region, the contributions of sweeps are dominant over the ejections. These observations are consistent with the observations of \citet{skaare1994turbulent} who report that the sweeps are the dominant motions in a strong APG-TBL, whereas in the ZPG-TBL both the ejections and sweeps are equally important when analysed using the quadrant-splitting of \citet{wallace1972wall}.

\textcolor{black}{Since the contributions of both Q1 and Q4 are amplified under APG, the increase in the contribution of Q4 is greater than in Q1 and this demonstrates that when an APG is imposed, the outer peak in the Reynolds shear stress profile is formed due to the energisation of sweep motions of high-momentum fluid in the outer region of the boundary layer.}

%#***-------------------------------------------------------------------***#
\section{Turbulence production}
\label{sec:turbulence_production}

\begin{figure}[!b]
\begin{center}
\begin{overpic}[width=0.45\textwidth]{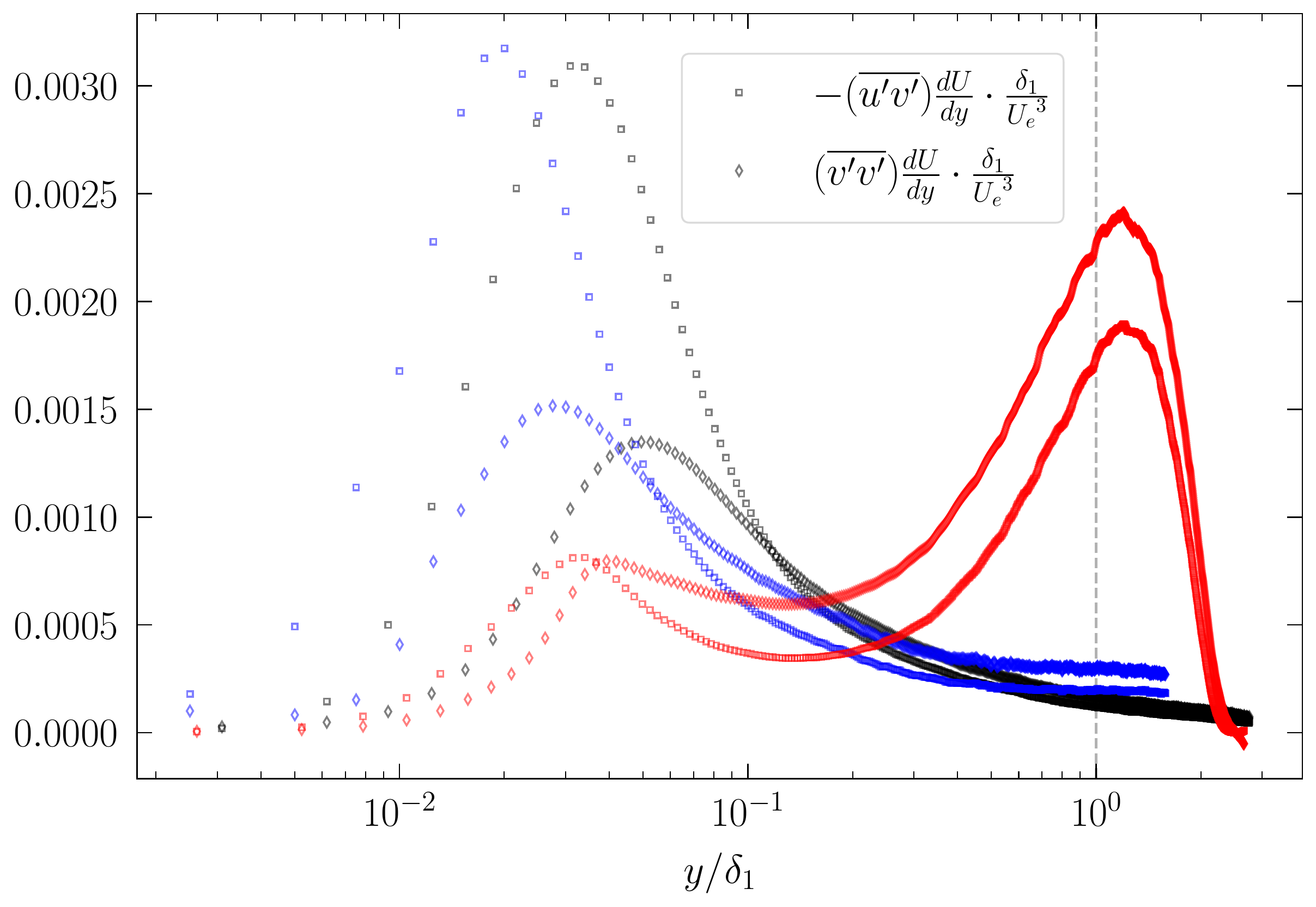} \end{overpic}
\end{center}
\vspace*{-0.2in}\caption{Profiles of the dominant turbulence production terms $-\overline{u^\prime v^\prime} \partial U/\partial y$ and $\overline{{v^\prime}^2} \partial U/\partial y$. Black symbols: ZPG-TBL, blue symbols: mild APG-TBL, and red symbols: strong APG-TBL.
\label{fig:turbulence_production}}
\end{figure}

The terms that significantly contribute to the turbulence production in the current TBLs are $-\overline{u^\prime v^\prime} \partial U/\partial y$ and $\overline{{v^\prime}^2} \partial U/\partial y$. These terms are present in the transport equations for $\overline{{u^\prime}^2}$ and $-\overline{u^\prime v^\prime}$, respectively. Figure \ref{fig:turbulence_production} shows the profiles of these terms scaled with $U_e$ and $\delta_1$. Because the turbulence production terms are prominent in the near-wall region of the ZPG-TBL and the mild APG-TBL, only the near-wall HSR measurements of these TBLs have been included. In both TBLs, the term $-\overline{u^\prime v^\prime} \partial U/\partial y$ is the dominant contributor, while $\overline{{v^\prime}^2} \partial U/\partial y$ contributes about 50\% of the other term. Furthermore, the maximum values of these terms are about 3\% and 11\% stronger in the mild APG-TBL than in the ZPG-TBL, respectively. Also, the peaks in the APG-TBL are closer to the wall than in the ZPG-TBL. No significant outer peaks are found in both TBLs within the range of the HSR data. However, the contribution of $\overline{{v^\prime}^2} \partial U/\partial y$ is greater than that of $-\overline{u^\prime v^\prime} \partial U/\partial y$ in the outer region.

\textcolor{black}{The strong APG-TBL shows the weaker inner peaks and stronger outer peaks, which demonstrate that with a strong increase in APG, the turbulence production is shifted from the near-wall region towards the outer region.}
Also, the overall contribution of $-\overline{u^\prime v^\prime} \partial U/\partial y$ is reduced and that of $\overline{{v^\prime}^2} \partial U/\partial y$ is enhanced. Compared to the peak in the $-\overline{u^\prime v^\prime} \partial U/\partial y$ profile of the ZPG-TBL, the weak inner peaks are about 75\% smaller and the stronger outer peaks are 39\% and 22\% smaller for the profiles of $-\overline{u^\prime v^\prime} \partial U/\partial y$ and $\overline{{v^\prime}^2} \partial U/\partial y$, respectively. \textcolor{black}{For the strong APG-TBL, the wall-normal location of the outer peaks in the turbulence production terms matches the outer peak locations in the Reynolds shear stress profiles and the outer inflection point in the velocity profile (shown in figure \ref{fig:stats_comparison_allTBL}(a)).} This is consistent with the observation of \citet{skaare1994turbulent} in a strong APG-TBL at $\beta \approx 20$, who also found two distinct peaks in the inner and outer regions for each of the above-plotted terms. They could not resolve the locations of the inner peak because their hotwire data did not have sufficient spatial resolution in the wall-normal direction. However, the outer peaks were found at the locations of maximum stress values, {\em i.e.} $y/\delta = 0.45$ which corresponds to $y = 1.69\delta_1$. As the APG increases, the outer peaks move closer to the displacement thickness height. Since, the current TBL has a stronger APG ($\beta \approx 30$), it has the peak locations at $y = 1.3\delta_1$. This shows the consistency of the current measurements with the observations made in previous studies.

%#***-------------------------------------------------------------------***#
\section{Third-order moments}

\begin{figure}
\begin{center}
\begin{overpic}[width=0.45\textwidth]{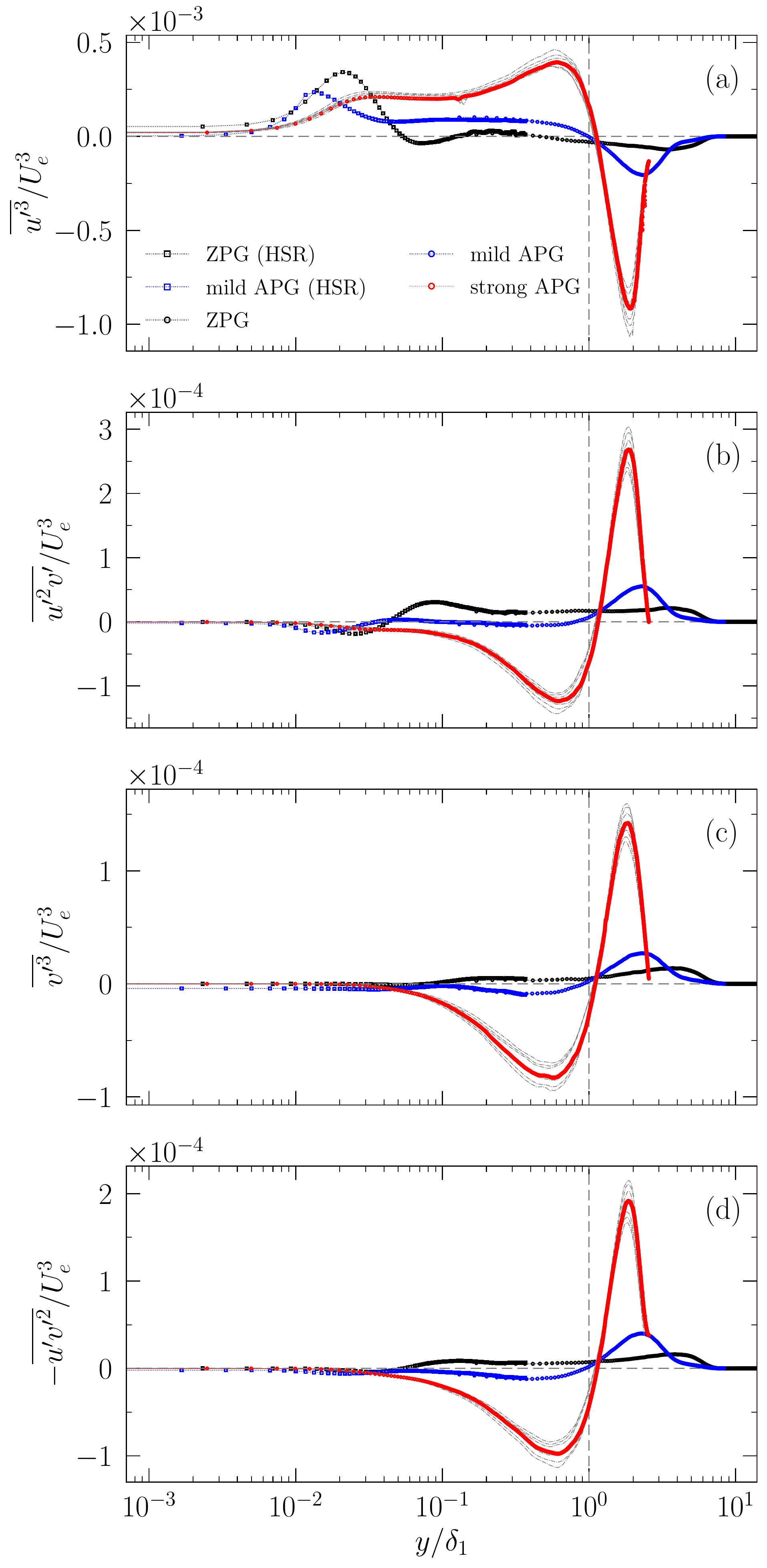} \end{overpic}
\end{center}
\vspace*{-0.2in}\caption{Third-order moments scaled with the outer variables.
\label{fig:third_order_moments}}
\end{figure}

Figure \ref{fig:third_order_moments} shows the outer-scaled profiles of the third-order moments: $\overline{{u'}^3}$, $\overline{{u'}^2 v'}$, $\overline{{v'}^3}$, and $-\overline{u' {v'}^2}$. The first term represents the turbulent transport of $\overline{{u'}^2}$ in the streamwise direction. The following three quantities represent the turbulent transport of $\overline{{u'}^2}$, $\overline{{v'}^2}$, and the turbulent work of $\overline{u'v'}$ along the wall-normal direction, respectively. For the strong APG-TBL, the third-order statistics are shown in the middle of the FOV. The figure also includes the strong APG-TBL profiles at several equidistant streamwise locations within the FOV, shown by the dot-dashed lines. It is obvious that these profiles collapse with the outer scaling, except at the maxima, where their distribution is similar to the distributions at peak locations in the corresponding Reynolds stress profiles. The maximum magnitude of $\overline{{u'}^3}$ is 3.4, 6.4, and 4.8 times larger than the maximum magnitudes of $\overline{{u'}^2 v'}$, $\overline{{v'}^3}$, and $-\overline{u' {v'}^2}$, respectively. The locations of the zero-crossing of the corresponding third-order moment profiles are approximately the same as the locations of the outer peaks in the corresponding Reynolds stress profiles. Below these locations, energy is transported downward from Reynolds stresses $\overline{{u'}^2}$, $\overline{{v'}^2}$, and $-\overline{u'v'}$. Beyond these locations, the energy is diffused towards the boundary layer edge. These observations are consistent with the findings of \citet{skaare1994turbulent} for a strong APG-TBL in $\beta \approx 20$ and $Re_{\delta_2} = 44,420$.

The zero crossings of the third-order moments in the mild APG-TBL are at $y \approx \delta_1$. There is an inner peak in the $\overline{u'^3}$ profile that is almost as strong as the inner peak in the strong APG-TBL, but slightly closer to the wall. The outer negative peak, however, is about five times weaker than in the strong APG-TBL and lies further away from the wall. There is no outer positive peak visible in the mild APG-TBL. In the profiles of $\overline{{u'}^2 v'}$, $\overline{{v'}^3}$, and $-\overline{u' {v'}^2}$ as well, the outer positive peaks are about five times weaker than in the strong APG-TBL. \textcolor{black}{This demonstrates that a decrease in the APG reduces the energy diffusion towards the boundary layer edge from the locations of the outer peaks in the Reynolds stress profiles, and this phenomenon is shifted closer to the boundary layer edge.}

In the ZPG-TBL, the inner positive peak in $\overline{{u'}^3}$ profile is stronger than in both APG-TBLs and its zero crossing is at the location of the inner peak in the $\overline{{u'}^2}$ profile. Beyond this point, $\overline{{u'}^3}$ fluctuates between positive and negative values. Contrary to APG-TBL, the profiles of $\overline{{u'}^2 v'}$, $\overline{{v'}^3}$, and $-\overline{u' {v'}^2}$ in the ZPG-TBL have positive values beyond their zero crossings in the inner region. The outer negative peak in the $\overline{{u'}^3}$ profile and the outer positive peaks in the profiles of $\overline{{u'}^2 v'}$, $\overline{{v'}^3}$, and $-\overline{u' {v'}^2}$ are about 20 times weaker than in the strong APG-TBL and are found further from the wall than in the mild APG-TBL. 

Another interesting observation is that the wall-normal profiles of the third-order moments of all TBLs meet at a single point, $y \approx 1.2 \delta_1$. The significance of this point is yet to be explored.

%#***-------------------------------------------------------------------***#
\section{Skewness and Flatness}

\begin{figure*}
\begin{center}
\begin{overpic}[width=0.8\textwidth]{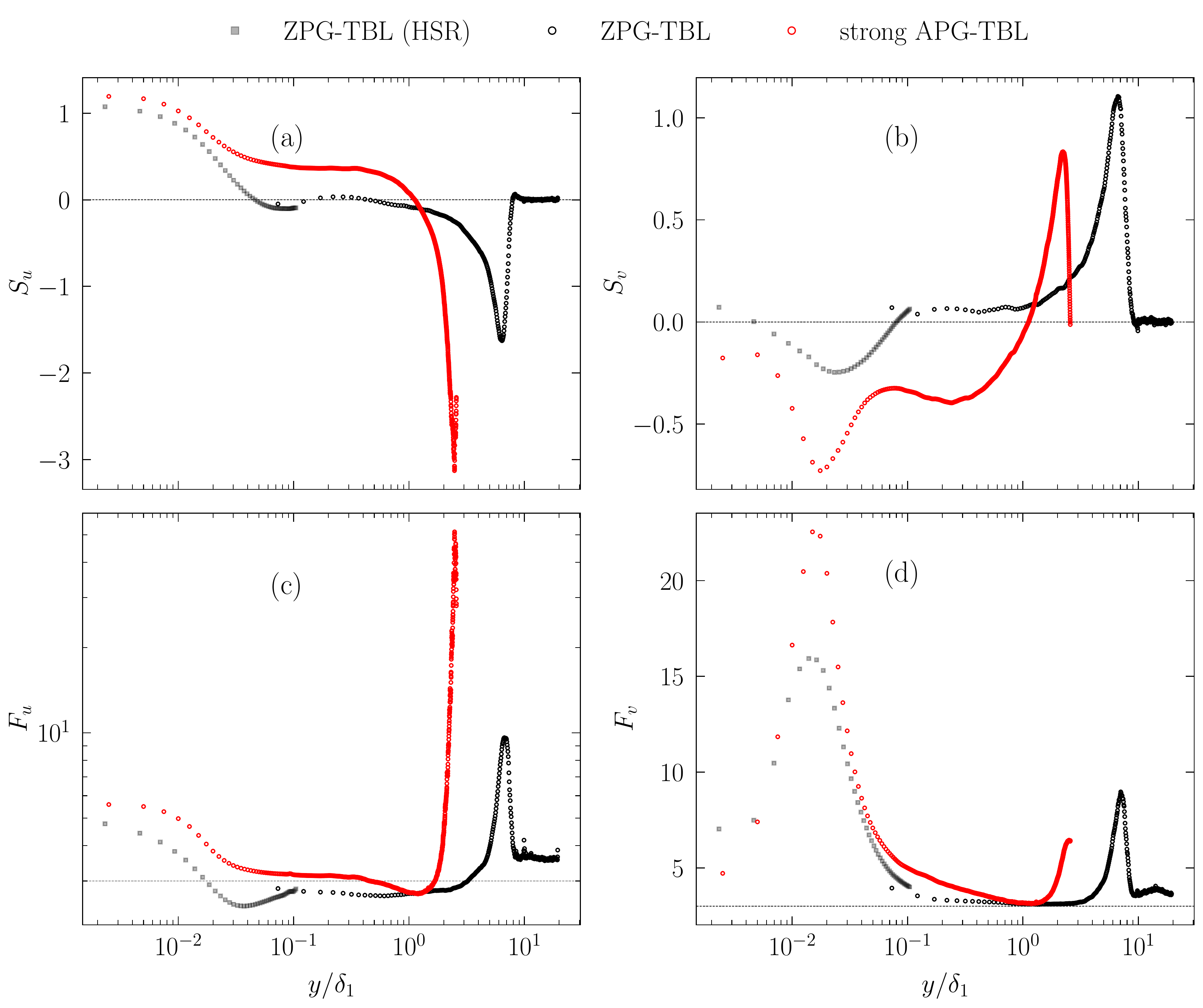} \end{overpic}
\end{center}
\vspace*{-0.2in}\caption{Skewness ($S_u$, $S_v$) and flatness ($F_u$, $F_v$) profiles for the ZPG-TBL and the strong APG-TBL.} 
\label{fig:skewness_flatness}
\end{figure*}

Skewness $S$ is a measure of the asymmetry of the probability density function (PDF) of a variable, whereas flatness $F$ represents the deviation of the PDF from the Gaussian distribution. For a variable $p$, these quantities are defined as:

\begin{equation}
S_p = \frac{\overline{p^3}}{(\overline{p^2})^{3/2}} \,\, , \,\,\,\,\,
F_p = \frac{\overline{p^4}}{(\overline{p^2})^2}
\end{equation}

$S_p = 0$ is expected for a symmetric PDF of $p$. In wall-bounded turbulence, velocity fluctuations ($u'$ and $v'$) often have $S = 0$ and $F = 3$, which represents a Gaussian distribution \citep{lin2007assessment}.

Figure \ref{fig:skewness_flatness} shows the skewness and flatness factors of the streamwise and wall-normal velocity fluctuations in the ZPG-TBL and the strong APG-TBL. These factors exhibit a plateau in the logarithmic region in both TBLs. In this region, the ZPG-TBL has $S_u \approx 0$ and $F_u \approx 3$ which shows that $u'$ has a nearly Gaussian distribution. $S_u > 0$ near the wall, while it has a negative peak in the outer layer. \textcolor{black}{This shows the asymmetric distributions of $u'$ in the near-wall and the outer layers. For $S_v$, this trend is the opposite, but the outer peak is not as strong as in $S_u$. In the ZPG-TBL, the first zero crossing in $S_u$ is around $0.04 \delta_1$ and in $S_v$, around $0.08\delta_1$.}

\begin{figure*}
\begin{center}
\begin{overpic}[width=0.8\textwidth]{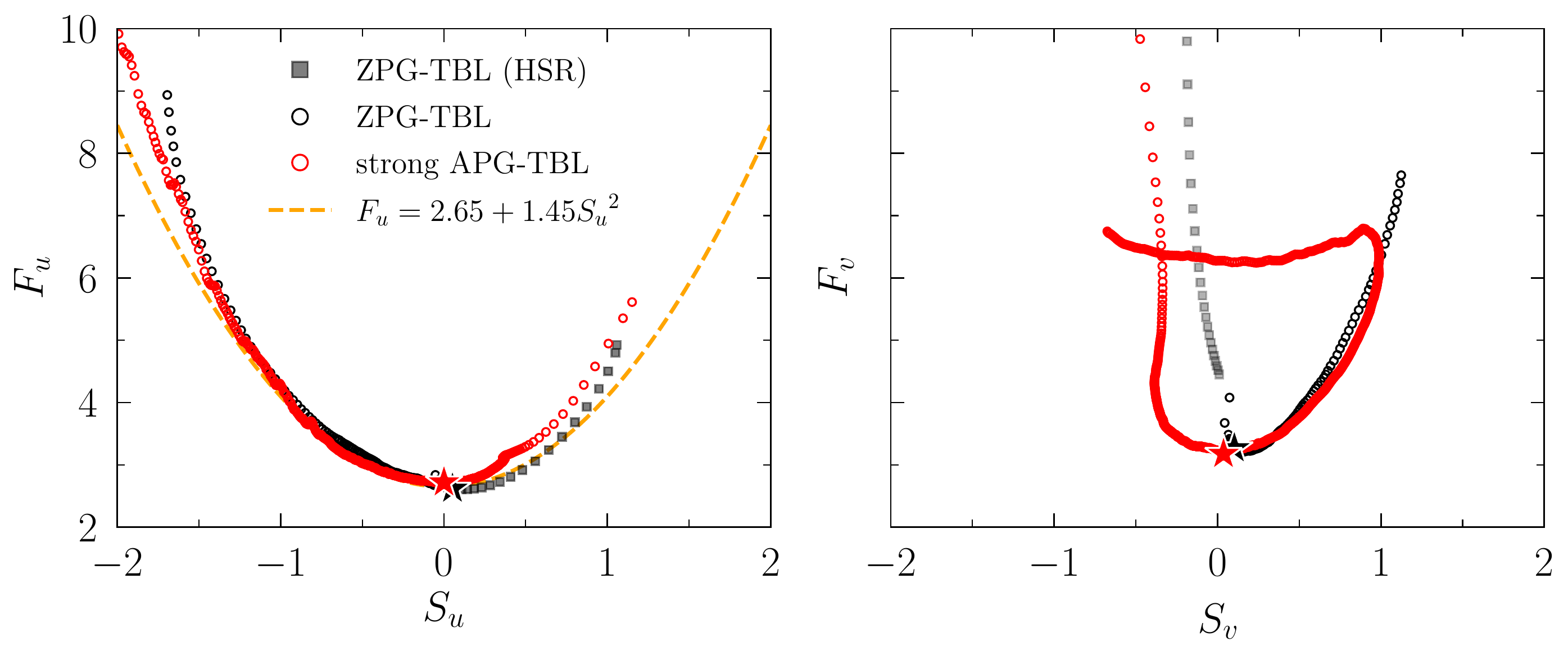} \put(10,12){(a)} \put(61,12){(b)} \end{overpic}
\end{center}
\vspace*{-0.2in}\caption{Skewness ($S_u$, $S_v$) vs. flatness ($F_u$, $F_v$) profiles of the ZPG-TBL and strong APG-TBL. The black and red stars show the locations of the maximum values in the corresponding Reynolds stress profiles of the ZPG-TBL and the strong APG-TBL, respectively. The dashed orange represents the formulation $F_u = 2.65 + 1.45 S_u^2$ that was reported by \citet{orlu2016high} as a curve fit to their $F_u$ vs. $S_u$ profile of a ZPG-TBL. 
\label{fig:skewness_vs_flatness}}
\end{figure*}

In the strong APG-TBL, the near-wall distribution of $S_u$ is similar but up to 15\% larger compared to the ZPG-TBL, but the negative peak in the outer layer is almost two times stronger. This means that in the outer region, the strong APG-TBL has streamwise fluctuations more skewed towards the smaller values than in the ZPG-TBL. The plateau in the logarithmic region is at $S_u\approx 0.4$ and the zero crossing is far away from the wall, in the outer region. The amplitude of $S_u$ in the logarithmic region increases with increasing APG, which pushes the zero-crossing away from the wall. The amplitude of $F_u$ is also observed to increase with APG. Similar distributions of the $S_u$ and $F_u$ profiles for the ZPG- and APG-TBLs are reported in \citet{mathis2015inner} and \citet{vila2020experimental}. This shows that the energy of large-scale motions in the log-layer increases with increasing APG.

\citet{hutchins2007large} showed that the large-scale structures of the outer region modulate the near-wall small scales when they superimpose on the near-wall region in turbulent boundary layers. \citet{mathis2009large} quantified this modulation by applying the Hilbert transform to the small-scale components of the streamwise velocity fluctuations. They found that the wall-normal profile of the amplitude modulation (AM) correlation coefficient between the large scales of the outer region and the envelope of the near-wall small scale has a strong resemblance to the skewness profile of $u'$. \citet{schlatter2010quantifying} and \citet{mathis2011relationship} assessed this relationship and showed that the Reynolds number trend in the skewness profiles of $u'$ is strongly related to the AM effect. Later, several studies used the dominant components of the skewness term in the study of the amplitude modulation of small scales by the large-scale structures \citep{bernardini2011inner,eitel2014simulation}. The skewness profiles in the current study demonstrate that the amplitude modulation of the near-wall small scales increases with increasing APG. This has also been indicated by \citet{mathis2009large,bernardini2011inner,eitel2014simulation}.

In the log-layer of the ZPG-TBL, $S_v \approx 0$ and $F_v \approx 3$ which shows that similar to $u'$, PDF of $v'$ also has a Gaussian distribution in this region. In the strong APG-TBL, $S_v$ has a predominantly negative distribution from the wall to the outer region. The outer layer peak in the strong APG-TBL case is found to be around $y=2.4\delta_1$ compared to $y=6.4\delta_1$ in the ZPG-TBL case. 

$F_v$ in the strong APG-TBL shows a larger inner peak than in the ZPG-TBL and a relatively smaller outer peak. In the log layer, $F_v$ in the strong APG-TBL is greater than in the ZPG-TBL. \citet{skaare1994turbulent} reported that $F_v$ closely follows $F_u$ in their strong APG-TBL ($\beta \approx 20)$, which does not agree with the current strong APG-TBL ($\beta \approx 30)$. $F_u$ is slightly below $F_v$ as also observed by \citet{skaare1994turbulent}, indicating that $u'$ has a narrower range compared to $v'$. However, in the wake region, the peak in $F_u$ is almost 10 times stronger than in $F_v$.

The correlation between the zero crossing of skewness, minimum flatness and the locations of the maxima in the corresponding Reynolds stress profiles can be visualised by plotting the flatness as a function of the skewness as shown in Figure \ref{fig:skewness_vs_flatness}. In $F_u$ vs. $S_u$, there is a good collapse between the ZPG-TBL and the strong APG-TBL for $S_u<0$, while for $S_u>0$ (near-wall region), the profiles are slightly distant from each other. In $F_v$ vs. $S_v$, the ZPG-TBL and the mild APG-TBL profiles collapse in the wake region, which is not the case near the wall. The dashed orange line is a representation of the parabola $F_u = 2.65 + 1.45 S_u^2$ which was reported in \citet{orlu2016high} as the best fit to their $F_u$ vs. $S_u$ profile for the ZPG-TBL. This also fits the current ZPG-TBL profile in the range of $-1<S_u<+1$. 

The black and red star markers show the locations of the maximum values in the corresponding Reynolds stress profiles for the ZPG-TBL and the strong APG-TBL, respectively. The collapse among the location of the peak in the Reynolds streamwise stress, and the locations of the zero skewness and the minimum flatness was previously observed by \citet{eitel2014simulation} for a ZPG-TBL and later confirmed for a ZPG-TBL by \citet{orlu2016high} and several mild APG-TBLs by \citet{vila2020experimental}. A similar collapse is observed in the current ZPG-TBL and strong APG-TBL for the streamwise velocity fluctuations at $S_u \sim 0$ and $F_{u_{min}} \sim 3$. \textcolor{black}{This shows that at the location of the outer peak in the Reynolds streamwise stress, the streamwise velocity fluctuations have more like a Gaussian distribution and are not skewed towards the larger or the smaller values. Interestingly, the wall-normal velocity fluctuations show a similar collapse among the locations of the peak in the Reynolds stress, the near-zero skewness and the minimum flatness.}

%=========================================================================
\section{Decomposition of mean skin friction}

In order to apply various flow control techniques to an APG-TBL flow to prevent or delay flow separation, the contributions of coherent structures to the skin friction distribution, and hence drag, need to be understood. \citet{fukagata2002contribution} proposed a method to decompose mean skin friction $C_f$ into three components (hereby referred to as FIK decomposition). 

The three components of the FIK decomposition are given in equation \ref{eq:FIK}.

\begin{equation}
\label{eq:FIK}
\begin{gathered}
C_{f_{FIK}} = \underbrace{ \frac{4(1-\delta_1/\delta_\Omega)}{Re_{\delta_1}} }_{C_{f_1}} \,\,\,+\,\,\,
		 \underbrace{ 4 \int_{0}^{1} \frac{- \overline{u' v'}}{U_2^2} \bigg(1 - \frac{y}{\delta_\Omega} \bigg)}_{C_{f_2}} \,\,\,+ \\
		 \underbrace{ 2 \int_{0}^{1} - \bigg(1 - \frac{y}{\delta_\Omega} \bigg)^2 \bigg(\frac{1}{\rho} \frac{\partial \overline{P}}{\partial x} + \overline{I_x} + \frac{\partial U}{\partial t} \bigg) d\bigg(\frac{y}{\delta_\Omega} \bigg)}_{C_{f_3}}
\end{gathered}
\end{equation}

\noindent where $\delta_\Omega$ represents the wall-normal distance at which the mean spanwise
vorticity $\Omega_z$ is 0.2\% of the mean vorticity at the wall and 
$$\overline{I_x} = \partial U^2 / \partial x \,\, + \partial(UV)/\partial y \,\, - \nu \partial^2 U/\partial x^2 \,\, + \partial(\overline{u'u'})/\partial x \, .$$

\begin{figure*}
\begin{center}
\includegraphics[width=0.8\textwidth]{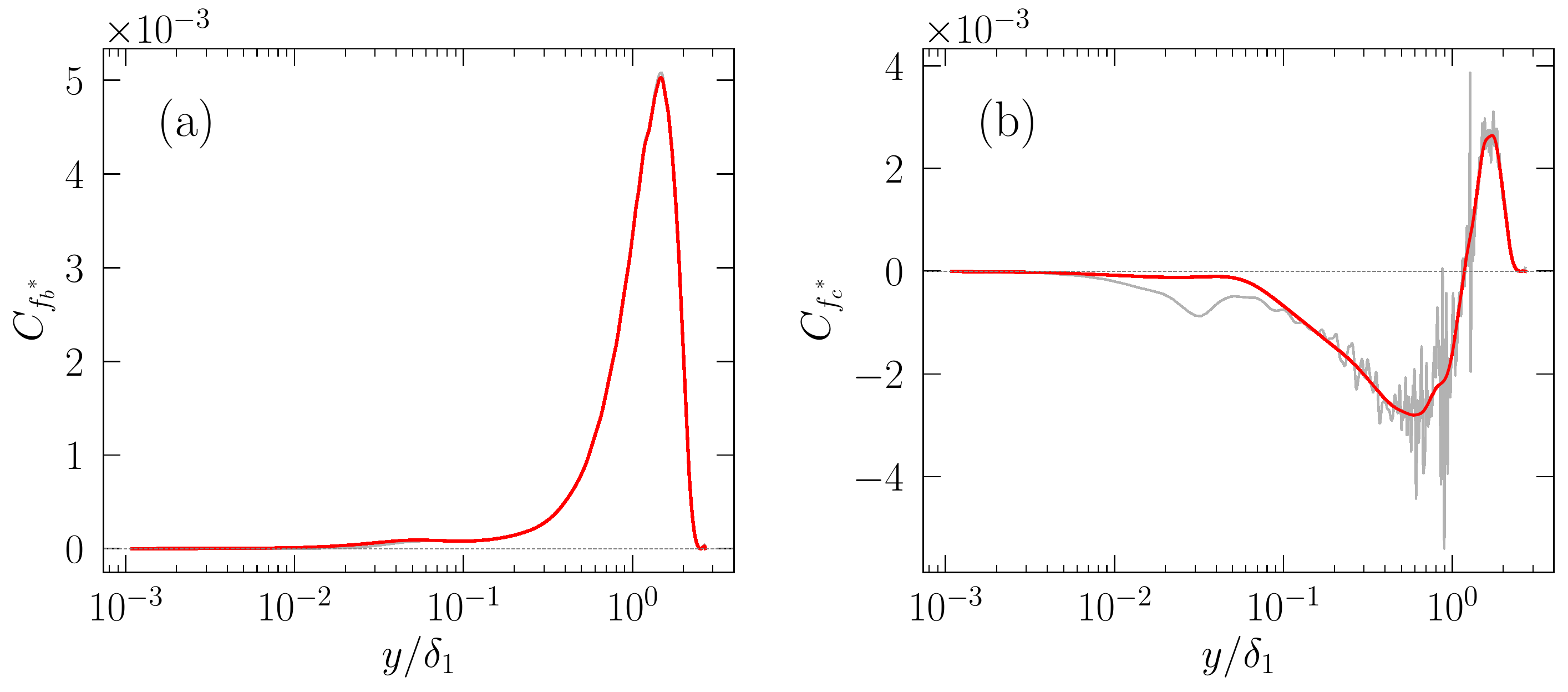}
\end{center}
\vspace*{-0.2in}\caption{Premultiplied integrands of (a) $C_{f_b}$ and (b) $C_{f_c}$. The red lines represent the profiles computed using the Gaussian filtered $\overline{u'v'}$ and $U$ in the wall-normal direction, whereas the grey lines represent the profiles computed without any filtering.
\label{fig:RD_decomposition_premultiplied_filter_check}}
\end{figure*}

The FIK decomposition is based on the contribution from the Reynolds shear stress where the stress is weighted by a linear ($1-y/\delta$) function. The linear nature of this function has no simple explanation in terms of physical processes \citep{renard2016theoretical}. The identity is obtained by a complex derivation which involves three integrations (by parts) of a momentum budget in a wall reference frame. In this decomposition method, the role of the wake region in skin friction generation is larger than the log-layer \citep{renard2016theoretical}.

More recently, a theoretical decomposition of mean skin friction has been devised by \citet{renard2016theoretical}. It is referred to here as the RD decomposition and its components as the RD components. The RD decomposition is based on the turbulent-kinetic-energy budget and is calculated by a single integration of the local energy budget over all wall-normal distances. This technique incorporates the turbulent fluctuations even from above the boundary layer edge ($y>\delta$) and suggests that $C_f$ at high Reynolds numbers is dominated by turbulent kinetic energy production in the outer region. This is consistent with the well-established importance of the logarithmic region. The flow control strategies based on the RD decomposition would control the turbulent kinetic energy rather than the Reynolds shear stress using the FIK decomposition \citep{renard2016theoretical}. The RD decomposition can provide further insights into the role of the logarithmic region in the generation of mean skin friction in high Reynolds number APG-TBL flows for flow control strategies that modify the turbulent dynamics in the logarithmic layer. 

The RD decomposition is presented as 

\begin{equation}
\begin{gathered}
C_{f_{RD}} = \underbrace{\frac{2}{U_e ^3} \int_{0}^{\infty} \nu \bigg(\frac{\partial U}{\partial y}\bigg) ^2 dy }_{C_{f_a}} \,\,\, + \,\,\,  
		\underbrace{\frac{2}{U_e ^3} \int_{0}^{\infty} - \overline{u' v'} \,\, \frac{\partial U}{\partial y} dy }_{C_{f_b}} \,\,\, + \,\,\, \\
		\underbrace{\frac{2}{U_e ^3} \int_{0}^{\infty} (U - U_e) \frac{\partial}{\partial y} \bigg( \frac{\tau}{\rho} \bigg) dy }_{C_{f_c}}
\end{gathered}
\end{equation}

\noindent where $$ \frac{\tau}{\rho} = \nu \frac{\partial U}{\partial y} - \overline{u' v'}. $$

$C_{f_b}$ is explicitly dependent on the turbulent fluctuations and is characterised by an integral of the positive turbulence production ($-\overline{u' v'} \,\, \partial U /\partial y$) normalised with the free-stream or edge velocity. $C_{f_a}$, $C_{f_b}$, and $C_{f_c}$ are also referred to as RD components. The premultiplied integrands of these contributions are

\begin{equation}
C_{{f_a}^*} = y \frac{2\nu}{U_e ^3} \bigg(\frac{\partial U}{\partial y}\bigg) ^2,  
\end{equation}

\begin{equation} 
C_{{f_b}^*} = y \frac{-2 \overline{ u' v' }}{U_e ^3} \frac{\partial U}{\partial y}, \text{ and} 
\end{equation}

\begin{equation}
C_{{f_c}^*} = y \frac{2 (U - U_e)}{U_e ^3} \frac{\partial}{\partial y} \bigg( \frac{\tau}{\rho} \bigg) dy .
\end{equation}

Regarding the application of the RD decomposition to the experimental measurements of a turbulent boundary layer, data noise is the major problem, especially in the calculation of $C_{f_c}$ where a single wall-normal differentiation of $\overline{u' v'}$ and a double differentiation of $U$ are involved. Figure \ref{fig:RD_decomposition_premultiplied_filter_check} presents the premultiplied integrands $C_{{f_b}^*}$ and $C_{{f_c}^*}$ for the strong APG-TBL, plotted against the wall-normal distance normalised by the displacement thickness, whereas the $C_{{f_c}^*}$ profile shown in grey is very noisy. This contrasts with the $C_{{f_b}^*}$ profile, which involves a single differentiation of $U$. In order to reduce the noise in the experimental measurements, especially with regards to the differentiation of the second-order statistics, it is proposed to use the Gaussian filtering in the wall-normal direction, which smoothens the $C_{{f_c}^*}$ profile to an extent where it can be used to calculate $C_{{f_c}}$. The red lines in figure \ref{fig:RD_decomposition_premultiplied_filter_check} represent the profiles computed after the Gaussian filtering of $\overline{u' v'}$ and $U$ using a kernel size of 20 viscous units ($\sim$ 21 data points) in the wall-normal direction. In order to assess the effect of Gaussian filtering on creating bias, the $C_{{f_b}^*}$ is also plotted using filtered $\overline{u' v'}$ and $U$. Since the red line collapses very well with the grey line, the filtering does not change the component that is measurable without filtering. Hence, it is deduced that filtering creates only a negligible bias in the calculation of $C_{{f_b}}$ and $C_{{f_c}}$ but rather reduces the measurement error by smoothing the $C_{{f_c}^*}$ profile.

\begin{figure*}
\begin{center}
\includegraphics[width=0.8\textwidth]{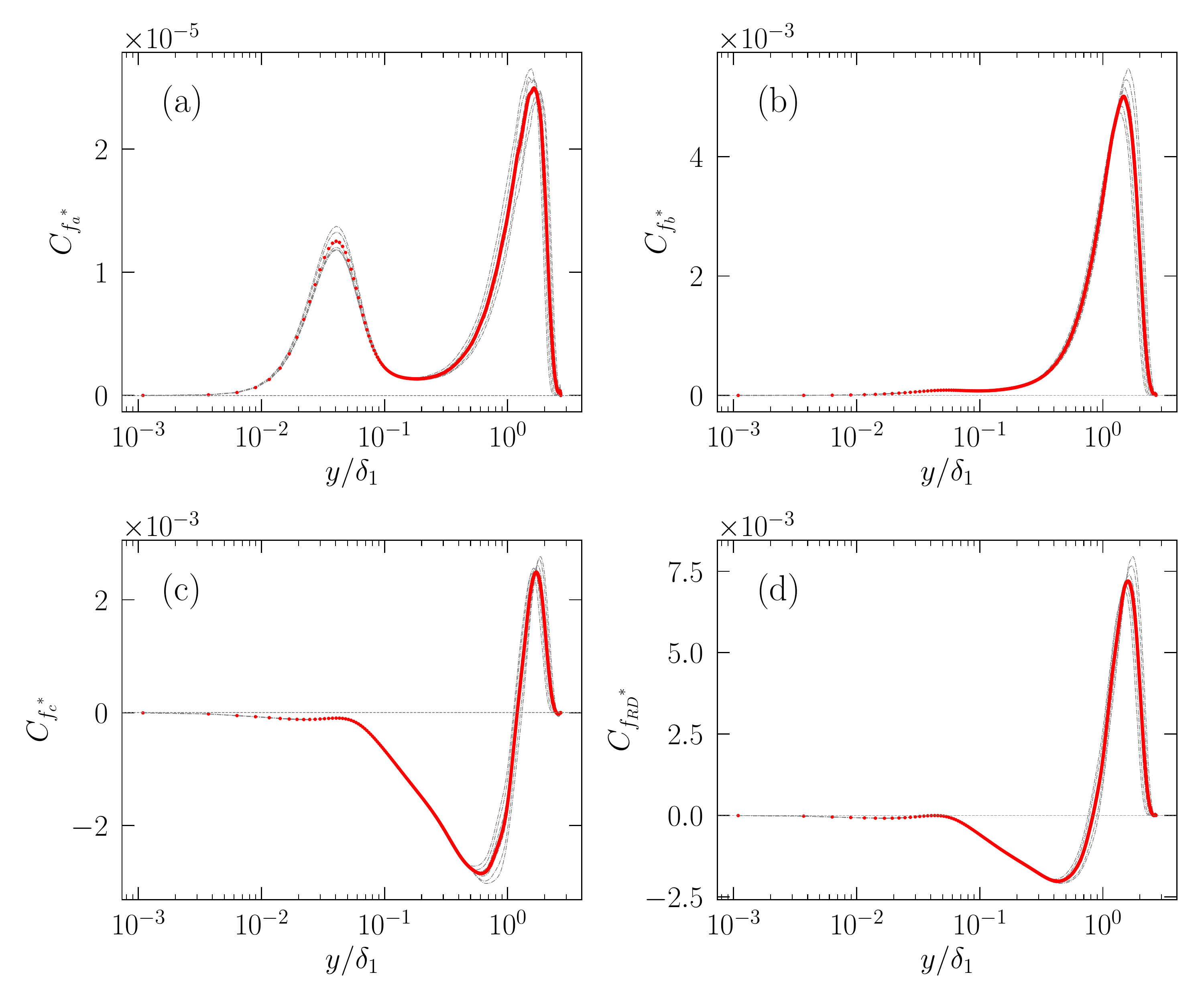}
\end{center}
\vspace*{-0.2in}\caption{Premultiplied integrands of the RD decomposition of mean skin friction. The grey lines in the background are at various equidistant locations along $x$ in the measured domain, while the red dots represent the profiles in the middle of the $x$ domain.
\label{fig:RD_decomposition_premultiplied}}
\end{figure*}

\begin{figure}
\begin{center}
\includegraphics[width=0.45\textwidth]{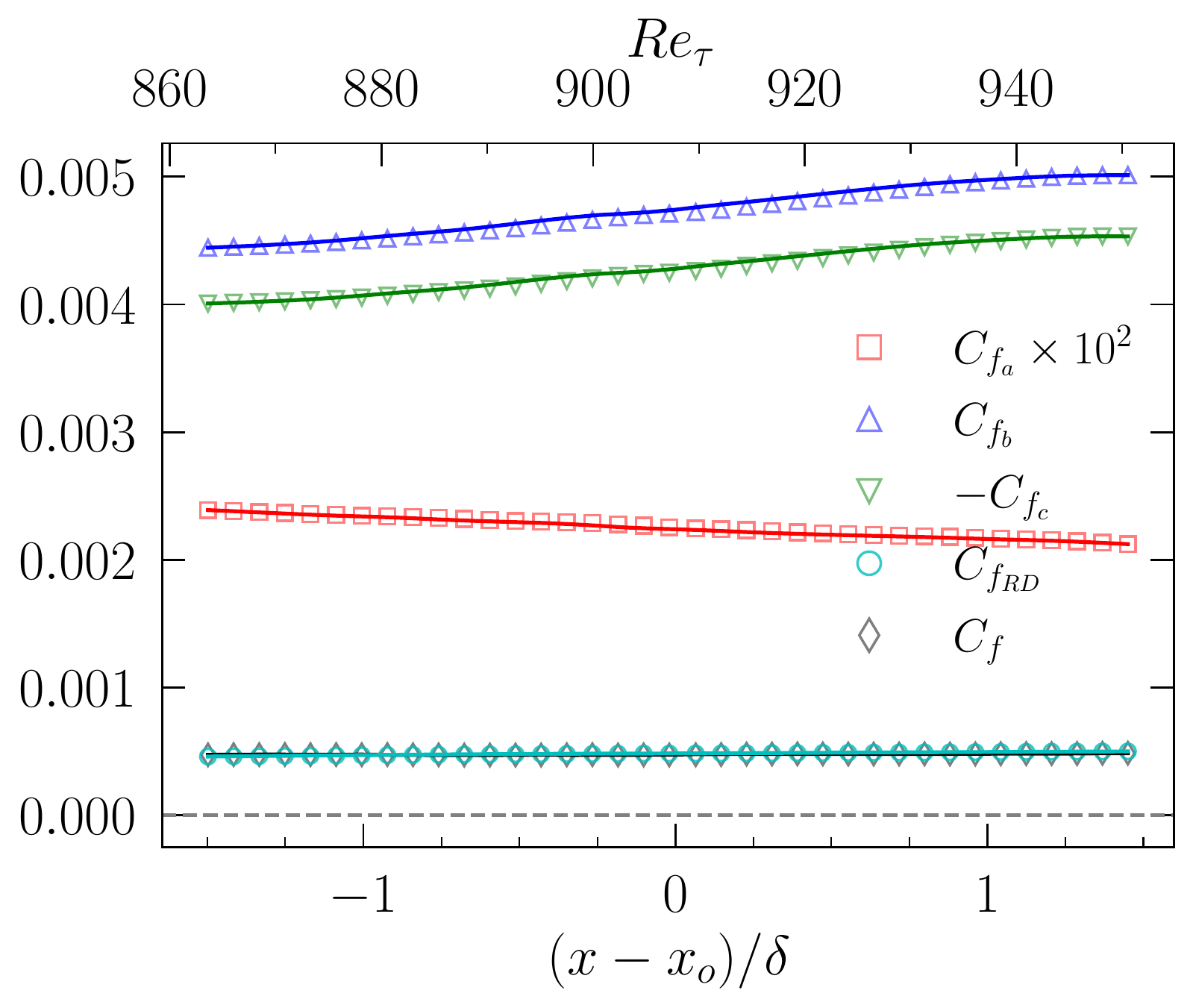}
\end{center}
\vspace*{-0.2in}\caption{RD decomposition of mean skin friction in the strong APG-TBL. The symbols represent the plotting against $(x-x_o)/\delta$, whereas the lines show the plotting against $Re_\tau$. 
\label{fig:RD_decomposition}}
\end{figure}

Figure \ref{fig:RD_decomposition_premultiplied} shows the premultiplied integrands of the RD components of $C_f$. The grey lines in the background are at various equidistant streamwise locations within the measured domain, while the red dots represent the profiles at the streamwise middle of the domain. \textcolor{black}{As can be seen, the decomposition using Gaussian filtering reduces the noise to a reasonable level. The wall-normal location of the outer peak in the $C_{{f_b}^*}$ profile and the zero-crossing of the $C_{{f_c}^*}$ profile coincide with the wall-normal location of the outer peak in the Reynolds shear stress profile. This is consistent with the observation of \citet{senthil2020analysis} for a strong APG-TBL at $\beta=39$. \citet{senthil2020analysis} also showed that $C_{f_b}$ is enhanced by 256\% whereas $C_{f_c}$ is reduced by 1,400\% when $\beta$ changes from 0 to 39. The values of $C_{f_b}$ and $C_{f_c}$ for the current strong APG-TBL ($\beta \approx 30$) are comparative to those reported by \citet{senthil2020analysis} for $\beta=39$ case. This demonstrates that the role of the outer layer is enhanced in skin friction generation when a TBL is imposed with a strong APG.}

Figure \ref{fig:RD_decomposition} represents the streamwise variation of the calculated RD components in the strong APG-TBL. $C_{f_a}$ decreases linearly with $x$ and $Re_\tau$ within the measured domain. $C_{f_b}$ and $-C_{f_c}$ increase almost linearly, while $C_{f_b}$ is slightly larger than $-C_{f_c}$, hence, the most of the effect of $C_{f_b}$ is balanced out by the negative values of $C_{f_c}$. \textcolor{black}{This shows that the positive contribution of the Reynolds shear stress to mean skin friction is reduced by the negative contribution of the strong APG. The streamwise average and the standard deviation of the RD components of $C_f$ are given in table \ref{tab:mean_SD_RD_components} represents. The values of these components are comparable to those of \citet{senthil2020analysis} who report the streamwise averaged values of $C_f = \num{0.000377}$, $C_{f_a} = \num{0.000064}$, $C_{f_b} = \num{0.004370}$, and $C_{f_v} = \num{-0.004057}$ for a strong self-similar APG-TBL ($\beta = 39$) of \citet{kitsios2017direct}. Since the current TBL has a weaker APG ($\beta \approx 30)$, its $C_f = \num{0.000482}$ is a little larger than the value reported by \citet{senthil2020analysis}.}

\begin{table}
\begin{center}
\caption{Streamwise average and standard deviation of the RD components of the $C_f$.}
\resizebox{\columnwidth}{!}{%
\begin{tabular}{lccc}
\hline \hline\noalign{\medskip}
 	&Mean $\times 10^3$ 	& Standard deviation $\times 10^3 $		& Mean/
Mean($C_{f_{RD}}$)		\\ 
\noalign{\smallskip}\hline\noalign{\medskip}

$C_{f_a}$		& 	0.0226	& 	 0.0008	& 	 0.0465 		 \\	
$C_{f_b}$		&	4.7259	& 	 0.1901	& 	 9.7830 		 \\	
$C_{f_c}$		&	$-4.2656$	& 	 0.1783	& 	 $-8.8298$ 	 \\	

\noalign{\smallskip}\hline\noalign{\medskip}
\label{tab:mean_SD_RD_components} 
\end{tabular}
}
\end{center}
\vspace{-0.2em}
\end{table}

%=======================================================================
\subsection{RD decomposition conditioned on the quadrant contributions}
\label{sec:RD_decomposition_quadrants}

\begin{figure*}
\begin{center}
\includegraphics[width=0.9\textwidth]{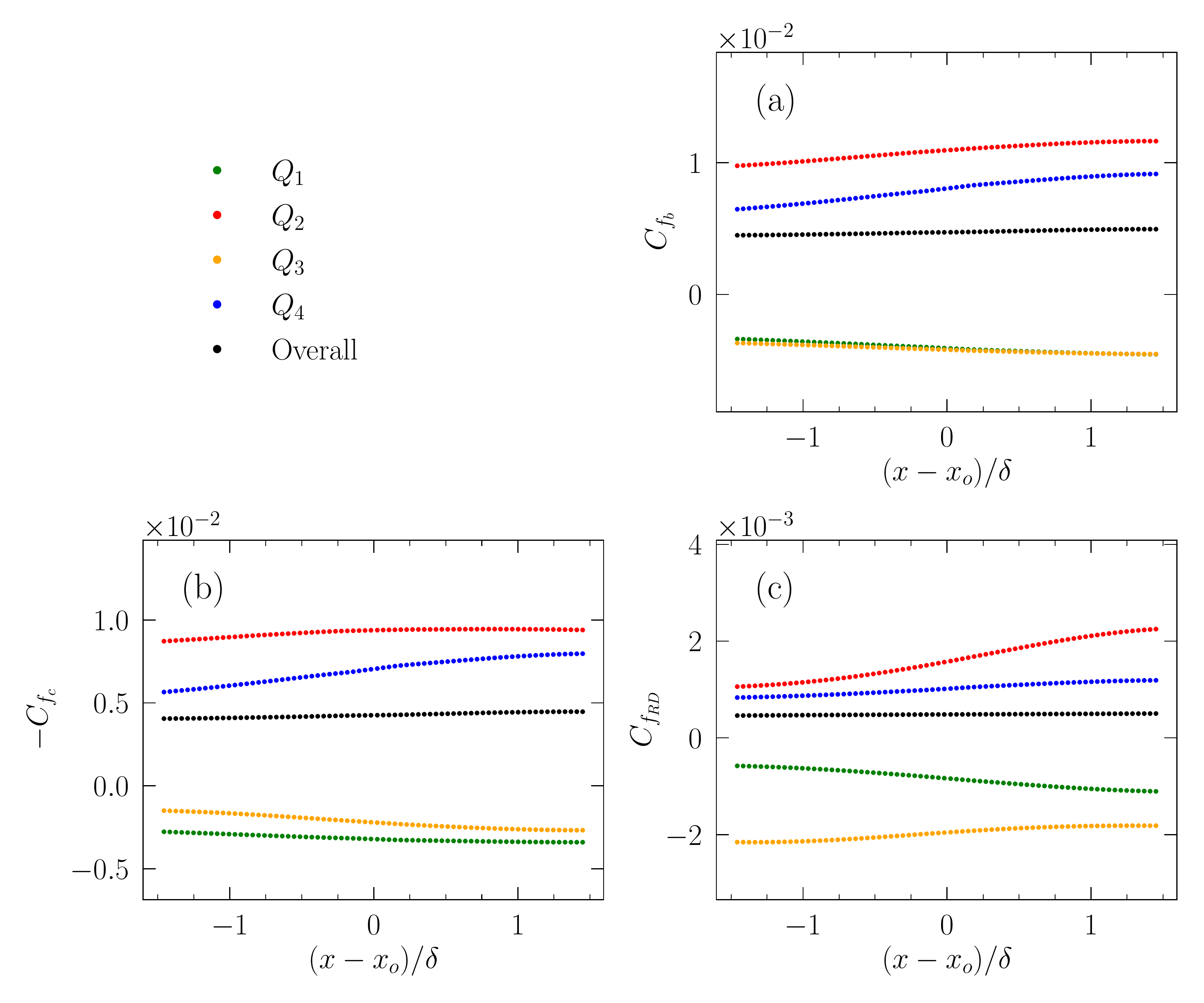}
\end{center}
\vspace*{-0.2in}\caption{RD decomposition of mean skin friction in the strong APG-TBL, conditional on the four quadrants. 
\label{fig:RD_decomposition_quadSplit}}
\end{figure*}

In order to assess the contributions of the four quadrants to the distribution of $C_{f_b}$, $C_{f_c}$, and $C_{f_{RD}}$, the contributions of the four quadrants to the Reynolds shear stress are used in the RD decomposition. The results for the strong APG-TBL are presented in figure \ref{fig:RD_decomposition_quadSplit}. Note that the same mean streamwise velocity is used in the conditional and unconditional RD decomposition. Therefore, $C_{f_a}$ for all quadrants, is the same as in figure \ref{fig:RD_decomposition}(a) and need not be presented again. It is observed that the Q2 and Q4 events are the positive contributors of $C_{f_b}$, $-C_{f_c}$ and $C_{f_{RD}}$. On the other hand, the contributions of Q1 and Q3 are always negative. Q1 and Q3 contribute almost equally to $C_{f_b}$ whereas Q3 contributes more to $-C_{f_c}$ than Q1. In the case of $C_{f_{RD}}$, Q1 contributes more than Q3. Overall, the positive and major contributions of Q2 and Q4 show that in a turbulent boundary layer, mean skin friction is mainly associated with the Q2 and Q4 events. This is consistent with the findings of \citet{chan2021skin} who studied the modification of the skin friction by a large-eddy breakup device in a ZPG-TBL. It is also in agreement with the observations of \citet{fuaad2019enhanced} whom, in their DNS study on drag reduction in a channel flow over superhydrophobic surfaces, conclude that a drop in ejections and sweeps reduces the turbulent contribution to the wall-shear-stress.

\begin{figure}[!t]
\begin{center}
\begin{overpic}[width=0.45\textwidth]{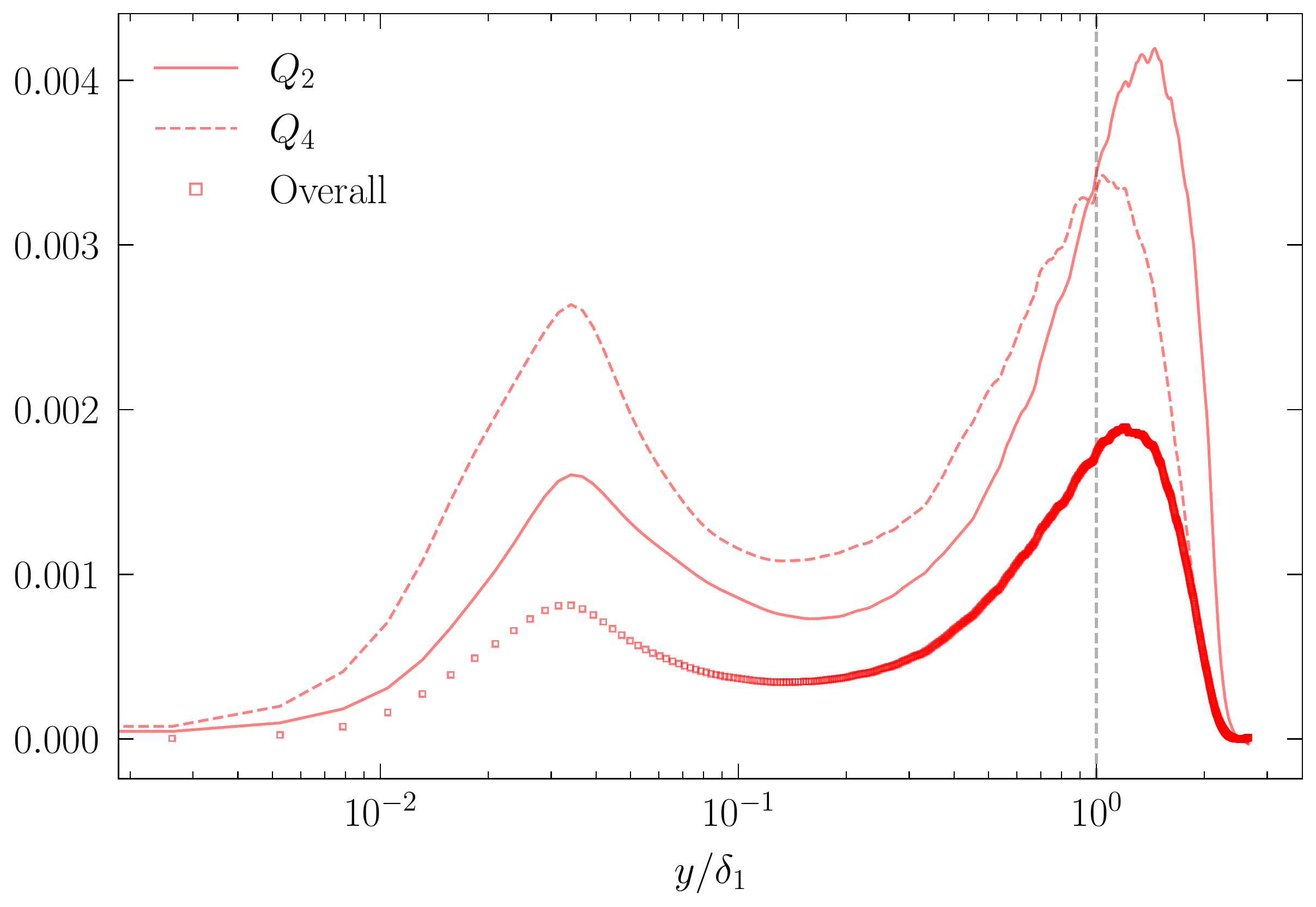} \end{overpic}
\end{center}
\vspace*{-0.2in}\caption{Profile of $-\overline{u^\prime v^\prime} \partial U/\partial y$ in the strong APG-TBL, conditioned on ejections and sweeps, compared with the unconditioned overall profile.
\label{fig:turbulence_production_quadSplit}}
\end{figure}

\textcolor{black} {Ejections (Q2 events) always contribute more to $C_{f_b}$, $-C_{f_c}$ and $C_{f_{RD}}$ than sweeps (Q4 events). Figure \ref{fig:turbulence_production_quadSplit} demonstrates that ejections have larger overall wall-normal contributions to the turbulence production term $-\overline{u^\prime v^\prime} \partial U/\partial y$ than sweeps. Since $C_f$ at high Reynolds numbers is dominated by the turbulence production within the logarithmic layer, the larger contribution of the ejections to the $C_f$ than the sweeps is expected.}

 %%%%%%%%%%%%%%%%%%%%%%%%%%%%%%%%%%%%%%%
 \section{Conclusion}

A self-similar APG-TBL flow is developed in the LTRAC wind tunnel and the flow velocity is measured using 2C-2D PIV with a field of view spanning $3.3\delta \times 1.1 \delta$ while achieving the wall-normal spatial resolution of less than the viscous length. The similarity coefficients $C_{uu}$, $C_{vv}$, $C_{uv}$, and $C_\nu$ for the current strong APG-TBL ($\beta \sim 30$) are not strictly constant along the $x$ direction, but their standard deviation is very small and comparable to those reported in \citet{kitsios2017direct} for a self-similar APG-TBL flow at $\beta=39$ and a Reynolds number comparable to the current case.  

It has been observed that the outer-layer statistics do not scale with viscous units in a strong APG-TBL, even if it is self-similar. Oppositely, in a self-similar strong APG-TBL, the profiles of the mean streamwise velocity and Reynolds stresses collapse when they are scaled with the outer variables, $U_e$ and $\delta_1$. The outer scaling also shows a remarkable collapse of the defect profiles, demonstrating the self-similar nature of the flow. This collapse is better than what is achieved using the local friction velocity and the defect thickness $\Delta = \delta_1 (2/C_f)^{1/2}$ \citep{skote1998direct} in various APG-TBLs \citep{lee2017large}. Near the wall, the defect velocity normalised with the local $U_e$ has larger values for larger $\beta$. The ZPG-TBL has the longest wall-normal defect region which decreases in its extent with increasing APG. 
 
When the streamwise and wall-normal velocity fluctuations are observed using the quadrant analysis of \citet{wallace1972wall}, it is found that the contributions of both Q1 and Q4 are amplified under APG and the increase in the Q4 contribution is larger than in the Q1. This demonstrates that when an APG is imposed, the outer peak in the Reynolds shear stress profile is formed due to the energisation of sweep motions of high-momentum fluid in the outer region of the boundary layer. 
 
The profiles of the dominant terms of turbulence production ($-\overline{u^\prime v^\prime} \partial U/\partial y$ and $\overline{{v^\prime}^2} \partial U/\partial y$) show weaker inner peaks and stronger outer peaks in the strong APG-TBL, instead of only inner peaks in the ZPG-TBL. This demonstrates that with a strong increase in APG, the turbulence production is shifted away from the near-wall region towards the outer region. For the strong APG-TBL, the wall-normal location of the outer peaks in the turbulence production terms matches the outer peak locations in the Reynolds shear stress profiles and the outer inflection point in the mean streamwise velocity profile. This location matches with the zero-crossings of profiles of the third-order moments $\overline{u'^3}$, $\overline{{u'}^2 v'}$, $\overline{{v'}^3}$, and $-\overline{u' {v'}^2}$ in the strong APG-TBL. The outer positive peaks in the profiles of $\overline{{u'}^2 v'}$, $\overline{{v'}^3}$, and $-\overline{u' {v'}^2}$ demonstrate that a decrease in APG results in the reduced energy diffusion towards the boundary layer edge from the locations of the outer peaks in the Reynolds stress profiles, and this phenomenon is shifted closer to the boundary layer edge.

The profiles of the skewness and flatness factors show that the streamwise and wall-normal velocity fluctuations have a nearly Gaussian distribution in the log-layer of the ZPG-TBL while having asymmetric distributions in the near-wall and the wake regions. In the near-wall region, the streamwise velocity fluctuations are skewed toward the larger values, while the wall-normal velocity fluctuations are skewed towards the smaller values. This trend is reversed in the wake region. In the strong APG-TBL, however, the log region shows skewness to larger values of streamwise velocity fluctuations and the smaller values of the wall-normal fluctuations and the zero-crossings of the skewness profiles move further from the wall. The further zero-crossings of the skewness profiles from the wall show that the amplitude modulation of the near-wall small scales increases with increasing APG, according to the interrelation of the skewness of $u'$ and the amplitude modulation coefficient demonstrated by \citet{schlatter2010quantifying} and \citet{mathis2011relationship}. The extent of asymmetry in the near-wall and wake regions increases with the increasing APG. Furthermore, in the strong APG-TBL, the locations of the zero-crossings in the skewness profiles match the locations of the maxima in the profiles of the Reynolds stress and turbulence production. This demonstrates that these locations have symmetric distributions of streamwise and wall-normal velocity fluctuations. 

The locations of the maximum Reynolds streamwise stress, the zero skewness and the minimum flatness collapse for the ZPG-TBL which has been reported in several previous studies. The current study also observes the same collapse in the strong APG-TBL. Interestingly, the wall-normal velocity fluctuations also exhibit the collapse among the locations of the peak in the Reynolds stress, the near-zero skewness and the minimum flatness in the ZPG-TBL and the strong APG-TBL.

With regards to the application of the identity of \citet{renard2016theoretical} in the decomposition of mean skin friction into three components (referred to as the RD components) in TBLs measured using 2C-2D PIV, it is shown that the noise in the computation of the third component, which involves the second derivative of the $U$ and the first derivative $-\overline{u'v'}$ is considerably reduced by Gaussian filtering of 21 viscous units in the wall-normal direction. It is observed that the wall-normal locations of the outer peak in the profiles of the premultiplied integrands of $C_{{f_b}}$ and $C_{{f_c}}$ coincide with the wall-normal location of the outer peak in the Reynolds shear stress profile in the strong APG-TBL. This is consistent with the observations of \citet{senthil2020analysis} for a strong APG-TBL at $\beta=39$. 

The contribution of the turbulent kinetic energy in skin-friction generation (represented by $C_{{f_b}}$) is enhanced with increasing APG because of the emergence of the outer peaks and the diminishing of the inner peaks in the turbulence production profiles. This is matched by the enhanced negative contribution due to APG (represented by $C_{{f_c}}$), and the overall $C_f$ decreases with an increase in APG. In the strong APG-TBL, $C_{f_a}$ decreases whereas the $C_{{f_b}}$ and $-C_{{f_c}}$ increase almost linearly in the streamwise direction. 
The findings are in agreement with the observations of \citet{senthil2020analysis} who performed a similar analysis using the DNS data of a strong APG-TBL ($\beta=39$) of \citet{kitsios2017direct}.

Using the quadrant analysis of \citet{wallace1972wall}, ejections are always shown to contribute more than sweeps, to the $C_{f_b}$, $-C_{f_c}$ and the overall $C_{f}$ of the RD decomposition. Since $C_f$ at high Reynolds numbers is dominated by turbulence production in the outer region and ejections have a larger overall contribution to turbulence production than sweeps, the larger contribution of ejections to $C_f$ is expected.

The methodology of \citet{skaare1994turbulent} is found good at bringing a boundary layer to the verge of separation and then relaxing the APG in order to make the flow self-similar. In future, such boundary layer flow could be produced with a similar methodology in a larger wind tunnel facility, where the flow is allowed to develop over a longer streamwise range and the second-order statistics are converged. The convergence of the Reynolds stresses' profiles at the location of the outer peaks in the strong APG-TBL may be checked as a function of the number of the statistically independent velocity fields as well as the streamwise distance from the trip wire, within the self-similar region.

A thorough analysis of the importance of displacement thickness height in the boundary layers could is imperative. The following quantities show similar values around the displacement thickness height. 

\begin{itemize}
	\item The outer-scaled Reynolds streamwise and shear stresses in the three TBLs.
	\item The relative contributions of ejections and sweeps to the Reynolds shear stress and the turbulence production term $-\overline{u'v'} \partial U/\partial y$ in the mild and strong APG-TBLs.
	\item The outer-scaled third-order moments $\overline{{u'}^3}$, $\overline{{u'}^2 v'}$, $\overline{{v'}^3}$, and $-\overline{u' {v'}^2}$ in the three TBLs. 
	\item The skewness and flatness profiles. 	
\end{itemize}

When APG increases, the outer peaks in the Reynolds stress profiles move from the outer region to a vicinity closer to the displacement thickness height. The reason for this shift in turbulent activity is yet to be investigated.

%%%%%%%%%%%%%%%%%%%%%%%%%%%%%%%%%%%%%%%
 \section*{Acknowledgements}
The authors would like to acknowledge the support of the Australian Government of this research through an Australian Research Council Discovery grant. Muhammad Shehzad acknowledges the Punjab Educational Endowment Fund
(PEEF), Punjab, Pakistan for funding his PhD research.  The research was also benefited from computational resources provided by the Pawsey Supercomputing Centre and the NCI Facility both supported by the Australian Government with access provided via a NCMAS grant.

%%%%%%%%%%%%%%%%%%%%%%%%%%%%%%%%%%%%%%%
\bibliography{ref.bib}% Produces the bibliography via BibTeX.

\end{document}